# One-third of Sun-like stars are born with misaligned planet-forming disks

Lauren I. Biddle[1], Brendan P. Bowler[2], Marvin Morgan[1,2], Quang H. Tran[3], Ya-Lin Wu[4]

[1*]Department of Astronomy, The University of Texas at Austin, 2515 Speedway, Stop C1400, Austin, 78751, TX, USA.
[2]Department of Physics, University of California Santa Barbara, Broida Hall, University of California, Santa Barbara, 93106, CA, USA.
[3]Department of Astronomy, Yale University, 266 Whitney Avenue, New Haven, 06511, CT, USA.
[4]Department of Physics, National Taiwan Normal University, Taipei, Taiwan.

[*]Corresponding author. E-mail: lbiddle@utexas.edu

**Exoplanets are organized in a broad array of orbital configurations[1,2] that reflect their formation along with billions of years of dynamical processing through gravitational interactions[3]. This history is encoded in the angular momentum architecture of planetary systems—the relation between the rotational properties of the central star and orbital geometry of planets. A primary observable is the alignment (or misalignment) between the rotational axis of the star and the orbital plane of its planets, known as stellar obliquity. Hundreds of spin-orbit constraints have been measured for giant planets close to their host stars[4], many of which have revealed planets on misaligned orbits. A leading question that has emerged is whether stellar obliquity originates primarily from gravitational interactions with other planets or distant stars in the same system, or if it is 'primordial'—imprinted during the star-formation process. We present a comprehensive assessment of primordial obliquities between the spin axes of young, isolated Sun-like stars and the orientation of the outer regions of their protoplanetary disks. The majority of systems are consistent with angular momentum alignment, but about one third of isolated young systems exhibit primordial misalignment. This suggests that some obliquities identified in planetary systems at older ages—including the Sun's modest misalignment with planets in the Solar system—could originate from initial conditions of their formation.**

It is long established that planets form out of protoplanetary disks of gas and dust[5], but there have been limited attempts so far to directly measure primordial stellar obliquities. The largest effort[6] investigated the orientations of only seven single Sun-like stars; however, the resulting obliquity constraints were broad, preventing robust conclusions about the underlying obliquity distribution. Also, nine other stars in that sample had stellar companions that could affect the inferred stellar obliquities in those systems. Other studies[7–9] have attempted to provide insights into the origin of spin–orbit obliquity by measuring stellar orientations with respect to debris disks—second-generation disks of dust around stars with intermediate ages (approximately 10–500 Myr). These works found general agreement with alignment, although the combined sample amounts to only 21 single Sun-like stars.

Directly measuring primordial obliquities for a sufficiently large sample of young stars is the natural next step, which has only recently become possible with the sensitivity and angular resolution of the Atacama Large Millimeter/submillimeter Array (ALMA) to measure disk inclinations coupled with nearly all-sky precision light curves from space-based facilities such as the Transiting Exoplanet Survey Satellite (TESS) and K2 to determine stellar inclinations. Here we carry out a statistical investigation into primordial stellar obliquity of single young stars with respect to their outer protoplanetary disks to form a more complete picture of the initial angular momentum architecture of protoplanetary systems. Our approach compares the inclination of the stellar spin axis with disk inclination measurements from resolved millimeter observations sensitive to the dust in the outer disk (≥10 au) to obtain a measurement of the minimum star–disk obliquity for a sample of 49 systems.

**Sample Description**

Our sample consists of T Tauri stars with spatially resolved protoplanetary disks and well-characterized disk inclinations ($i_{\rm disk}$), stellar radii ($R_*$), projected rotation velocities ($v\sin i_*$), and rotation periods ($P_{\rm rot}$). Stellar inclinations are determined by merging projected rotational velocities with stellar radii and rotation periods. With our adopted measurements of $R_*$, $v\sin i_*$, and $P_{\rm rot}$, we determine the stellar inclination ($i_*$) following a Bayesian probabilistic framework[10], which properly accounts for the correlation between $v\sin i_*$ and the equatorial rotation velocity ($v_{\rm eq}$, where $v_{\rm eq} = 2\pi R_*/P_{\rm rot}$). In particular, we use the analytical expression for the posterior distribution of $i_*$[11] (see Methods). We also compile spectral types, masses ($M_*$), and effective temperatures ($T_{\rm eff}$) to more broadly assess potential trends in the resulting obliquity distribution. In selecting our sample, we reject stars with spectral types earlier than F6 to avoid pulsating variables[12] with periodic brightness signatures that are difficult to distinguish from rotation[13]. The majority of our sample consists of stars with spectral types between K0 and M5 with masses spanning approximately 0.5–1.5 $M_\odot$. Last, we focus our investigation on gravitationally isolated systems and exclude those with known binary companions, which can induce torques on protoplanetary disks and erase initial conditions[14].

The true star–disk obliquity ($\Psi$) is the angle between the star's rotational angular momentum vector and the angular momentum vector of the disk. For this work, this angle cannot be determined in full because the spatial orientation of the star's rotational axis (the polar position angle) and spin direction (clockwise or counterclockwise) are unknown. Similarly, spatially resolved kinematic measurements of the disk and information about its orientation (the longitude of ascending versus descending nodes)—which is needed to identify whether its angular momentum vector is angled toward or away from Earth—are not always available. We therefore report the absolute difference between the sky-projected inclination of the star and disk ($\Delta i$), which corresponds to a lower limit on the true obliquity angle, $\Psi$.

**Stellar Obliquities in Planet-Forming Systems**

The majority of systems in the sample show a tendency for low minimum obliquities, but there is also a spread in $\Delta i$ out to about 60° (Fig. 1). We found no signs that this spread significantly correlates with spectral type, $M_*$, $T_{\rm eff}$, $R_*$, $P_{\rm rot}$, $v\sin i_*$, or $i_*$. The average $\Delta i$ of the sample is 17°. It is apparent that a notable portion of stars are misaligned with respect to their protoplanetary disks, at least at the characteristic spatial scales sampled by ALMA observations of tens to hundreds of au. We classify a star as misaligned if the maximum *a posteriori* (MAP) value of its $\Delta i$ probability distribution is at least two times the lower ('–') $\Delta i$ uncertainty, corresponding to a departure from 0° at 95.4% confidence. In the sample, we find a misalignment fraction of 16/49, corresponding to a primordial misalignment rate of $33^{+7}_{-6}$%. The remaining 33/49 systems show no significant evidence of misalignment (see Supplementary Discussion).

Next, we assess the behavior of the distribution of $\Delta i$ at a population level with two approaches. Our first approach is the computation of a fixed-width Gaussian kernel density estimate (KDE) of $\Delta i$ to characterize the overall shape of the distribution. The KDE is a non-parametric model constructed from individual $\Delta i$ probability distributions in the sample. The resulting KDE shows a broad spread in minimum stellar obliquity, peaking between 0° and 15° and tapering off beyond 60°. We also model the underlying distribution of $\Delta i$ using a hierarchical Bayesian Modeling (HBM) approach. HBM is a method to reconstruct the population-level behavior of a sample by constraining the hyperparameters of a parametric model based on constraints for many individual objects. This method is well suited to model the underlying distribution in $\Delta i$ given the marked variation among individual probability distributions of $\Delta i$ across the sample. Here we explore three population-level models to represent the underlying distribution in $\Delta i$: a Rayleigh distribution, a Gaussian distribution, and a Truncated Gaussian distribution. The Rayleigh and Gaussian best-fit models of the underlying distribution of $\Delta i$ yield similar results. The best-fit Truncated Gaussian model shows more concentrated power toward lower obliquities. Together, the model fits of the $\Delta i$ distribution indicate that it is broad with a characteristic peak value between about 10° and 25°.

The primary source of potential systematics in this study arises from the possible overestimation of stellar rotation periods and underestimation of stellar radii. In particular, unknown systematic errors in estimates of $P_{\rm rot}$ or $R_*$ could bias the parameter inferences of $i_*$, which, in turn, could propagate to the resulting distribution of $\Delta i$. To assess whether any unanticipated biases in $P_{\rm rot}$ or $R_*$ could impact the conclusions from this study, we conduct a series of tests in which $i_*$ is computed with amended values of $R_*$ and $P_{\rm rot}$ that represent multiplicative corrections to the fiducial values. A total of 25 individual tests were performed for all combinations of –30%, –15%, 0%, +15%, and +30% offsets applied to each of the nominal values of $R_*$ and $P_{\rm rot}$, resulting in a new distribution of $\Delta i$ for each permutation. The results of each test are displayed in Figs. 2 and 3. What we learn from these tests is that even if there are significant deviations from the nominal values of $R_*$ and $P_{\rm rot}$, the effect on $i_*$ yields a broad distribution in $\Delta i$ whereby the misalignment fraction remains substantial. From this test, we also find that the misalignment probability (Fig. 3) is higher in grid cells with large $P_{\rm rot}$ and low $R_*$, which both act to increase $i_*$. This enhancement in misalignment probability is likely the result of the increasing frequency of equator-on stars for certain combinations of $P_{\rm rot}$ and $R_*$, which also appear to show a greater number of non-physical $v_{\rm eq}$; so, although misaligned obliquities in these cells appear more probable, they are not likely to be realistic. The conclusion from this work nevertheless remains: most young stars are consistent with being aligned with their planet-forming disks, and modest misalignment in primordial stellar obliquity is also common.

**Causes and Consequences of Misalignment**

This sample consists of single stars that reside in low-density environments[16], so stellar flybys that could torque the outer disk are unlikely to induce misalignment. The isolated nature of the stars in our sample means that it represents a population of effectively pristine systems[17]. More likely scenarios from which misalignment could have occurred in these systems include chaotic processes during cloud core collapse[18,19] as well as late-stage gas accretion onto young disks from the surrounding envelope through large-scale streamers, which may impart additional angular momentum onto the outer disk, inducing misalignment in the outer region[20,21]. Such streamers have primarily been detected feeding Class I systems[22], although a few recent discoveries demonstrate that this process can occur as late as the Class II evolutionary stage[23,24].

The scale of obliquity misalignments observed here can be generated by several possible scenarios. For example, hydrodynamic simulations of cloud core collapse indicate that obliquity angles reach about 40° on average, but can be as high as 80° as a result of turbulence in the formation environment[18]. Giant planets on moderately inclined orbits can tilt the inner disk, usually by a few degrees[25]. However, sufficiently massive planets can disconnect the inner and outer disks, and in some cases, broken inner-outer disks can induce considerable obliquity misalignments spanning approximately 20°–150° by torquing the host star and reorienting its spin axis[26]. While the timescale for star-disk torquing agrees with stellar ages in our sample (a few Myr), there does not appear to be a trend in morphological characteristics in the disk sample with respect to misalignment. Late accretion infall also has the potential to affect the orientation of outer protoplanetary disks with misalignment angles up to about 80° (ref. 27).

A consequence of misalignment is that planets on wide orbits could form with very inclined orbits[18]. If these disks maintain the same inclinations at smaller separations of 1–10 au where most giant planets form[28], then the outer disk orientations may offer clues about primordial alignment of gas giants. In this context, it is noteworthy that the distribution in star-outer disk $\Delta i$ appears generally similar to the distributions of stellar spin-orbit obliquities observed in transiting hot- and warm-Jupiters around Sun-like stars (Fig. 1). Although the majority of these systems appear to be aligned[15,29], a significant fraction of obliquity misalignments could therefore be primordial in nature. In this work, the fraction of systems showing moderate misalignments (16/49) represents the lower limit of the true obliquity angle, $\Psi$ (ref. 30). It is therefore possible that some systems could have extreme star-disk misalignments much higher than the projected $\Delta i$ value and may be connected to hot Jupiters with polar or retrograde orbits[31]. Without complete knowledge of the spin-orbit angular momentum geometry of the star and disk, the prevalence of primordial retrograde disks cannot be addressed with $\Delta i$ alone (although the latter configurations are expected to be unlikely outcomes of single-star formation[32]). In addition, primordial obliquity has the potential to explain the misalignments of some co-planar multiplanet systems (such as Kepler-30 (ref. 15)), which do not have a known outer planetary or stellar companion that could be responsible for exciting the inclination of the inner system.

The Solar System has a slight but unambiguous misalignment of about 6° relative to the Sun's spin axis[33,34]. This feature is even more appreciable when considering the flatness of the Solar System; the mutual inclinations of the gas and ice giants is 0.3° on spatial scales extending out to 30 au, and this flatness continues out to the dynamically cold Kuiper Belt at nearly twice this distance. The primordial stellar obliquity distribution found in this work may nevertheless provide context for the Sun's obliquity. If the current Solar obliquity reflects its primordial obliquity, we can more precisely place it in context with the star-disk systems in this sample by estimating its spectral type at a comparably young age (about 5 Myr). Based on evolutionary models[35], a 1 $M_{\odot}$ star at 5 Myr has an effective temperature of about 3800 K. Comparing this effective temperature with the spectral type calibration appropriate for pre-main sequence stars[36], we estimate the pre-main sequence solar spectral type to be approximately K9. Placed into context, the Sun's obliquity in its pre-main sequence phase is in agreement with the majority of the sample, which shows a preference for minimum obliquities within about10° (the typical precision of $\Delta i$). Considering the relatively common occurrence of more severe misalignments in protoplanetary disk systems, the slight (yet precise) misalignment angle of the Sun is not at all odd. Although the origin of the Solar obliquity is still a topic of debate[37–39], these results may indicate that a slightly inclined planet-forming disk is a natural explanation for the observed Solar misalignment without the need to invoke post-formation torques.

**Considering Inner-Outer Disk Misalignment**

Several recent studies have investigated the relative alignment of the inner and outer regions of protoplanetary disks. For instance, inner disk misalignment has been inferred from the existence of 'dipper' stars (a common class of young variable stars which undergo strong dips in brightness) that host a protoplanetary disk. A population-level analysis of the outer disk inclinations among dipper stars[40] found that the dipping phenomenon is probably unrelated to the outer disk, suggesting further that inner-outer disk misalignments may be common. Supporting evidence of inner-outer disk misalignments has also emerged in observations of shadows cast onto outer disks as a result of an inclined inner disk that blocks light from the central star[41]. Additionally, scattered-light observations of inner dust disks at sub-au separations have shown direct evidence of inner-outer disk misalignment. Recent observations with VLTI/GRAVITY identified 6 out of 11 systems with significant inner-outer disk misalignment[42]. A handful of systems in our sample have shown evidence of inner-outer disk misalignment in the form of dips, shadowing, or scattered light imaging; however, the sample size of systems with complete stellar, inner disk, and outer disk inclination measurements is limited and prevents a broader analysis of this broader angular momentum architecture (see Supplementary Discussion).

**The 'Primordial vs. Post-Formation' Question**

Direct imaging of exoplanets is sensitive to separations spanning tens to hundreds of au[43], and recently, minimum obliquities of this population have been explored by comparing the stellar inclination to planetary orbital planes traced out with patient orbit monitoring[11,44]. This emerging population consists of eleven planets across six systems[44–49], which show a trend toward alignment with the central star. As more planets are discovered at wide separations with direct imaging across a broad range of ages, a direct comparison can be made between the primordial and long-period planet obliquity distributions to determine the role of dynamics in exciting orbital inclinations over time.

The fundamental differences across space and time between the protoplanetary disk sample and planet samples complicate the question of whether the primordial obliquities of Sun-like stars are the primary basis for the obliquity distribution of mature short-period giant planets. This gap in parameter space, the possibility of inner-outer disk misalignments, and uncertainty regarding the dynamical evolutionary pathways following the epoch of planet formation prevent firm conclusions on this topic. However, this gulf in age and orbital distance can soon be addressed with the Gaia astrometric survey's fourth data release, which is expected to yield thousands of 3D Keplerian orbit fits of cold Jupiters—including their inclinations—at orbital distances up to about 7 au[50].

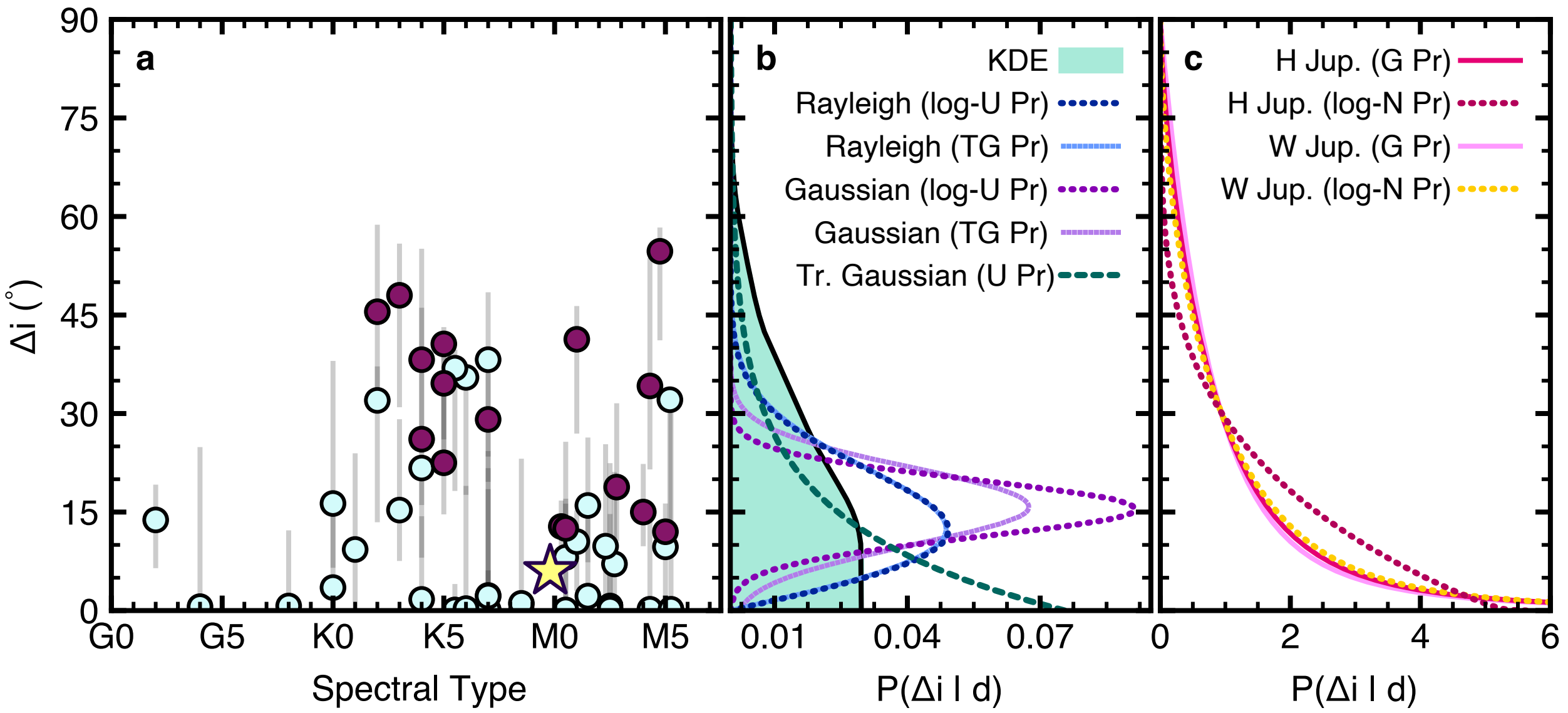

**Fig. 1 | Minimum stellar obliquity, Δi, as a function of spectral type. a**, Point colors represent the misalignment status of each system (n=49). Dark purple circles are misaligned at the >2$\sigma$ level, and light blue points show no evidence of misalignment. The yellow star represents the location of the pre-main sequence Sun (which would have corresponded to a spectral type of about K9 at an age of approximately 5 Myr) assuming that the origin of its current obliquity is primordial. Error bars are shown at 1$\sigma$ confidence. **b**, Reconstructed distributions of Δi. The light green area shows the Gaussian KDE of the distribution of Δi. The HBM results for the Rayleigh distribution generated with a log-Uniform prior (labeled as 'log-U Pr') are represented by the dotted dark blue line, and the same model fit with a Truncated Gaussian prior ('TG Pr') is shown by the solid light-blue line. The Gaussian distribution model fits produced with a log-Uniform prior and Truncated Gaussian prior are plotted as the purple dashed and solid lines, respectively. The Truncated Gaussian model fit generated with a Uniform prior ('U Pr') is shown in dark teal. **c**, Best-fit HBM results using a Beta distribution for the minimum obliquities of a sample of 25 hot ('H') and 22 warm ('W') Jupiter systems[15] also determined using a similar Δi methodology, shown here for comparison. Here, 'log-N Pr.' represents model fits produced with a log-Normal prior.

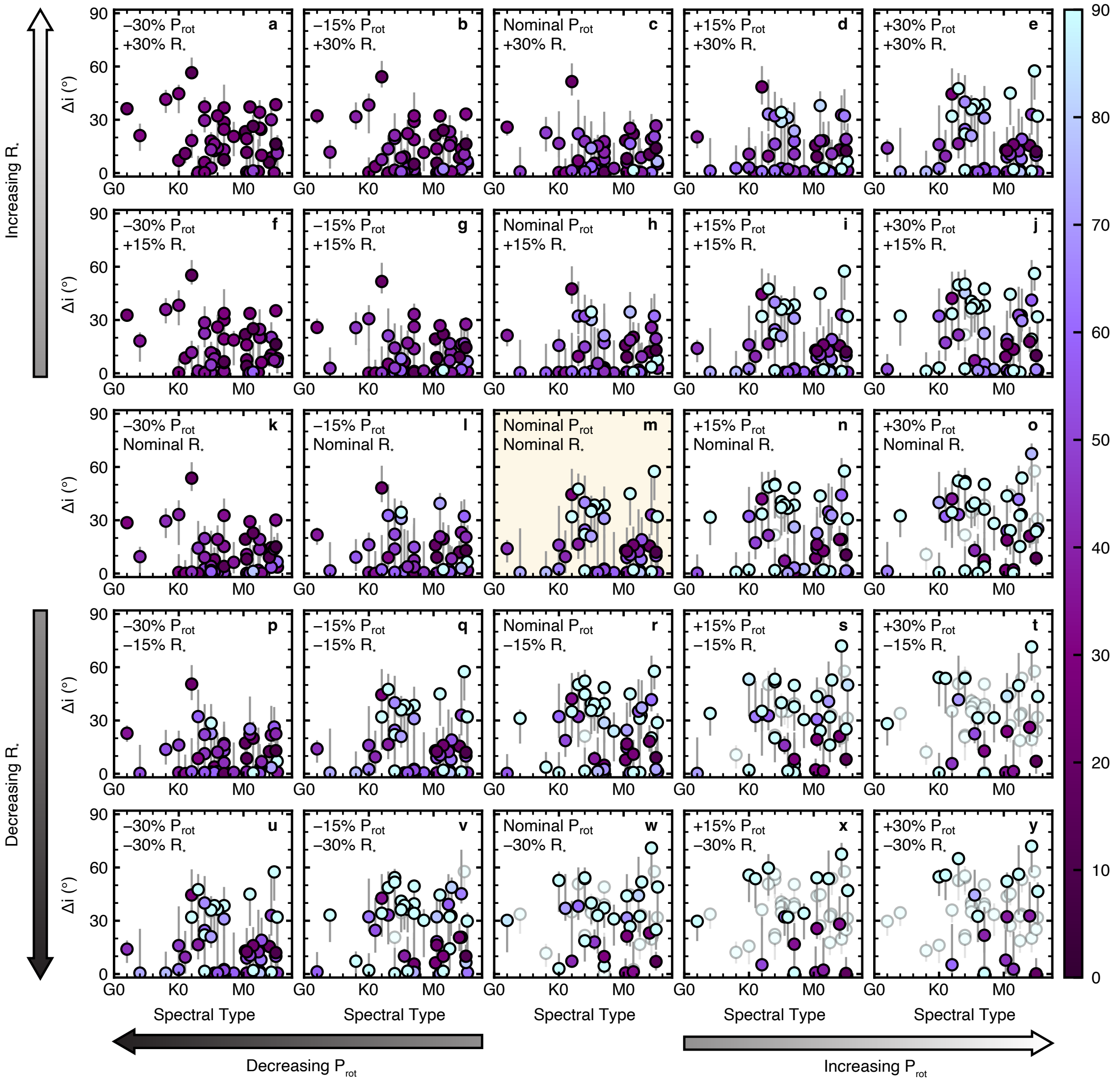

**Fig. 2 | The results of 25 independent tests to probe the effect that potential biases in $P_{rot}$ and $R_*$ could have on the computation of $i_*$. a–y,** Each cell represents an individual test where Δi is determined using larger or smaller values of $P_{rot}$ and/or $R_*$ compared to the nominal adopted values. Cells show Δi as a function of Spectral Type, with data points shaded with respect to $i_*$. The nominal results of our analysis are shown in the center of the grid, identified by a shaded background. Grid cells to the right of center show outcomes for incrementally greater rotation periods, and grid cells to the left of center show outcomes for incrementally decreased rotation. Nominal stellar radii were increased for cells above the center row and were decreased for cells in the bottom two rows. In several cases, the combination of test values of $P_{rot}$ and $R_*$ produced equatorial velocities that were less than $v\sin i_*$ by more than $2\sigma$ and are therefore not physically viable. The non-physical results are indicated by the faded points, and are excluded from further analysis. Error bars are shown at $1\sigma$ confidence.

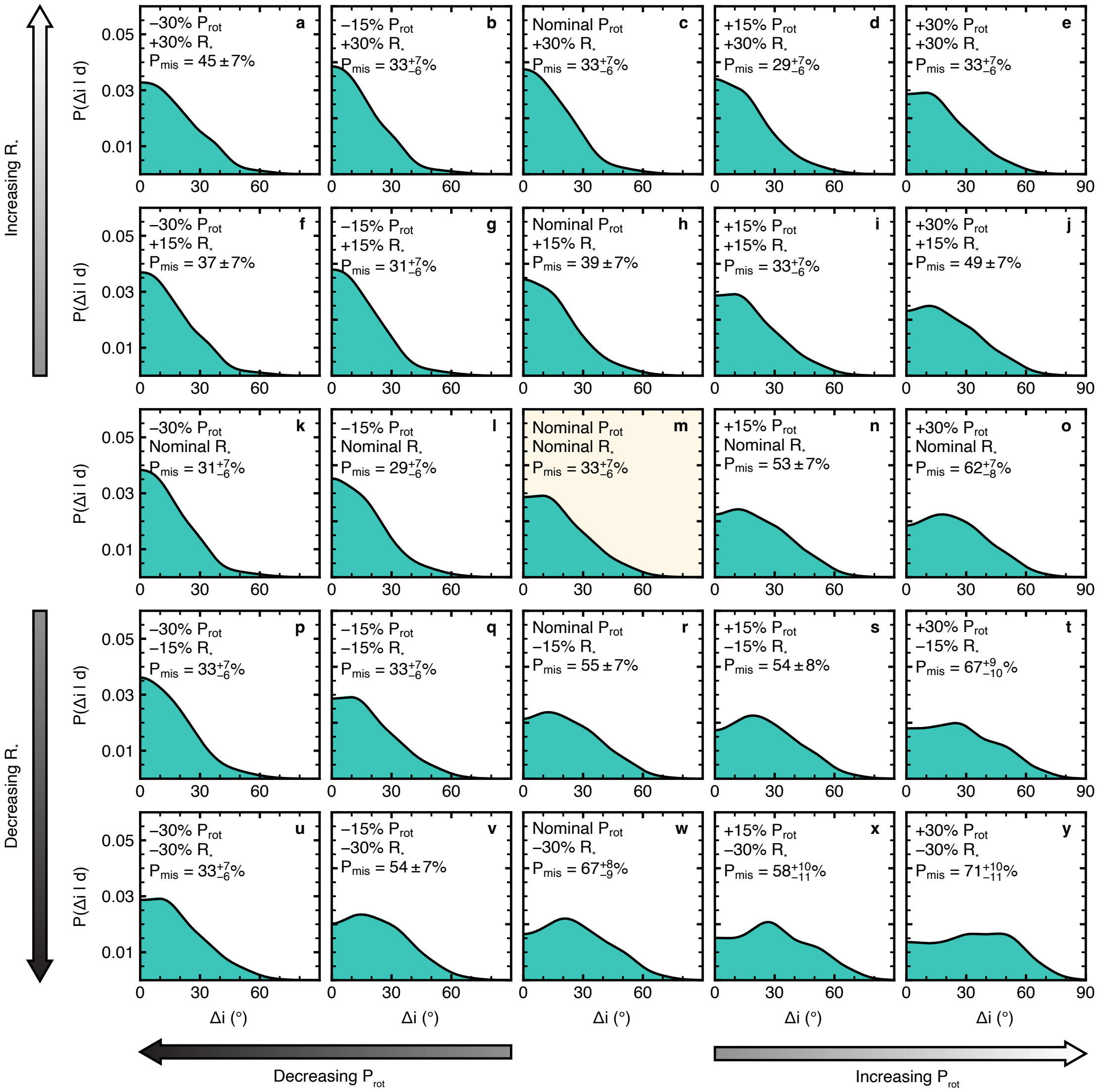

**Fig. 3 | The resulting Δ*i* distributions for corresponding to grid cells in Fig. 2. a–y,** Each cell contains the KDE representation of the distribution of Δ*i* in the corresponding cell in Fig. 2. The misalignment probability ($P_{\mathrm{mis}}$) of each Δ*i* distribution is printed in its respective cell.

# Methods

## Protoplanetary Disk Inclinations

We assemble a catalog of resolved protoplanetary disks by first referencing the list of resolved disks in the Catalog of Circumstellar Disks (https://www.circumstellardisks.org/), which identifies 227 disks around pre-main sequence stars as of its last update on August 13, 2021. We supplement these with entries from another recent compilation of ALMA-detected disks[51], as well as resolved disks from individual studies compiled from the recent literature. This amounted to a sample of 690 unique systems. For this study, we isolate our sample to systems with well-determined inclinations ($\sigma_{\mathrm{idisk}}$<10°) derived from resolved submillimeter observations from the Atacama Large Millimeter/submillimeter Array ($\sigma_{\mathrm{idisk}}$ = 2.6°, on average). Our adopted measurements of $i_{\mathrm{disk}}$ therefore describe the orientation of dust in the outer disk at characteristic spatial scales of tens to hundreds of au. After filtering out disks that do not meet the uncertainty threshold, we are left with 172 systems. We then remove known binaries and stars with spectral types earlier than F6, yielding a sample of 94. The majority of the stars belong to well-studied star-forming regions, primarily the Taurus-Auriga complex, the Lupus complex, and $\rho$ Ophiuchus, with ages ranging approximately 1–3 Myr[52–56]. For many disks, multiple independent inclination measurements have been made. For our analysis, we adopt the measurement of $i_{\mathrm{disk}}$ with the highest reported precision. We note that some disk inclinations are reported with asymmetric uncertainties. In these cases, we adopt the mean of the upper and lower limits.

## Stellar Properties

For each object in the initial sample of 94 single, low mass, protoplanetary disk-bearing stars, we compile $v\sin i_*$, $P_{\mathrm{rot}}$, and $R_*$ measurements available in the literature. Rotation periods are either adopted from previous measurements or are newly determined using light curves from *TESS* or *K2*. After selecting for stars that satisfy the requirement that $v\sin i_*$, $P_{\mathrm{rot}}$, and $R_*$ have been reliably measured (detailed below), we arrive at a final sample of 49 systems. A minimum uncertainty of 5% is imposed on all adopted stellar parameters if the reported values are smaller than this. Adopted values of $T_{\mathrm{eff}}$, $M_*$, $v\sin i_*$, $P_{\mathrm{rot}}$, and $R_*$ are given in Extended Data Table 1.

## Projected Rotation Velocities

For each object in the sample, we compile published independent measurements of the stellar $v\sin i_*$ that were obtained with high-resolution spectra, and adopt the weighted mean. Because our measurements are drawn from a diverse array of sources, our adopted $v\sin i_*$ could be subject to systematic effects related to data acquisition, processing, and uncertainty estimation. To account for this, we impose a 5% lower limit on the precision of $v\sin i_*$ to mitigate the possibility of a single measurement unfairly weighting the mean. In addition, if an object has two or more measurements with no reported uncertainties, we combine them into a single average value and adopt the standard deviation as an estimate of the uncertainty. If only one $v\sin i_*$ is reported without an uncertainty, it is excluded.

## Rotation Periods

We compiled stellar rotation periods in the literature drawn from independent datasets and adopt the weighted mean and uncertainty. To supplement these, we uniformly analyze *TESS*[57,58] and *K2*[59] light curves for targets in our compilation of resolved disks, avoiding the duplication of $P_{\mathrm{rot}}$ measurements from previous analyses of the same *TESS* and *K2* datasets we consider here[60–67]. We add new rotation periods for 43 objects.

We use the Python package, lightkurve[68], to download the *TESS* 2-minute and *K2* 30-minute cadence Pre-search Data Conditioning Simple Aperture Photometry (PDCSAP) light curves[69–72], and for seven objects we analyze the *TESS* light curves reduced with the MIT Quick-Look Pipeline[57]. For *K2* light curves where instrumental systematics appear to be present in the PDSCAP data, we assess the *K2* light curves reduced with either the EPIC Variability Extraction and Removal for Exoplanet Science Targets[73] pipeline (one object), or the K2 Systematics Correction[74] pipeline (two objects). Because of the potential for crowding in the field, we visually inspect each full frame image to verify that there is no visibly apparent contamination from nearby stars.

We prepare the data products from each *TESS* sector or *K2* campaign by removing any 'NaNs' from the data series and dividing the light curve by the median flux. Some objects were observed over multiple sectors or campaigns. If an object was observed over two or more consecutive *TESS* sectors, we stitch the data together to form one light curve. We do not merge light curves that are separated by one or more sectors because the spot evolution that occurs during long gaps in time may cause the rotation signature to no longer be coherent over the duration of the time series, making it difficult to properly assess the periodogram[75] and the phase curve. Next, we remove features unrelated to rotational variability from the data (i.e., flares and other photometric outliers) by applying a high-pass Savitzky–Golay filter[76] and selecting only the points in the original dataset that fall within three standard deviations of the flattened time series. After detrending the data, we compute generalized Lomb-Scargle periodograms[77] of the prepared light curves with a sampling resolution of 0.01 days over a search window from 0.2 days to 1/3 the total length of time in the light curve. We identify the stellar rotation period as that which corresponds to the maximum signal in the periodogram. The uncertainty on $P_{\rm rot}$ is estimated as the half-width of the peak profile in the power spectrum. If one object has more than one *TESS* or *K2* light curve, each light curve is analyzed separately and we adopt the weighted mean of the individual results.

Doppler imaging of young stars demonstrates that spots can exist at high and intermediate latitudes rather than at the equator[78–86]. This can impact the measured rotation periods if the young stars in our sample experience differential rotation. Periodic signals detected in the light curves may not actually trace the equatorial rotation period used to compute the stellar inclination angle, which would bias the results toward higher values of $i_*$. We account for the effect of possible differential rotation of spots at unknown latitudes by inflating the uncertainties of the rotation periods with an error term $\sigma_{\rm shear}$ that represents half the maximum difference in rotation between the pole and equator[11]. This term relates the star's absolute shear $\Delta\Omega$ (a metric to quantify differential rotation), and the equatorial rotation period, $P_{\rm max}$, determined from the light curve:

$$\sigma_{\rm shear} \approx \frac{1}{2}\left(P_{\rm max} - \left(\frac{\Delta\Omega}{2\pi} + \frac{1}{P_{\rm max}}\right)^{-1}\right). \quad (1)$$

We assume a Sun-like shear of $\Delta\Omega = 0.07$ rad day$^{-1}$ and add $\sigma_{\rm shear}$ in quadrature to the $P_{\rm rot}$ uncertainties from the periodogram analysis in our sample. Given that the exact spot distributions are unknown, we do not apply any explicit corrections to the measured period value.

## Stellar Radii

We adopt $R_*$ estimates from the *TESS* Input Catalog[87] (TIC). Although the overall distribution of the TIC radii in our sample does not differ significantly from the weighted mean values of other radii found in the literature (Extended Data Fig. 1), adopting radii from the TIC ensures consistency across the sample. TIC radius estimates are determined via the Stefan-Boltzman relation based on Gaia-determined distances, extinction-corrected *G*-band magnitudes, and *G*-band bolometric corrections[88]. Moreover, the accuracy of $R_*$ in the TIC is well characterized; stellar radii in the catalog are found to be typically within 7% the values measured for the same stars with asteroseismology[89]. We therefore inflate the uncertainties on $R_*$ from the TIC by adding a 7% error term in quadrature with the nominal uncertainty[11]. For the majority of stars in the sample, there is no uncertainty reported with the TIC radius. For these objects, we assume a conservative uncertainty of 16%, which corresponds to the 95% quantile of the entire TIC radius uncertainty distribution. To this, we then add an additional 7% systematic error in quadrature.

The stellar radius and rotation period should always yield an equatorial rotation velocity ($v_{eq}$, where $v_{eq} = 2\pi R_*/P_{rot}$) that is greater than or equal to $v\sin i_*$. The majority of the sample adheres to this to within $2\sigma$ with the exception of nine stars (2MASS J04202555+2700355, 2MASS J04360131+1726120, AA Tau, DoAr 25, FT Tau, IQ Tau, Sz 73, T Cha, and WSB 52). This disagreement could be the result of potentially overestimated rotation periods, overestimated $v\sin i_*$, or underestimated radii. The average $P_{rot}$ and $v\sin i_*$ of this subset ($P_{rot}$ = 5.0±2.7 d and $v\sin i_*$ = 18.7±9.3 km $s^{-1}$) compared to non-discrepant systems ($P_{rot}$ = 4.8±2.0 d and $v\sin i_*$ = 16.6±11.3 km $s^{-1}$) do not indicate that an overestimation of either parameter is the cause of the disagreement. However, the TIC radii of the stars in this subset are on average 36% lower than the mean of all non-TIC radii compiled for these same stars, whereas the rest of the objects in the sample that do not yield discrepant $v_{eq}$ and $v\sin i_*$ values are on average only 3% lower than the mean of their non-TIC counterparts. Some TIC radii may therefore be underestimated, leading to the disagreement in viable values of $v_{eq}$.

For the discrepant stars, we adopt radii from alternative sources, most of which originate from a separate catalog of radius estimates[90] using spectra from the APOGEE[91], GALAH[92], and RAVE[93] surveys, validated with CHARA interferometry[94], Hubble Space Telescope flux standards[95], and asteroseismology[96], which we refer to as the Y23 catalog. The Y23 catalog has a characteristic accuracy within 5% of asteroseismology measurements[96]. We thus add a conservative 5% systematic error term in quadrature to the uncertainties quoted in the Y23 catalog. Four stars do not have a radius estimate from Y23 (2MASS J04202555+2700355, Sz 73, V1094 Sco, and WSB 52), so we adopt the mean of the literature radii and estimate a conservative uncertainty equal to 2 times the standard deviation. After these adjustments to radii, the sample $v_{eq}$ and $v\sin i_*$ are consistent to within $2\sigma$ (Extended Data Fig. 2), the only exception being WSB 52, which is discrepant at the $2.4\sigma$ level. Five systems did not have reported TIC radii (2MASS J04334465+2615005, CIDA-7, Elias 2-24, MHO 6, and WSB 63), so for these systems we also adopt the radius estimate from the Y23 catalog, adding a 5% systematic error term in quadrature to the quoted uncertainties. All compiled values of $T_{eff}$, $M_*$, $v\sin i_*$, $P_{rot}$, and $R_*$ from the literature are provided in the Supplementary Methods.

**Stellar Inclination**

With our adopted measurements of $R_*$, $v\sin i_*$, and $P_{\rm rot}$, we determine the stellar inclination $i_*$ following a Bayesian probabilistic framework[10] which properly accounts for the correlation between $v\sin i_*$ and $v_{\rm eq}$. In particular, we use the analytical expression for the posterior distribution of $i_*$[11], which assumes an isotropic prior on $i_*$ and considers uniform priors on $v\sin i_*$, $R_*$, and $P_{\rm rot}$ (with a moderately precise measurement uncertainty of $P_{\rm rot}$<20%):

$$P(\, i_* \mid P_{\rm rot}, R_*, v\sin i_* \,) \propto \sin i_* \times \frac{e^{-\frac{\left(v\sin i_* - \frac{2\pi R_*}{P_{rot}}\sin i_*\right)^2}{2\left(\sigma^2_{v sin i_*} + \sigma^2_{v_{\rm eq}} \sin^2 i_*\right)}}}{\sqrt{\sigma^2_{v\sin i_*} + \sigma^2_{v_{\rm eq}} \sin^2 i_*}} \,, \qquad (2)$$

where

$$\sigma_{v_{\rm eq}} = \frac{2\pi R_*}{P_{\rm rot}} \sqrt{\left(\frac{\sigma_{R_*}}{R_*}\right)^2 + \left(\frac{\sigma_{P_{\rm rot}}}{P_{\rm rot}}\right)^2} \,, \qquad (3)$$

and $\sigma_{P{\rm rot}}$, $\sigma_{v\sin i*}$, and $\sigma_{R*}$ are the uncertainties on the rotation period, projected rotational velocity, and stellar radius, respectively. We note that the reported $v\sin i_*$ measurements for four objects in our sample (2MASS J16083070-3828268, GW Lup, Sz 114, and Sz 130) are upper limits. For these cases, we use the analytical expression

$$P(\, i_* \mid P_{\rm rot}, R_*, v\sin i_* \,) \propto \sin i_* \times \left( erf\left( \frac{l - \frac{2\pi R_*}{P_{\rm rot}} \sin i_*}{\sqrt{2}\sigma_{v_{\rm eq}} \sin i_*} \right) + \, erf\left( \frac{\sqrt{2}\pi R_*}{\sigma_{v_{\rm eq}} P_{\rm rot}} \right) \right), \qquad (4)$$

where $l$ is the upper limit of $v\sin i_*$. The $i_*$ posterior distributions are shown in the Supplementary Methods. For each star, we report the $i_*$ MAP value and 68% highest density interval in Extended Data Table 2.

**Star-Disk Minimum Obliquity, Δ*i***

To determine $\Delta i$ for each object, we randomly draw $10^6$ samples from the posterior distribution of $i_*$ and the probability distribution of $i_{\rm disk}$ and compute the absolute difference between each sampled pair. Our adopted value of $\Delta i$ is the distribution mode, and our reported uncertainty range is the 68% highest density interval. Resulting $\Delta i$ values and confidence ranges are provided in Extended Data Table 2, and $\Delta i$ probability distributions are shown in the Supplementary Methods. Extended Data Fig. 3 shows $\Delta i$ plotted as a function of system properties such as spectral type, $M_*$, $T_{\rm eff}$, $R_*$, $P_{\rm rot}$, $v\sin i_*$, $i_{\rm disk}$, and $i_*$. We compute the Pearson correlation coefficient, $r$, between $\Delta i$ and each system property to test for correlated dependencies and identify no strong relationship with any parameter. Correlation coefficients in this test range from -0.04 to 0.37. These tests further yield high $p$-values, suggesting that any correlation indicated in the resulting $r$ values is statistically indistinguishable from the null hypothesis of no correlation. Correlation coefficients are displayed in Extended Data Fig. 3. We note that the correlation coefficient for $\Delta i$ with respect to $i_*$ appears to be moderately correlated with $r = 0.37$ and $p = 0.01$, however, these results do not take into account the uncertainties of the individual data points. When we repeat this test 100 times drawing $\Delta i$ randomly from the individual probability distributions, the average Pearson-$r$ correlation coefficient is equal to 0.01 with an average $p$-value of 0.48. The moderate but significant correlation in $i_*$ vs. $\Delta i$ that is initially apparent does not hold when taking into account the uncertainties in $\Delta i$. Repeating this procedure for all other system parameters yields similar results. We therefore identify no significant correlation in $\Delta i$ with respect to spectral type, $M_*$, $T_{\rm eff}$, $R_*$, $P_{\rm rot}$, $v\sin i_*$, $i_{\rm disk}$, nor $i_*$.

Because potential trends in $\Delta i$ may not present only as a linear relationship, we conduct a second test to identify possible clustering of minimum obliquities at higher or lower values within subpopulations of dependent variables. Specifically, we separate the sample into two subgroups equal in size consisting of smaller or larger stellar parameters. We then determine the mean and standard deviation of each subgroup and compute the significance of the difference of the means. For every system parameter, the two subpopulations are consistent at a $<1\sigma$ level. We therefore find no evidence that $\Delta i$ clusters at low or high values for either subpopulation of sample parameters (Extended Data Fig. 3).

**Kernel Density Estimate of $\Delta i$ Distribution**

The KDE of the population level $\Delta i$ distribution is computed for 500 samples drawn from each individual $\Delta i$ probability distribution in the sample (resulting in total of 21,500 samples). We choose a kernel broadening parameter of 5.3°, which is the average deviation from the mean of the 68% confidence interval limits for each $\Delta i$ distribution. Next, we use the uncertainties of $i_{\rm disk}$, $v\sin i_*$, $P_{\rm rot}$, and $R_*$ to generate a family of KDE reconstructions of the $\Delta i$ distribution. For every object, we randomly sample 500 new values for each parameter, drawn from a normal distribution that reflects the parameter's adopted value and uncertainty. From the samples, we generate 500 new $\Delta i$ probability distributions for each object, which are then used to generate new KDEs following the same method to compute the nominal KDE. The diversity of the resulting family of KDE reconstructions are shown in Extended Data Fig. 4.

**Hierarchical Bayesian Modeling of $\Delta i$**

We estimate the underlying distribution of $\Delta i$ with HBM using the open-source Python software ePop![97]. Although the package was originally developed to model eccentricities, it can be generalized by mapping observations that span different ranges to a single domain from 0–1. ePop! uses an importance sampling methodology[98] and offers several choices for underlying parametric models, which have previously been used to characterize stellar obliquity distributions[15]. In the context of HBM, the model parameters are hyperparameters with posterior distributions determined with the affine-invariant Markov chain Monte Carlo sampler emcee[99]. Here, we explore three population-level models to represent the underlying distribution in $\Delta i$. We choose these models because they are flexible and consist of just one or two free parameters, easing the interpretation of the results. The first underlying model we test is the Rayleigh distribution, $\mathcal{R}(\Delta i|\nu)$, given by

$$\mathcal{R}(\Delta i|\nu) = \frac{\Delta i}{\nu^2} e^{-\frac{\Delta i^2}{2\nu^2}}. \tag{5}$$

The second is a Gaussian distribution, $\mathcal{N}(\Delta i|\mu, \sigma)$,

$$\mathcal{N}(\Delta i|\mu,\sigma) = \frac{1}{\sigma\sqrt{2\pi}} e^{-\frac{1}{2}\left(\frac{\Delta i-\mu}{\sigma}\right)^2}, \tag{6}$$

and the third is a Truncated Gaussian, $\mathcal{T}(\Delta i|\mu, \sigma, a, b)$,

$$\mathcal{T}(\Delta i|\mu,\sigma,a,b) = \frac{1}{\sigma}\frac{\frac{1}{\sqrt{2\pi}} e^{-\frac{1}{2}\left(\frac{\Delta i-\mu}{\sigma}\right)^2}}{\Phi\left(\frac{b-\mu}{\sigma}\right)-\Phi\left(\frac{a-\mu}{\sigma}\right)}, \quad a \leq \Delta i \leq b. \tag{7}$$

We convert each object's distribution in $\Delta i$ (originally ranging from 0°–90°) to a new, normalized variable $\Delta i' = \Delta i/90°$, spanning the interval [$a$ = 0, $b$ = 1] to satisfy the generalized domain space in ePop!. The resulting posterior is then readily remapped to the original scale. Two hyperpriors are tested on the Rayleigh and Gaussian distributions to evaluate the degree to which our choice of priors affects the posteriors. We choose hyperpriors that have demonstrated the ability to produce families of distributions with sufficiently diverse morphologies to yield the most robust results[56]. Our first hyperprior is a Truncated Gaussian with $\mu' = 0.69$, $\sigma' = 1.0$:

$$p(\Delta i|\mu',\sigma') = \frac{1}{\sigma'}\frac{\frac{1}{\sqrt{2\pi}} e^{-\frac{1}{2}\left(\frac{\Delta i-\mu'}{\sigma'}\right)^2}}{1-\frac{1}{2}\left(1+\mathrm{erf}\left(\frac{-\mu'}{\sigma'\sqrt{2}}\right)\right)} \tag{8}$$

and our second choice of hyperprior draws from a log-Uniform distribution ranging from 0.01 to 100:

$$p(\Delta i) = \frac{1}{\Delta i}. \tag{9}$$

For the Truncated Gaussian underlying model, we apply a uniform hyperprior:

$$f(\Delta i|x_0,x_1) = \frac{1}{x_1-x_0}, \quad x_0 \leq x \leq x_1, \tag{10}$$

with $\mu$ ranging from -1000 to 1000 and $\sigma$ from 0 to 1000. Each MCMC run consists of 80 walkers for $6\times10^4$ steps with a burn-in fraction of 50%. For both parametric models, we perform a visual inspection of the trace plots to ensure that the chains have properly converged. Each of the best-fit models of the underlying distribution of $\Delta i$ yields similar results, indicating that the stellar obliquity distribution is broad (Extended Data Fig. 4). We provide the best-fit parameters and confidence ranges for the Rayleigh, Gaussian, and Truncated Gaussian models in Extended Data Table 3.

**Characterization of Systematics and Biases**

The distribution of MAP values of $i_*$ shows that 14 out of 49 stars in our sample are equator-on with inclinations that are likely to be >80°. If the stellar orientation was randomly distributed—which is a reasonable assumption for isolated stars but may not be valid for this sample with resolved disks—the probability of $i_*$>80° is $\int_{i_*=80°}^{i_*=90°} sin i_* \, di_*$ = 0.174. This points to an expectation value of about 8 for a sample of 49 stars, suggesting that there may be a preference toward equator-on stars in the sample population above what would be expected by chance. Using binomial statistics, we can quantify the significance of this discrepancy by computing the probability of an event occurring that is at least as extreme as these measurements ($k$ = 14 out of $n$ = 49 systems). The probability of a success is $p$ = 0.174, and so the probability of observing at least 14 stars out of 49 with $i_*$>80° can be computed by taking 1 minus the probability of observing fewer than 14 stars with $i_*$ >80°: $P(k \geq 14 \mid p = 0.174, n = 49) = 1 - P(k<14 \mid p=0.174, n=49) = 1 - \sum_{k=0}^{13} \binom{n}{k} p^k (1-p)^{(n-k)} = 0.054$. There is thus a probability of about 5% of there being at least 14 equator-on stars in our sample. The over-representation of high-inclination stars is mild, exceeding the expectation value by only 6.

## Computational Tools Used

This research has made use of the VizieR[100] catalogue access tool, CDS Strasbourg, France[101], and the following open-source software: ePop![97], lightkurve[68], Astropy[102], Numpy[103], Scipy[104], and Matplotlib[105].

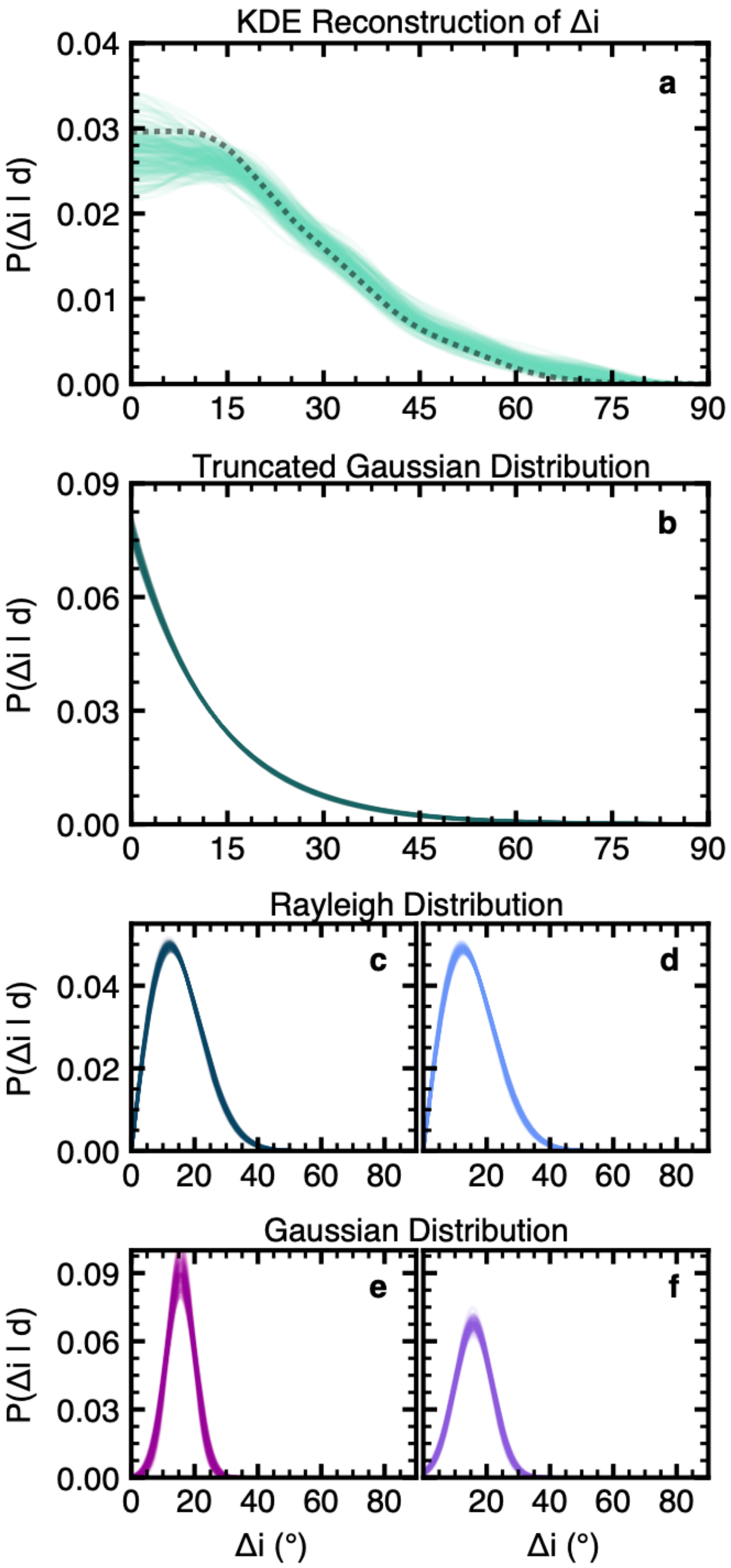

**Extended Data Fig. 1 | A summary of the properties of the sample. a–h,** Solid histograms show the distributions of the adopted system parameters: Spectral Type, $M_*$, $T_{\mathrm{eff}}$, $R_*$, $P_{\mathrm{rot}}$, $v\sin i_*$, $i_{\mathrm{disk}}$, and $i_*$. **d,** The distribution of radii extracted from the TIC (solid purple), which is the primary source of radius estimates used in our analysis. For comparison, the hatched histogram shows the distribution of the weighted mean of $R_*$ values compiled from the literature for each object (labeled 'Lit'). **h,** A comparison of the distributions of $i_*$ maximum a posteriori (MAP) values computed with $R_*$ from the TIC and from the literature, showing general agreement across the sample.

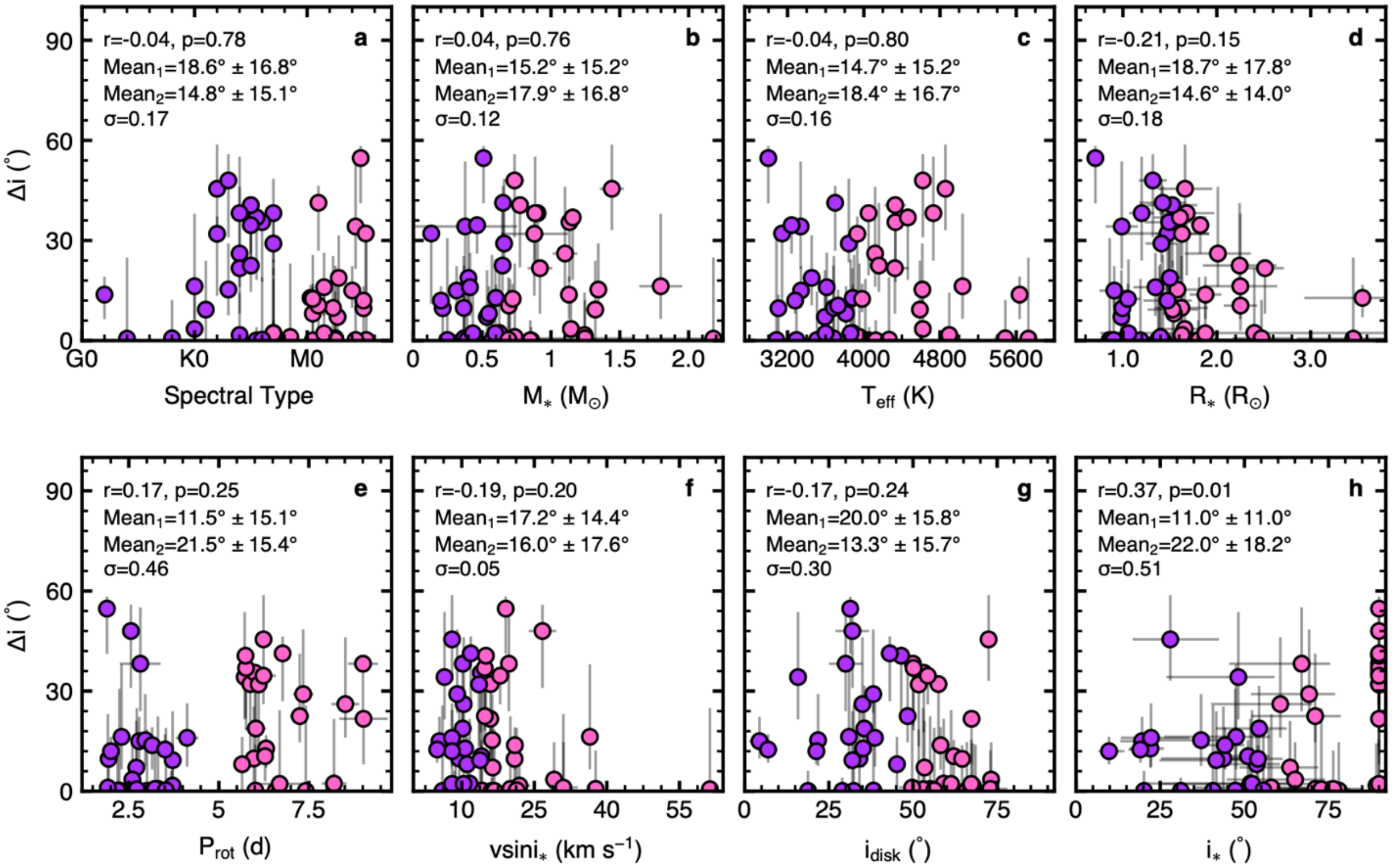

**Extended Data Fig. 2 | Residuals of the equatorial velocity, $v_{eq}$, and the measured projected rotational velocity, $v\sin i_*$, for every star in the sample.** The color of each point (n=49) is mapped to its rotation period indicated by the color bar on the right. Individual point sizes scale with the actual stellar radius. Dark error bars correspond to 1$\sigma$ confidence and light error bars correspond to 2$\sigma$. The shaded region where $v_{eq} - v\sin i_* < 0$ km s$^{-1}$ indicates where $v_{eq}$ and $v\sin i_*$ are non-physical (i.e., $v\sin i_*$ is greater than $v_{eq}$).

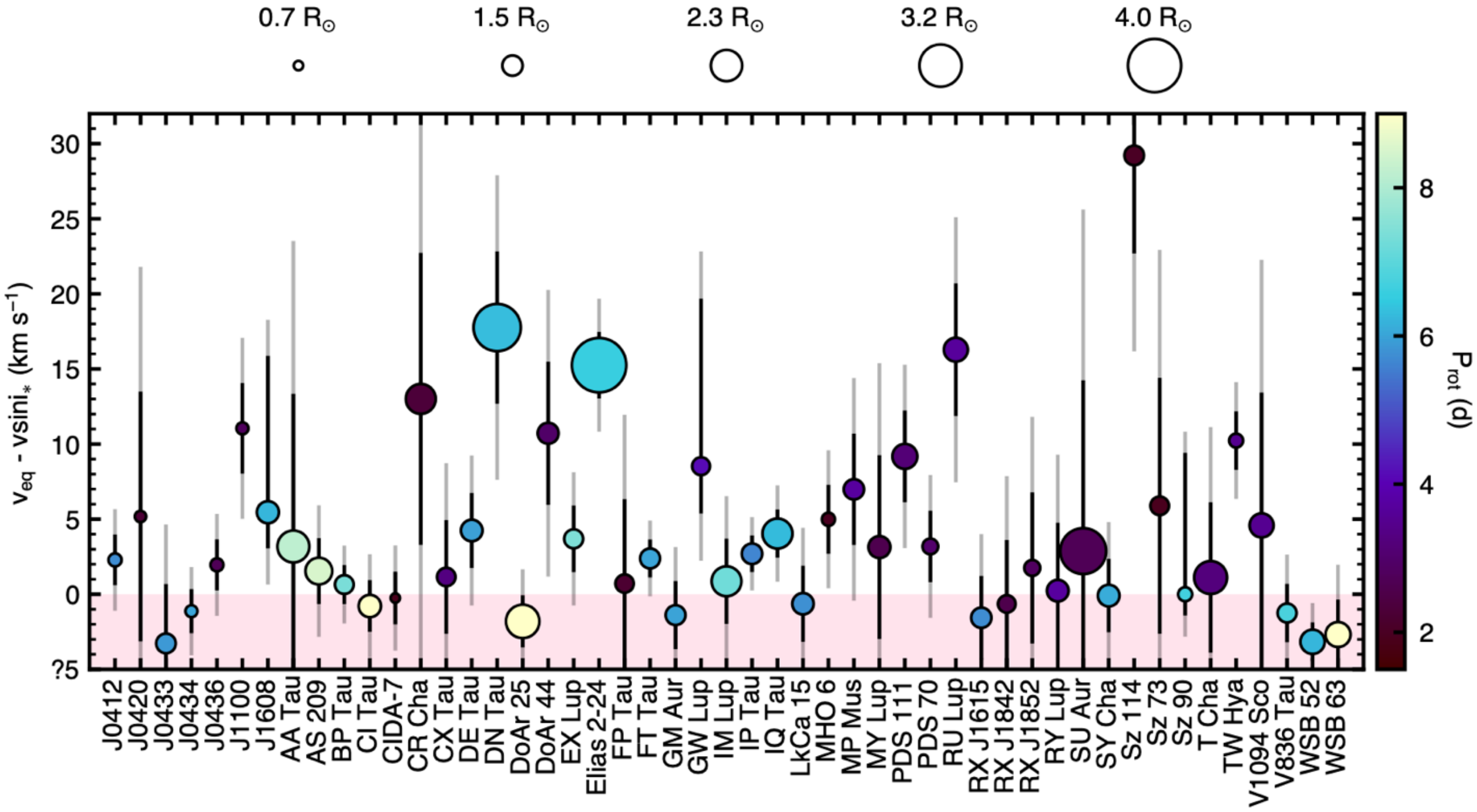

**Extended Data Fig. 3 | Δ*i* as a function of adopted system properties. a–h,** Δ*i* as a function of: Spectral Type, $M_*$, $T_{\rm eff}$, $R_*$, $P_{\rm rot}$, $v\sin i_*$, $i_{\rm disk}$, and $i_*$. Pearson-*r* correlation coefficients and their respective *p*-values are printed in the upper left of each cell. Data points within each panel are split into two sub-populations that are used to test for parameter-dependent clustering in Δ*i* (see Methods) and are shown here in purple (sub-group 1) and pink (sub-group 2). The mean values and standard deviations for groups 1 and 2 are printed in each panel, and the significance level of the difference between sub-groups ($\sigma$) is also printed in each panel. Error bars are shown at 1$\sigma$ confidence.

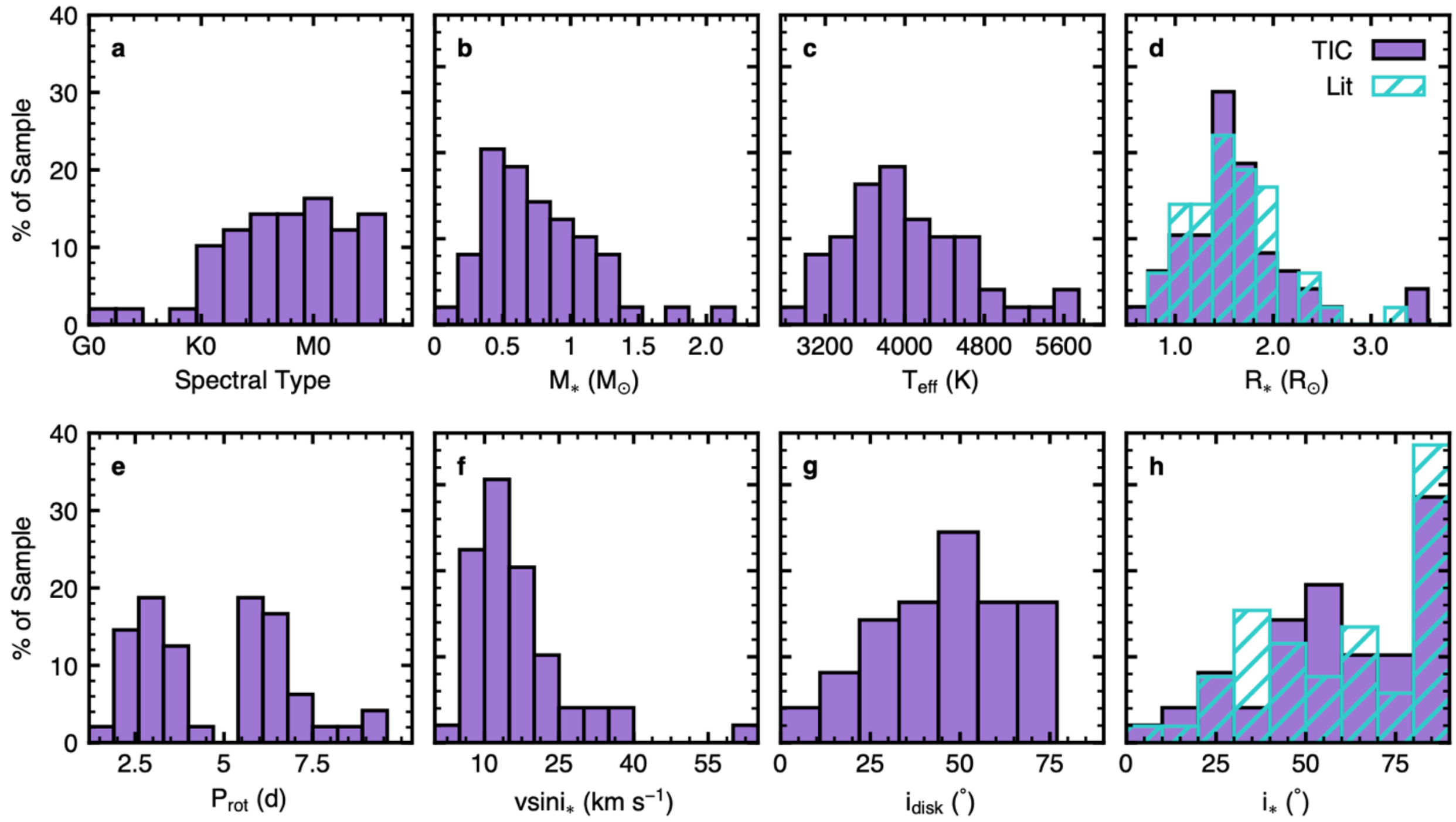

**Extended Data Fig. 4 | Breadth of the KDE and HBM representations of the distribution of Δ*i* across the sample.** **a**, KDE reconstructions of the Δ*i* distribution of the sample. Each solid line represents one KDE calculated with randomly selected values from probability distributions of $i_{\rm disk}$, $v\sin i_*$, $P_{\rm rot}$, and $R_*$. The nominal KDE is displayed as the dotted line. **b**, One hundred random draws from the family of resulting best fit Truncated Gaussian models of the underlying distribution of Δ*i*. **c–d,** Same as b, showing Rayleigh models of the underlying distribution of Δ*i*. Results are determined with a log-Uniform hyperprior and a Truncated Gaussian hyperprior, respectively. **e–f,** Same as c–d, but showing the family of results for the Gaussian model.

**Extended Data Table 1 | Adopted star properties.** The stellar parameters displayed in each column are the weighted mean of all relevant measurements found in the literature for each object. The compilation of effective temperatures, masses, $v\sin i_*$, rotation periods, stellar radii, disk inclinations, associated literature references, and the complete names of identifiers labeled with a '*' are provided in the Supplementary Methods.

| Identifier | Spectral Type | $i_{disk}$ (°) | $i_*$ (°) | Δi (°) | Misaligned? |
|---|---|---|---|---|---|
| 2M J0412 | M4.3 | 15.8 ± 0.8 | $48.2^{+21.4}_{-10.8}$ | $33^{+21}_{-11}$ | Y |
| 2M J0420 | M5.3 | 38.2 ± 0.3 | $47.3^{+29.7}_{-11.2}$ | $0^{+31}_{-0}$ | ··· |
| 2M J0433 | M5.2 | 57.6 ± 0.07 | $90.0^{+0.0}_{-21.8}$ | $31^{+0.3}_{-21}$ | ··· |
| 2M J0434 | M4.3 | 68.5 ± 0.2 | $90.0^{+0}_{-18.9}$ | $0^{+14}_{-0}$ | ··· |
| 2M J0436 | M2.7 | 53.4 ± 1.4 | $64.0^{+15.1}_{-8.4}$ | $8^{+13}_{-8}$ | ··· |
| 2M J1100 | M4.0 | 4.3 ± 3.1 | $19.6^{+6.5}_{-4.5}$ | $15^{+7}_{-6}$ | Y |
| 2M J1608 | K2.0 | 74 ± 5 | $28.0^{+11.2}_{-14.5}$ | $44^{+14}_{-11}$ | Y |
| AA Tau | K7.0 | 59.1 ± 0.3 | $51.8^{+32.1}_{-11.3}$ | $2^{+19}_{-3}$ | ··· |
| AS 209 | K4.0 | 35.0 ± 0.1 | $60.7^{+20.5}_{-9.5}$ | $26^{+20}_{-10}$ | Y |
| BP Tau | K7.0 | 38.2 ± 0.5 | $69.2^{+17.3}_{-7.7}$ | $32^{+16}_{-9}$ | Y |
| CI Tau | K7.0 | 50.0 ± 0.1 | $90.0^{+0.0}_{-20.2}$ | $38^{+1}_{-19}$ | ··· |
| CIDA-7 | M4.7 | 31.3 ± 0.3 | $90.0^{+0.0}_{-17.0}$ | $57^{+1}_{-17}$ | Y |
| CR Cha | K0.0 | 31.0 ± 1.4 | $47.5^{+21.3}_{-10.0}$ | $15^{+22}_{-9}$ | ··· |
| CX Tau | M2.5 | 55 ± 1 | $71.1^{+18.9}_{-7.5}$ | $0^{+22}_{-0}$ | ··· |
| DE Tau | M2.3 | 34 ± 1 | $43.7^{+18.5}_{-9.3}$ | $10^{+15}_{-10}$ | ··· |
| DN Tau | M0.3 | 35.2 ± 0.2 | $22.1^{+6.3}_{-3.9}$ | $12^{+4}_{-5}$ | Y |
| DoAr 25 | K4.0 | 67.4 ± 0.2 | $90.0^{+0.0}_{-14.5}$ | $21^{+1}_{-14}$ | ··· |
| DoAr 44 | K3.0 | 21.8 ± 0.9 | $37.1^{+13.8}_{-7.6}$ | $15^{+14}_{-8}$ | ··· |
| Elias 2-24 | K5.5 | 29.0 ± 0.3 | $31.1^{+3.2}_{-2.8}$ | $0^{+4}_{-0}$ | ··· |
| EX Lup | M0.5 | 32.4 ± 0.9 | $40.7^{+22.3}_{-12.7}$ | $0^{+26}_{-0}$ | ··· |
| FP Tau | M2.5 | 66.0 ± 4 | $78.0^{+11.9}_{-12.6}$ | $1^{+13}_{-1}$ | ··· |
| FT Tau | M2.8 | 35.5 ± 0.4 | $54.4^{+12.8}_{-8.2}$ | $18^{+13}_{-8}$ | Y |
| GM Aur | K6.0 | 53.21 ± 0.01 | $90.0^{+0.0}_{-19.1}$ | $35^{+1}_{-18}$ | ··· |
| GW Lup | M1.5 | 38.7 ± 0.3 | $22.2^{+8.9}_{-11.4}$ | $16^{+10}_{-9}$ | ··· |
| IM Lup | K5.0 | 48.4 ± 0.3 | $71.0^{+19.0}_{-7.6}$ | $21^{+19}_{-7}$ | Y |
| IP Tau | M0.5 | 45.2 ± 0.3 | $53.6^{+10.3}_{-6.9}$ | 8±8 | ··· |
| IQ Tau | M1.0 | 62.1 ± 0.2 | $51.1^{+9.0}_{-6.3}$ | $10^{+5}_{-7}$ | ··· |
| LkCa 15 | K5.5 | 50.16 ± 0.03 | $90.0^{+0.0}_{-21.5}$ | $37^{+2}_{-20}$ | ··· |
| MHO 6 | M45.0 | 64.56 ± 0.03 | $54.0^{+10.7}_{-7.3}$ | $9^{+4}_{-8}$ | ··· |
| MP Mus | K1.0 | 32 ± 1 | $41.6^{+16.9}_{-8.7}$ | $9^{+15}_{-9}$ | ··· |
| MY Lup | K0.0 | 73.2 ± 0.1 | $65.0^{+20.3}_{-9.0}$ | $0^{+14}_{-0}$ | ··· |
| PDS 111 | G2.0 | 58.2 ± 0.1 | $44.2^{+8.9}_{-6.3}$ | $13^{+6}_{-7}$ | ··· |
| PDS 70 | K7.0 | 51.7 ± 0.1 | $57.3^{+16.5}_{-8.9}$ | $0^{+18}_{-0}$ | ··· |
| RU Lup | K7.0 | 18.8 ± 1.6 | $20.2^{+5.8}_{-3.9}$ | $0^{+6}_{-0}$ | ··· |
| RX J1615 | K5.0 | 46.5 ± 0.5 | $90.0^{+0.0}_{-21.0}$ | $41^{+2}_{-20}$ | Y |
| RX J1842 | K3.0 | 32 ± 5 | $90.0^{+0.0}_{-22.7}$ | $47^{+9}_{-16}$ | Y |
| RX J1852 | K4.0 | 30 ± 5 | $67.0^{+21.5}_{-8.5}$ | $42^{+12}_{-19}$ | Y |
| RY Lup | K4.0 | 67.7 ± 0.6 | $89.2^{+0.8}_{-24.9}$ | $1^{+13}_{-1}$ | ··· |
| SU Aur | G4.0 | 53.0 ± 1.5 | $72.3^{+17.1}_{-8.7}$ | $0^{+25}_{-0}$ | ··· |
| SY Cha | K2.0 | 51.7 ± 1.2 | $90.0^{+0.0}_{-23.4}$ | $33^{+3}_{-20}$ | ··· |
| Sz 114 | M5.0 | 21.3 ± 1.3 | $9.8^{+3.2}_{-4.6}$ | $11^{+5}_{-3}$ | Y |
| Sz 73 | K8.5 | 49.7 ± 3.9 | $58.0^{+22.9}_{-10.4}$ | $0^{+23}_{-0}$ | ··· |
| Sz 90 | K7.0 | 61.3 ± 5.3 | $52.3^{+26.9}_{-20.2}$ | $0^{+24}_{-0}$ | ··· |
| T Cha | G8.0 | 73 ± 5 | $76.4^{+13.6}_{-8.7}$ | $0^{+12}_{-0}$ | ··· |
| TW Hya | K0.5 | 6.9 ± 2.9 | $19.0^{+3.5}_{-2.7}$ | $12^{+5}_{-4}$ | Y |
| V1094 Sco | K6.0 | 53.0 ± 0.2 | $55.6^{+26.1}_{-10.6}$ | $1^{+17}_{-1}$ | ··· |
| V836 Tau | M1.0 | 43.1 ± 0.8 | $90.0^{+0.0}_{-19.1}$ | $43^{+3}_{-17}$ | Y |
| WSB 52 | K5.0 | 54.4 ± 0.3 | $90.0^{+0.0}_{-9.2}$ | $34^{+1}_{-9}$ | Y |
| WSB 63 | M1.5 | 67.3 ± 0.5 | $90.0^{+0.0}_{-18}$ | $0^{+15}_{-0}$ | ··· |

**Extended Data Table 2 | Adopted spectral types, disk inclinations, stellar inclinations, and the resulting minimum obliquities.** Quoted values of the stellar inclinations and minimum obliquities represent the maximum a posteriori (MAP) value and 68% highest density interval determined from their associated posterior and probability distributions.

| Model | Hyperprior | Parameter | |
|---|---|---|---|
| | | $\nu$ | |
| Rayleigh | log-Uniform | $11.7^{\circ}{}^{+1.8^{\circ}}_{-1.8^{\circ}}$ | |
| | Truncated Gaussian | $12.6^{\circ}{}^{+1.8^{\circ}}_{-1.8^{\circ}}$ | |
| | | $\sigma$ | μ |
| Gaussian | log-Uniform | $4.5^{\circ}{}^{+1.8^{\circ}}_{-3.6^{\circ}}$ | $15.3^{\circ}{}^{+1.8^{\circ}}_{-1.8^{\circ}}$ |
| | Truncated Gaussian | $6.3^{\circ}{}^{+1.8^{\circ}}_{-1.8^{\circ}}$ | $16.2^{\circ}{}^{+1.8^{\circ}}_{-1.8^{\circ}}$ |
| | | $\sigma$ | μ |
| Truncated Gaussian | Uniform | $9.5^{\circ}{}^{+2.2^{\circ}}_{-3.0^{\circ}}$ | $-615.0^{\circ}{}^{+327.1^{\circ}}_{-266.5^{\circ}}$ |

**Extended Data Table 3 | Posteriors of HBM parameters from MCMC fitting.** The best-fit parameters are shown for a Rayleigh, Gaussian, and Truncated Gaussian model of the underlying distribution of $\Delta i$, given either a log-Uniform, Truncated Gaussian, or Uniform hyperprior.

| Identifier | RA (J2000) (h m s) | Dec (J2000) (° ′ ″) | $T_{eff}$ (K) | $M_*$ ($M_\odot$) | $v\sin i_*$ (km $s^{-1}$) | $P_{rot}$ (d) | $R_*$ ($R_\odot$) |
|---|---|---|---|---|---|---|---|
| 2M J0412* | 04 12 40.711 | +24 38 15.404 | 3337 ± 3 | 0.4 ± 0.4 | 6.4 ± 0.7 | 5.7 ± 0.2 | 1.0 ± 0.2 |
| 2M J0420* | 04 20 25.557 | +27 00 35.549 | 3083 ± 32 | 0.25 ± 0.01 | 14.2 ± 0.8 | 2.2 ± 0.1 | 0.9 ± 0.4 |
| 2M J0433* | 04 33 44.659 | +26 15 00.436 | 3143 ± 4 | 0.13 ± 0.03 | 15.9 ± 0.8 | 5.9 ± 0.2 | 1.5 ± 0.5 |
| 2M J0434* | 04 34 31.289 | +17 22 20.122 | 3294 ± 4 | 0.38 ± 0.10 | 8.7 ± 0.6 | 6.0 ± 0.2 | 0.9 ± 0.2 |
| 2M J0436* | 04 36 01.313 | +17 26 12.091 | 3592 ± 86 | 0.54 ± 0.06 | 16.4 ± 0.8 | 2.7 ± 0.1 | 1.0 ± 0.1 |
| 2M J1100* | 11 00 40.242 | −76 19 28.007 | 3340 ± 64 | 0.32 ± 0.03 | 5.4 ± 0.8 | 2.8 ± 0.1 | 0.9 ± 0.2 |
| 2M J1608* | 16 08 30.685 | −38 28 27.288 | 4856 ± 53 | 1.4 ± 0.1 | <8.0 | 6.2 ± 0.2 | 1.7 ± 0.3 |
| AA Tau | 04 34 55.427 | +24 28 52.671 | 3879 ± 172 | 0.62 ± 0.04 | 11.6 ± 0.3 | 8.2 ± 0.3 | 2.4 ± 1.6 |
| AS 209 | 16 49 15.294 | −14 22 09.057 | 4118 ± 87 | 1.1 ± 0.09 | 10.4 ± 0.4 | 8.5 ± 0.4 | 2.0 ± 0.4 |
| BP Tau | 04 19 15.846 | +29 06 26.471 | 3843 ± 14 | 0.66 ± 0.06 | 9.1 ± 0.4 | 7.3 ± 0.3 | 1.4 ± 0.2 |
| CI Tau | 04 33 52.026 | +22 50 30.094 | 4049 ± 74 | 0.90 ± 0.02 | 10.2 ± 0.3 | 9.0 ± 0.4 | 1.7 ± 0.3 |
| CIDA-7 | 04 42 21.022 | +25 20 34.306 | 2997 ± 77 | 0.51 ± 0.02 | 19.1 ± 1.0 | 1.90 ± 0.02 | 0.7 ± 0.1 |
| CR Cha | 10 59 06.853 | −77 01 40.270 | 5036 ± 61 | 1.80 ± 0.16 | 36.5 ± 0.9 | 2.3 ± 0.2 | 2.2 ± 0.4 |
| CX Tau | 04 14 47.873 | +26 48 10.620 | 3681 ± 42 | 0.37 ± 0.01 | 20.1 ± 0.4 | 3.3 ± 0.1 | 1.4 ± 0.2 |
| DE Tau | 04 21 55.649 | +27 55 05.708 | 3685 ± 7 | 0.37 ± 0.05 | 9.5 ± 0.3 | 5.9 ± 0.3 | 1.6 ± 0.3 |
| DN Tau | 04 35 27.385 | +24 14 58.545 | 3883 ± 32 | 0.60 ± 0.04 | 10.7 ± 0.3 | 6.3 ± 0.2 | 3.6 ± 0.6 |
| DoAr 25 | 16 26 23.681 | −24 43 14.346 | 4326 ± 144 | 0.92 ± 0.09 | 15.9 ± 0.8 | 9.0 ± 0.7 | 2.5 ± 0.2 |
| DoAr 44 | 16 31 33.456 | −24 27 37.582 | 4614 ± 48 | 1.3 ± 0.1 | 16.3 ± 0.5 | 3.0 ± 0.1 | 1.6 ± 0.3 |
| Elias 2-24 | 16 26 24.078 | −24 16 13.881 | 4056 ± 138 | 0.86 ± 0.11 | 16.2 ± 0.9 | 6.6 ± 0.4 | 4.1 ± 0.2 |
| EX Lup | 16 03 05.492 | −40 18 25.427 | 3898 ± 46 | 0.57 ± 0.07 | 6.0 ± 1.4 | 7.4 ± 0.3 | 1.4 ± 0.3 |
| FP Tau | 04 14 47.315 | +26 46 26.017 | 3508 ± 63 | 0.39 ± 0.01 | 31.0 ± 0.8 | 2.19 ± 0.03 | 1.4 ± 0.2 |
| FT Tau | 04 23 39.198 | +24 56 13.868 | 3453 ± 11 | 0.40 ± 0.01 | 10.2 ± 0.6 | 6.0 ± 0.3 | 1.5 ± 0.1 |
| GM Aur | 04 55 10.987 | +30 21 58.948 | 4332 ± 30 | 1.10 ± 0.02 | 14.0 ± 0.4 | 6.0 ± 0.2 | 1.5 ± 0.3 |
| GW Lup | 15 46 44.709 | −34 30 36.086 | 3682 ± 85 | 0.41 ± 0.03 | <8.0 | 4.1 ± 0.3 | 1.3 ± 0.2 |
| IM Lup | 15 56 09.189 | −37 56 06.541 | 4158 ± 74 | 0.65 ± 0.05 | 14.8 ± 0.5 | 7.2 ± 0.3 | 2.2 ± 0.4 |
| IP Tau | 04 24 57.093 | +27 11 56.075 | 3813 ± 26 | 0.55 ± 0.03 | 11.1 ± 0.3 | 5.7 ± 0.2 | 1.1 ± 0.2 |
| IQ Tau | 04 29 51.565 | +26 06 44.486 | 3730 ± 26 | 0.69 ± 0.02 | 14.1 ± 0.4 | 6.3 ± 0.2 | 2.2 ± 0.2 |
| LkCa 15 | 04 39 17.804 | +22 21 03.083 | 4461 ± 36 | 1.20 ± 0.03 | 14.8 ± 0.4 | 5.8 ± 0.2 | 1.6 ± 0.3 |
| MHO 6 | 04 32 22.110 | +18 27 42.650 | 3103 ± 57 | 0.21 ± 0.01 | 20.9 ± 1.0 | 1.94 ± 0.02 | 1.0 ± 0.1 |
| MP Mus | 13 22 07.414 | −69 38 12.569 | 4588 ± 7 | 1.30 ± 0.06 | 13.8 ± 0.3 | 3.7 ± 0.1 | 1.5 ± 0.3 |
| MY Lup | 16 00 44.502 | −41 55 31.340 | 4619 ± 37 | 1.10 ± 0.04 | 29.1 ± 2.0 | 2.6 ± 0.1 | 1.7 ± 0.3 |
| PDS 111 | 05 24 37.250 | −08 42 01.715 | 5635 ± 85 | 1.13 ± 0.06 | 21.0 ± 1.1 | 3.2 ± 0.2 | 1.9 ± 0.2 |
| PDS 70 | 14 08 10.108 | −41 23 52.994 | 4016 ± 60 | 0.70 ± 0.06 | 16.8 ± 0.5 | 3.0 ± 0.1 | 1.2 ± 0.1 |
| RU Lup | 15 56 42.294 | −37 49 15.880 | 4125 ± 69 | 0.85 ± 0.08 | 8.5 ± 0.6 | 3.7 ± 0.1 | 1.8 ± 0.3 |
| RX J1615* | 16 15 20.223 | −32 55 05.507 | 4329 ± 113 | 0.78 ± 0.06 | 15.0 ± 1.4 | 5.7 ± 0.2 | 1.5 ± 0.3 |
| RX J1842* | 18 42 57.983 | −35 32 43.298 | 4619 ± 82 | 0.74 ± 0.07 | 26.7 ± 2.9 | 2.6 ± 0.1 | 1.3 ± 0.2 |
| RX J1852* | 18 52 17.306 | −37 00 12.444 | 4729 ± 57 | 0.88 ± 0.05 | 19.7 ± 0.9 | 2.8 ± 0.6 | 1.2 ± 0.1 |
| RY Lup | 15 59 28.370 | −40 21 51.638 | 4896 ± 62 | 1.20 ± 0.07 | 21.9 ± 2.3 | 3.7 ± 0.1 | 1.6 ± 0.3 |
| SU Aur | 04 55 59.392 | +30 34 01.074 | 5726 ± 54 | 2.2 ± 0.5 | 61.3 ± 1.6 | 2.7 ± 0.1 | 3.5 ± 0.6 |
| SY Cha | 10 56 30.263 | −77 11 39.352 | 3931 ± 38 | 0.88 ± 0.24 | 13.6 ± 0.4 | 6.1 ± 0.2 | 1.6 ± 0.3 |
| Sz 114 | 16 09 01.834 | −39 05 12.835 | 3279 ± 25 | 0.20 ± 0.03 | <8.0 | 2.00 ± 0.03 | 1.5 ± 0.3 |
| Sz 73 | 15 47 56.923 | −35 14 35.187 | 3953 ± 31 | 0.74 ± 0.12 | 31 ± 3 | 1.9 ± 0.1 | 1.4 ± 0.3 |
| Sz 90 | 16 07 10.055 | −39 11 03.703 | 3861 ± 41 | 0.60 ± 0.08 | <8.0 | 6.7 ± 0.6 | 1.1 ± 0.2 |
| T Cha | 11 57 13.262 | −79 21 31.681 | 5485 ± 119 | 1.25 ± 0.35 | 37.7 ± 1.0 | 3.23 ± 0.06 | 2.5 ± 0.3 |
| TW Hya | 11 01 51.808 | −34 42 17.276 | 3986 ± 30 | 0.72 ± 0.03 | 4.9 ± 0.4 | 3.6 ± 0.1 | 1.1 ± 0.1 |
| V1094 Sco | 16 08 36.162 | −39 23 02.871 | 4264 ± 60 | 0.69 ± 0.07 | 21.4 ± 1.0 | 3.5 ± 0.1 | 1.8 ± 0.6 |
| V836 Tau | 05 03 06.603 | +25 23 19.303 | 3699 ± 38 | 0.65 ± 0.07 | 11.9 ± 0.4 | 6.8 ± 0.2 | 1.4 ± 0.3 |
| WSB 52 | 16 27 39.422 | −24 39 15.971 | 3242 ± 19 | 0.46 ± 0.03 | 17.9 ± 0.6 | 6.2 ± 0.4 | 1.8 ± 0.1 |
| WSB 63 | 16 28 54.069 | −24 47 44.729 | 3598 ± 33 | 0.43 ± 0.07 | 10.4 ± 2.1 | 12.3 ± 1.3 | 1.9 ± 0.1 |

## References for Main Article and Methods

# Supplementary Information

“One-third of Sun-like Stars are born with misaligned planet-forming disks”

# Contents

## Supplementary Discussion

## Supplementary Material

## References for Supplementary Information

## Abbreviated Identifiers:

**2M J0412** = 2MASS J04124068+2438157
**2M J0420** = 2MASS J04202555+2700355
**2M J0433** = 2MASS J04334465+2615005
**2M J0434** = 2MASS J04343128+1722201
**2M J0436** = 2MASS J04360131+1726120
**2M J1100** = 2MASS J11004022-7619280
**2M J1608** = 2MASS J16083070-3828268
**RX J1615** = RX J1615.3-3255
**RX J1842** = RX J1842.9-3532
**RX J1852** = RX J1852.3-3700

# Supplementary Discussion

## SD–1: Distribution of Sample Disk Orientations

Randomly oriented disks will have inclinations that follow a $\sin i$ distribution. However, the distribution of the orientation of the disks in our sample appears to favor moderate inclinations of $\sim 30°–75°$ (Extended Data Figure 1). The lack of inclinations near 90° (corresponding to an edge-on orientation) is the result of selection biases in imaging surveys, which historically focus on disks with more face-on orientations, while inclinations near 0° are simply less likely for a random distribution of $i_{\rm disk}$. The list of measured $i_{\rm disk}$ and associated references is provided in SM–Table 2.

## SD–2: Stellar Inclination Parameter Space

In this analysis, the stellar inclination $i_*$ is defined over the interval [0°, 90°]. This is because the direction of the star's rotation (which would allow us to determine whether its angular momentum vector points toward or away from Earth) cannot be determined with $v\sin i_*$, $P_{\rm rot}$, and $R_*$. As a result, we do not mirror $i_*$ nor $i_{\rm disk}$ about 90°. The missing angle that we do not generally have access to is the longitude of ascending node. This can sometimes be determined for the disk with resolved gas kinematics, but it is very challenging to measure for stars (e.g., ref. 1). The implications are that for our computation of minimum obliquity, we are effectively assuming the angular momentum vectors of the star and disk are pointed in the same hemisphere (toward Earth). This approach avoids the assumption that prograde and retrograde orbits are equally likely *a priori*——a prior that may not be well founded. This does not dictate that the systems are restricted to prograde motion because these are minimum obliquities.

## SD–3: Stellar Light Curves

With visual inspection, the periodic signature identified as rotation must produce a clearly periodic phase curve to be considered reliable. In addition, we avoid signals that are likely to be an alias of the true rotation signature. However, the light curves of disk-bearing T Tauri stars are diverse[2–4] with morphological features attributed to a combination of temporal changes in star spots, hot accretion shocks, and obscuring material in the star's immediate environment[5–7]. The manifestation of these events in the light curve can make identifying the star's rotation period challenging. Here, we consider a rotation period reliable if the light curve's periodogram produces a single strong peak such that the shape of the phase-folded light curve is consistent with that period and the amplitude is significant when compared to the scatter in the data. We note that a few objects (AA Tau, BP Tau, CI Tau, DE Tau, LkCa 15, and TW Hya) show additional complexity in the light curve, but are still considered reliable. In these cases, there exist at least two additional forms of evidence that the periodic signal traces stellar rotation. For example, the primary period is sustained over years-long timescales, confirmed via magnetic topological maps, and/or in agreement with periodic modulation in spectroscopic emission features that trace corotating magnetospheric accretion shocks such as HeI $\lambda$5876 Å and CaII $\lambda$8542 Å.

The resulting distribution of $P_{\rm rot}$ is consistent with the range of rotation periods commonly observed in disk-bearing T Tauri stars (typically between ~2-10 d; e.g., refs. 6, 8–11). All *TESS* and *K2* light curves analyzed in this work, as well as the associated periodograms and phased light curves, are shown in SD-Figure 1. The compilation of rotation periods assembled from the literature as well as those determined in this work are provided in SM–Table 3.

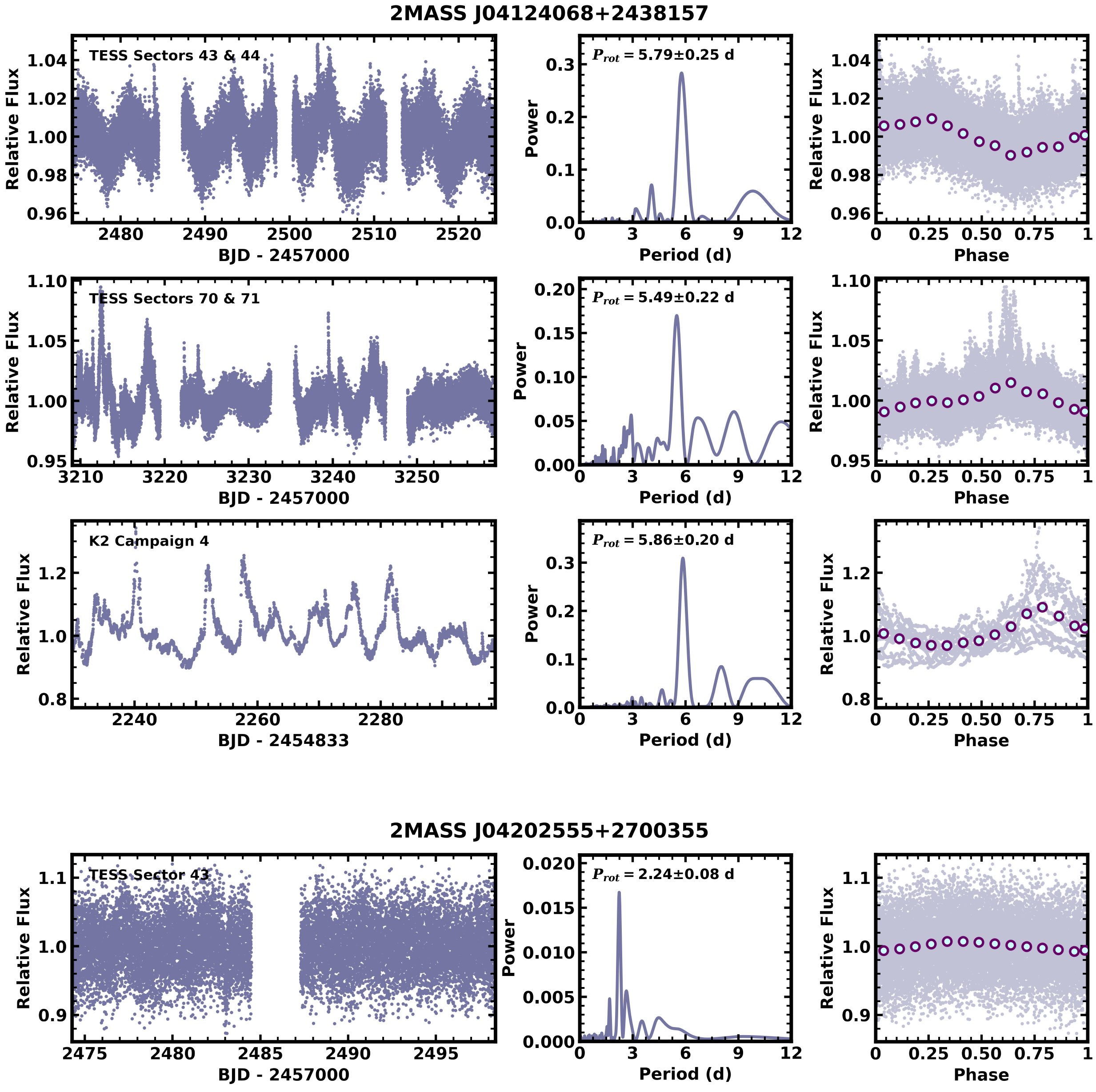

**SD–Figure 1.** TESS and K2 light curves, periodograms, and phase curves for objects in the sample. Notes on individual light curves are provided in the captions below, as needed.

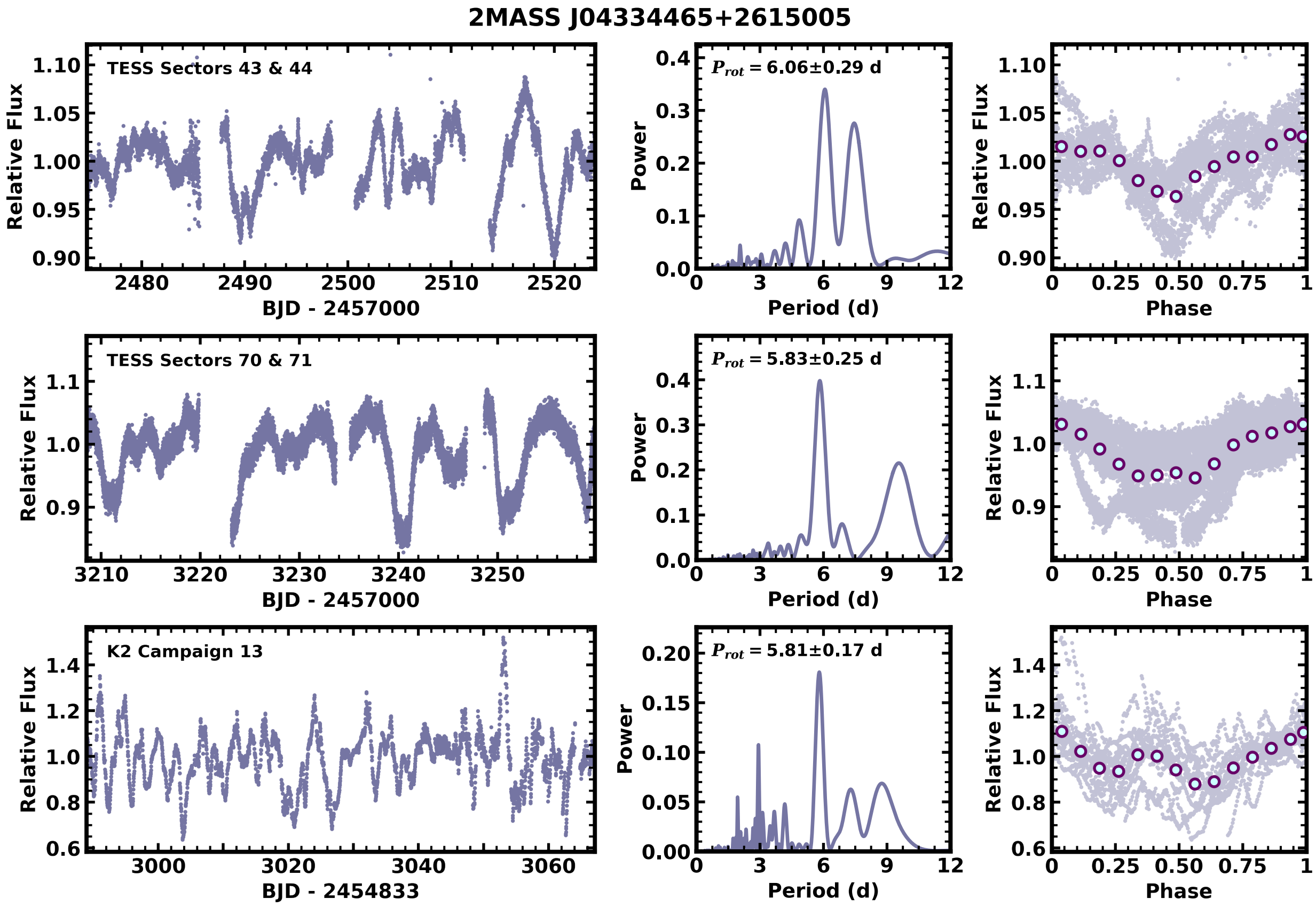

**SD–Figure 1.** *Continued.*

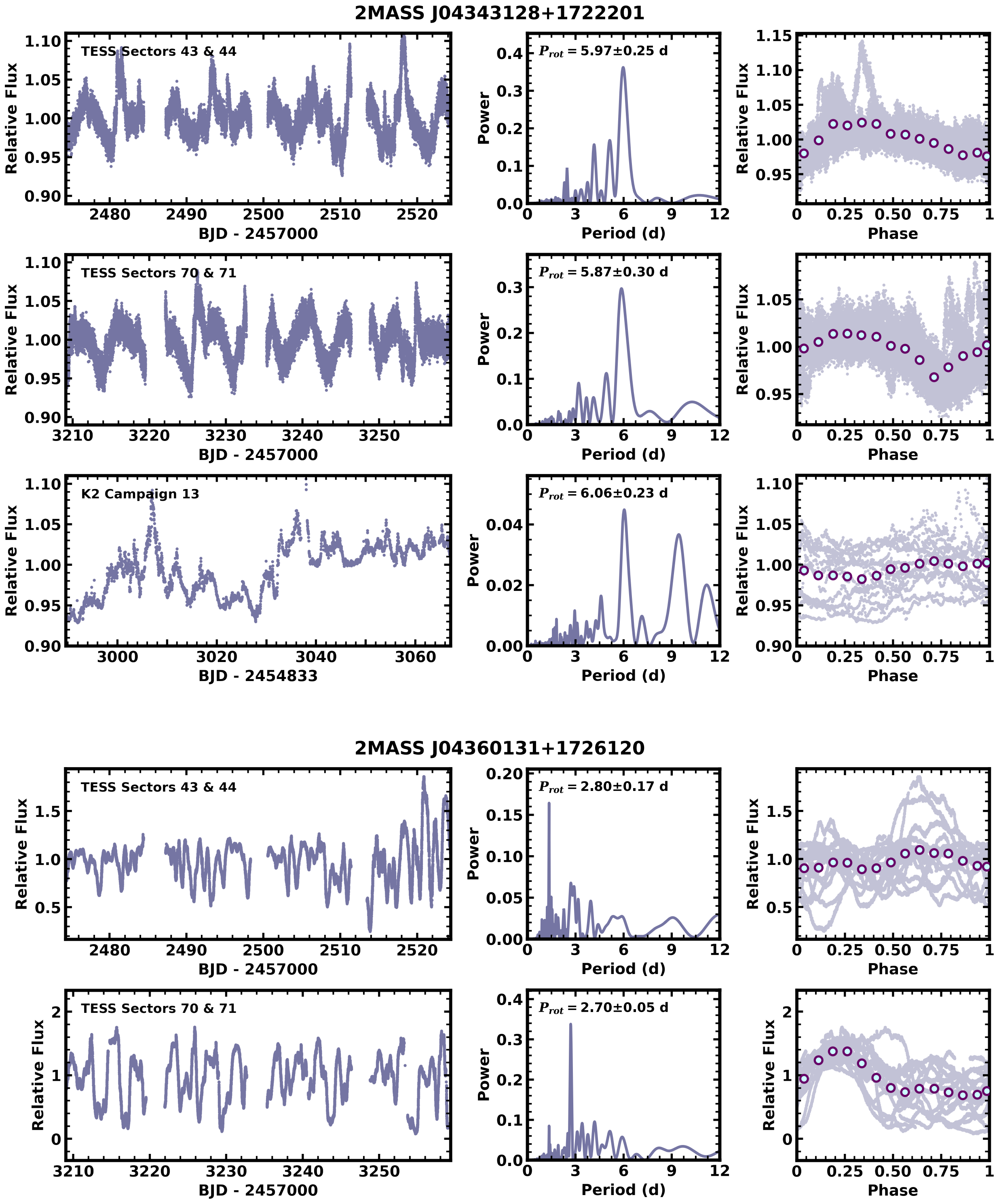

**SD–Figure 1.** *Continued*. For 2MASS J04360131+1726120, we adopt the rotation period identified in TESS Sectors 70 and 71.

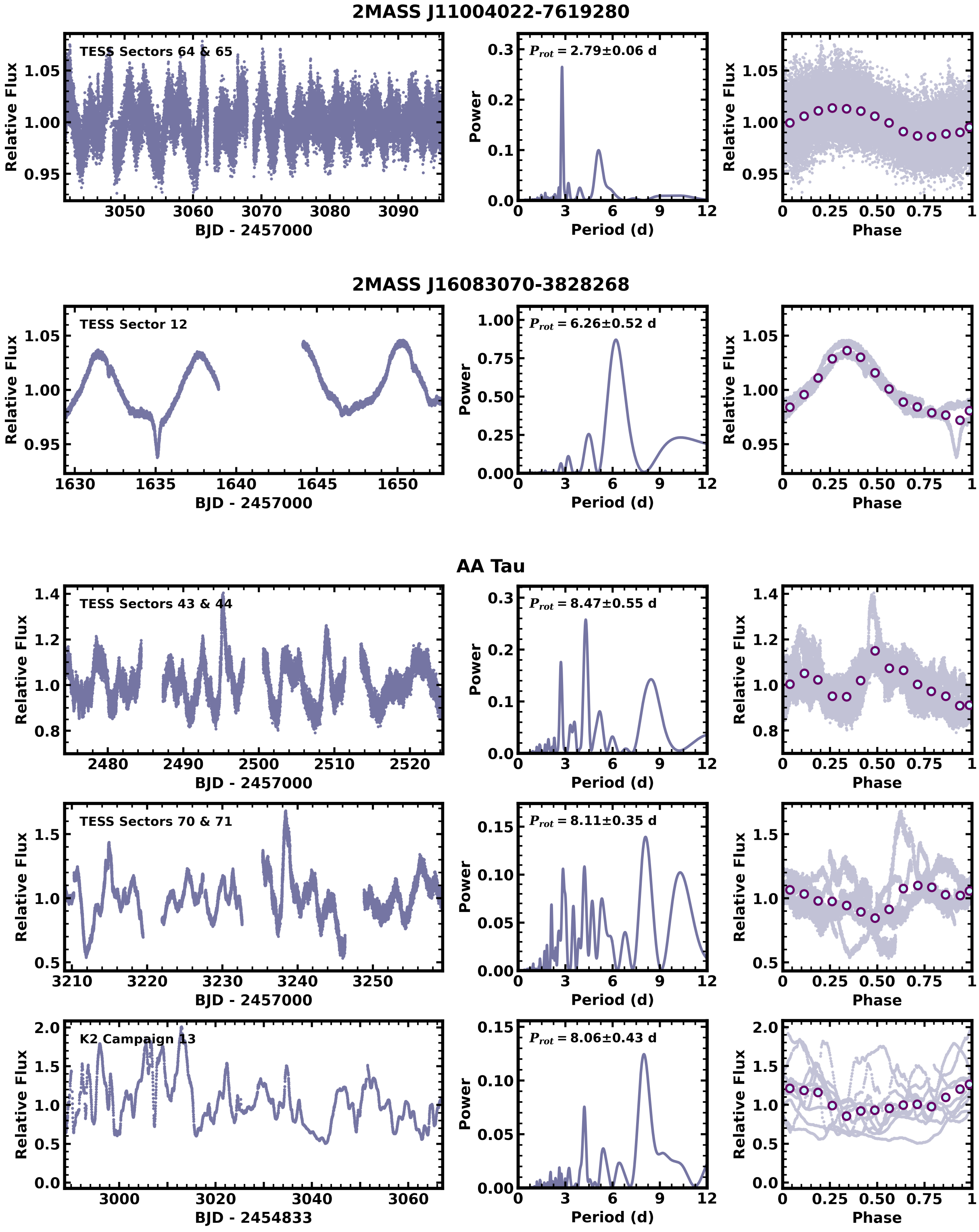

**SD–Figure 1.** *Continued.* We note that for AA Tau's TESS lightcurve of Sectors 43 and 44, the period corresponding to the highest power in the periodogram is ~4 days, which is approximately 1/2 the typical period measured for this object (See TESS Sectors 70 and 71, K2 Campaign 13, and other literature periods in SM-Table 3, and is therefore likely to be an effect of hot and/or cold spots located on opposite faces of the star. We therefore identify the rotation period in this lightcurve as that which corresponds to the next highest peak for periods greater than 4 days.

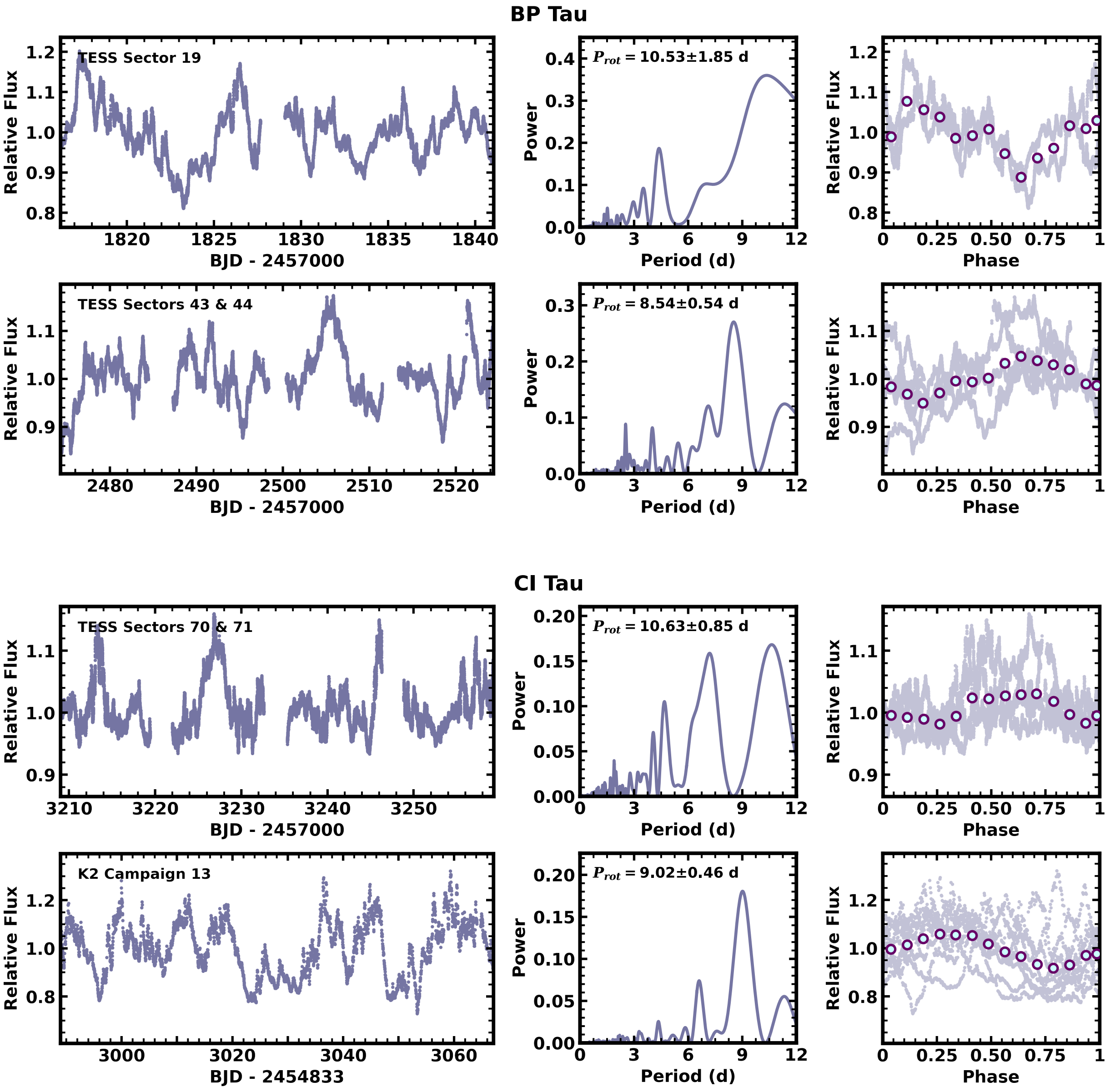

**SD–Figure 1.** *Continued.*

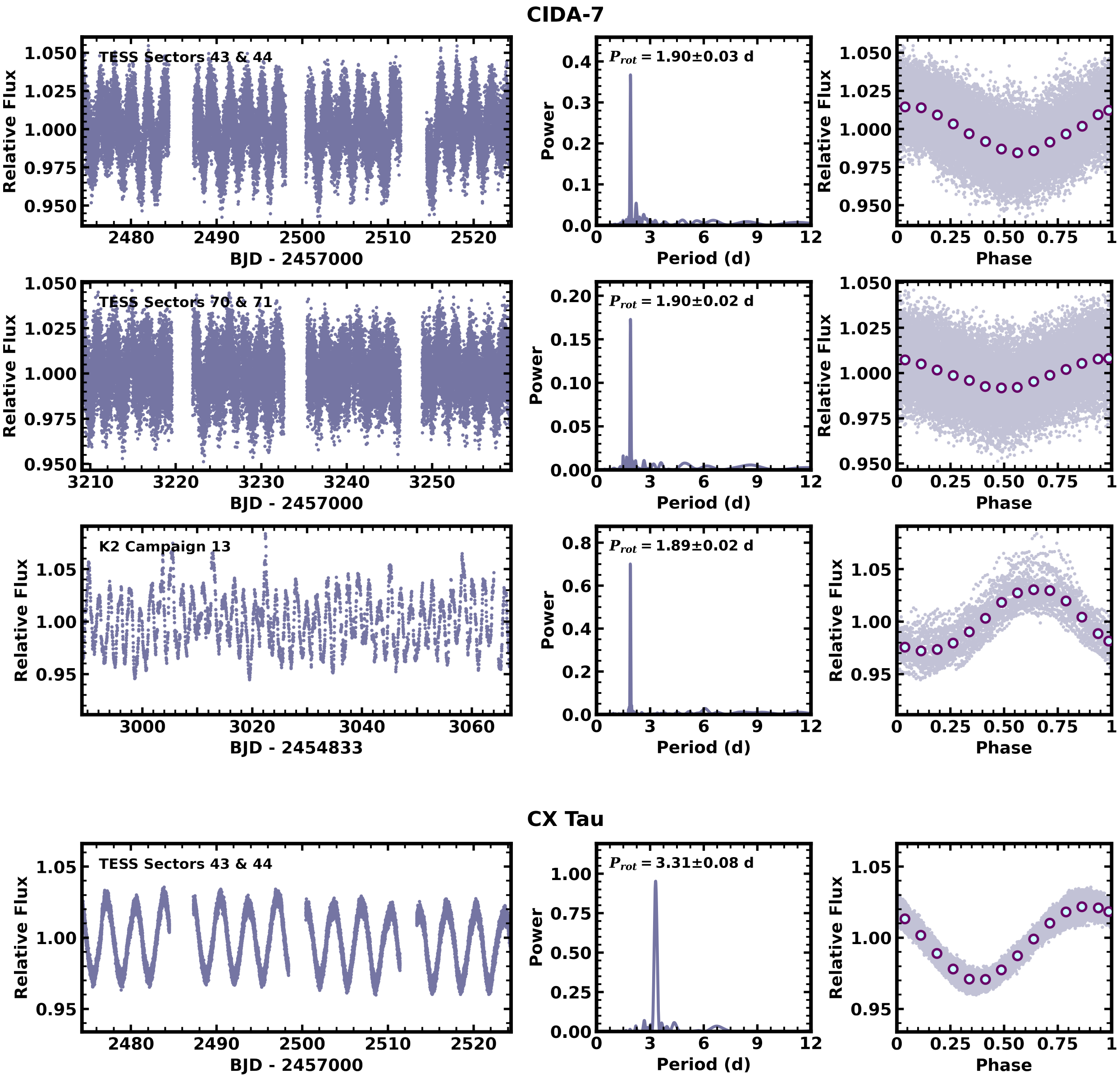

**SD–Figure 1.** *Continued.*

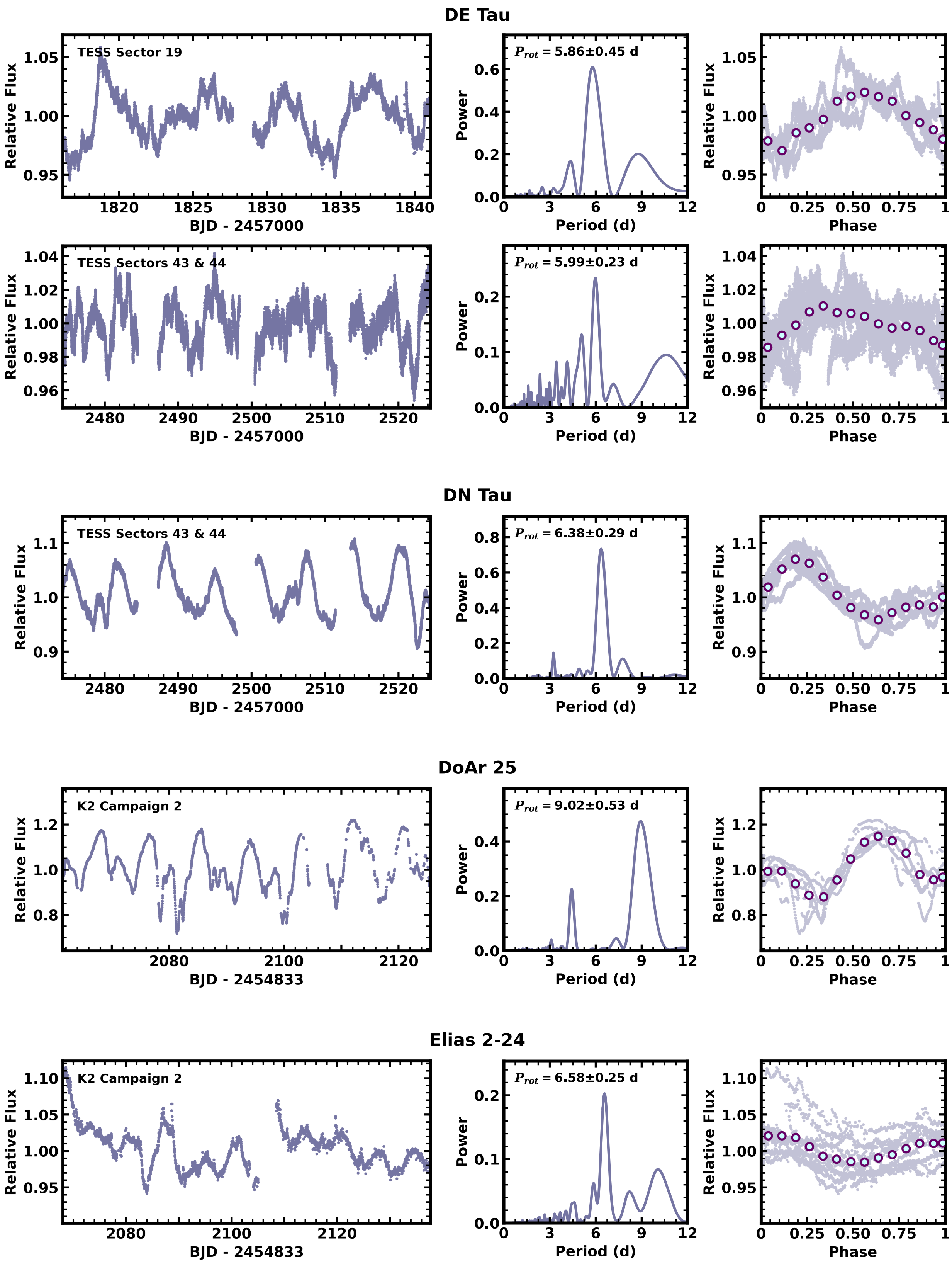

**SD–Figure 1.** *Continued.*

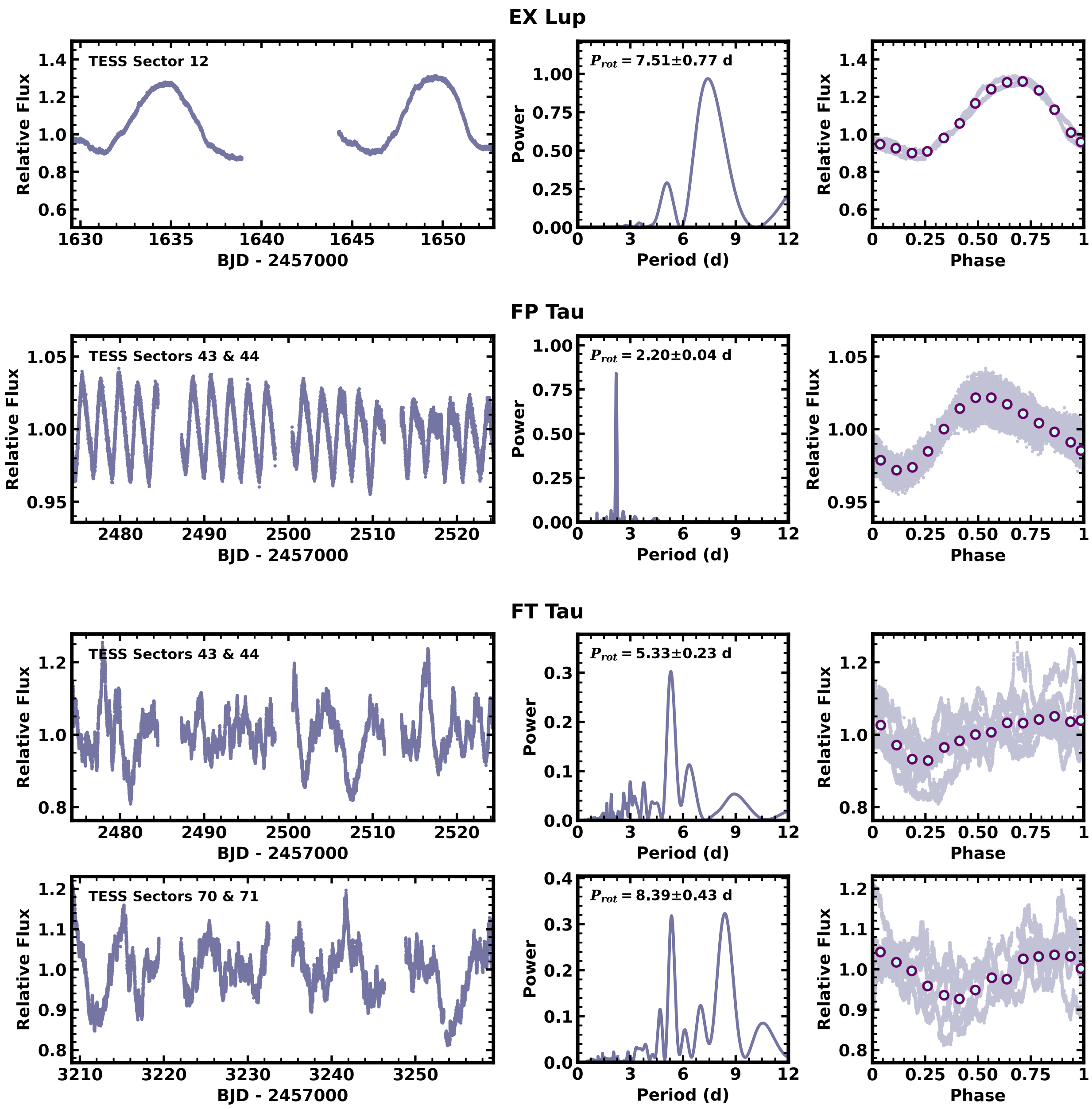

**SD–Figure 1.** *Continued.*

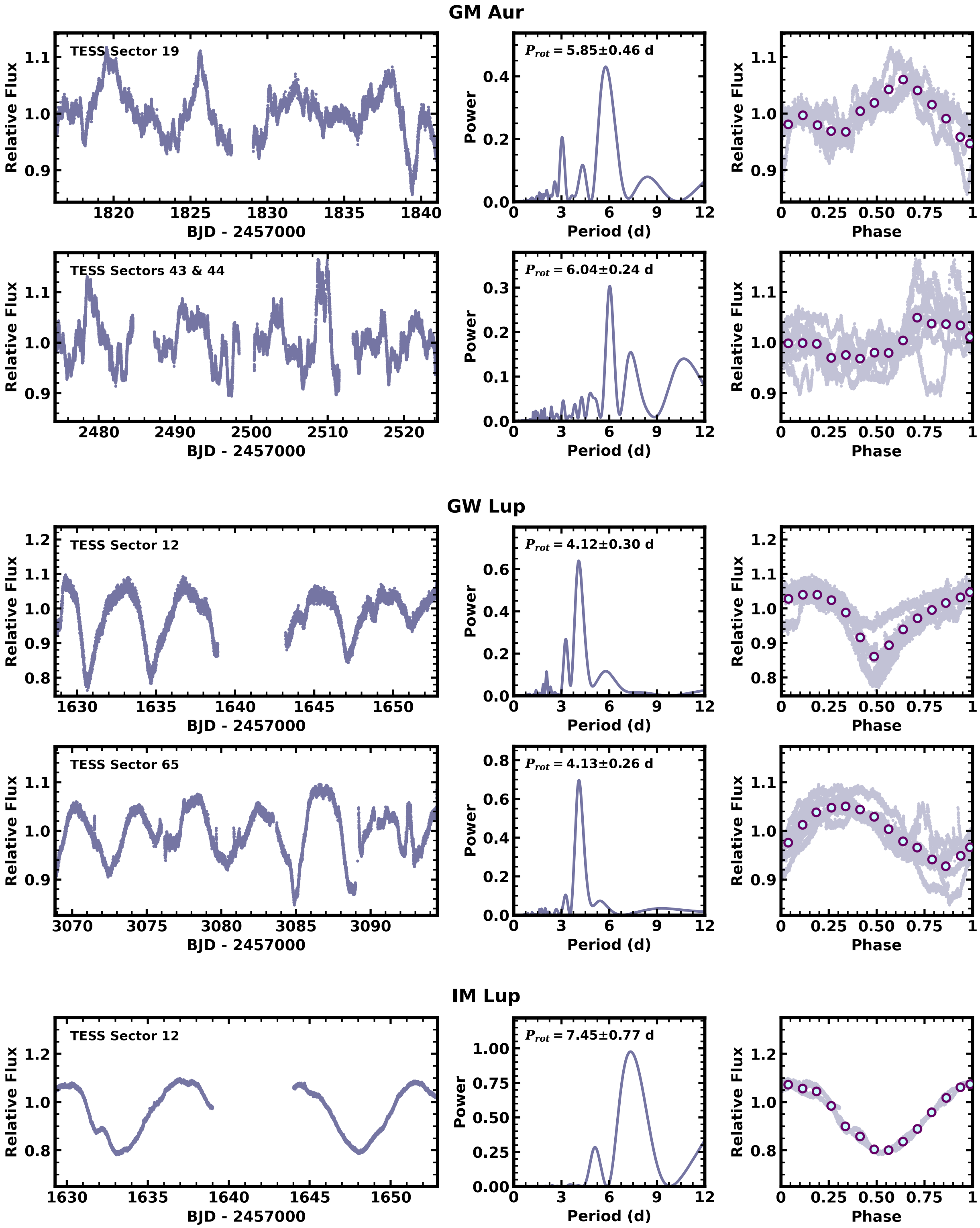

**SD–Figure 1.** *Continued.*

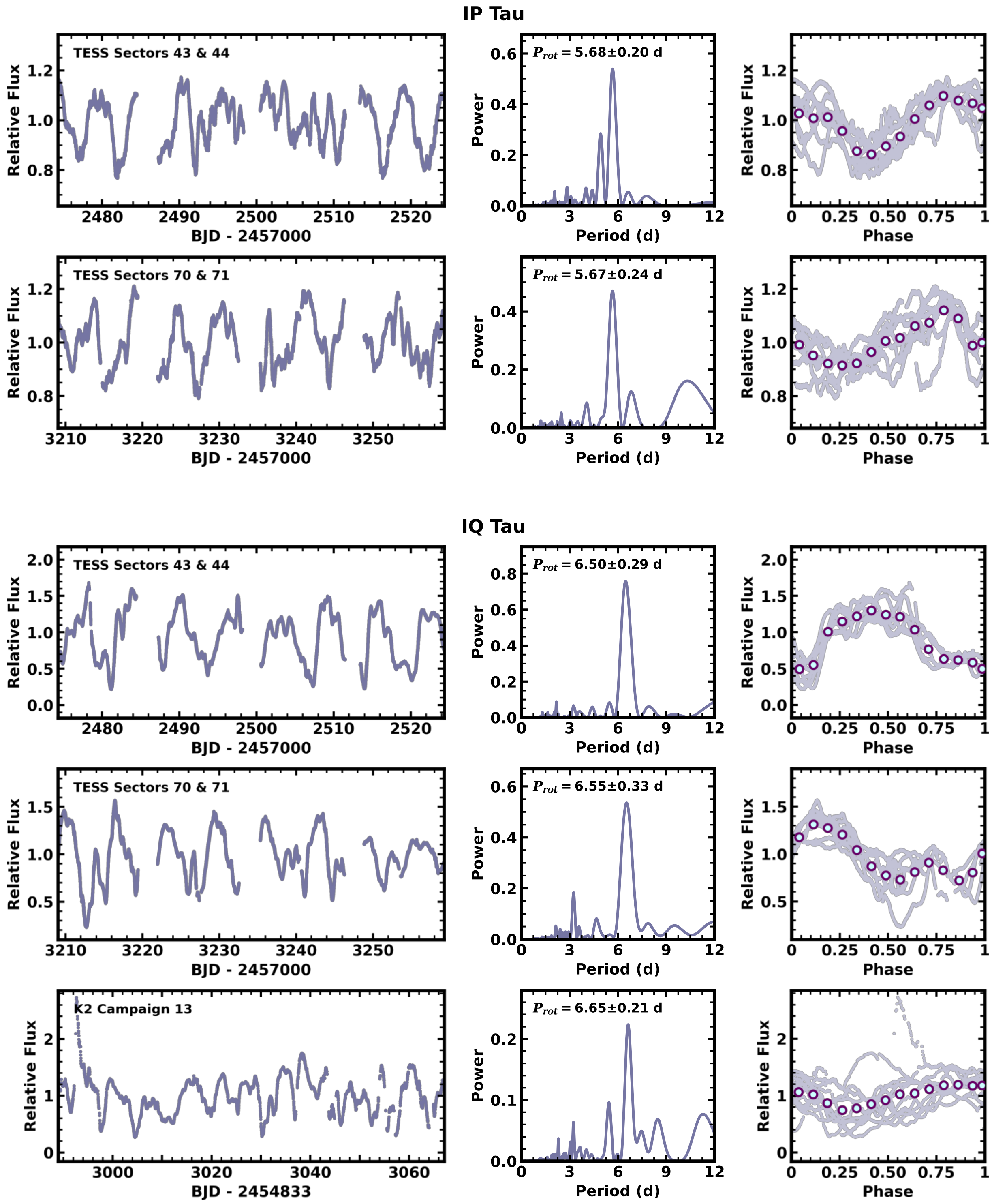

**SD–Figure 1.** *Continued.*

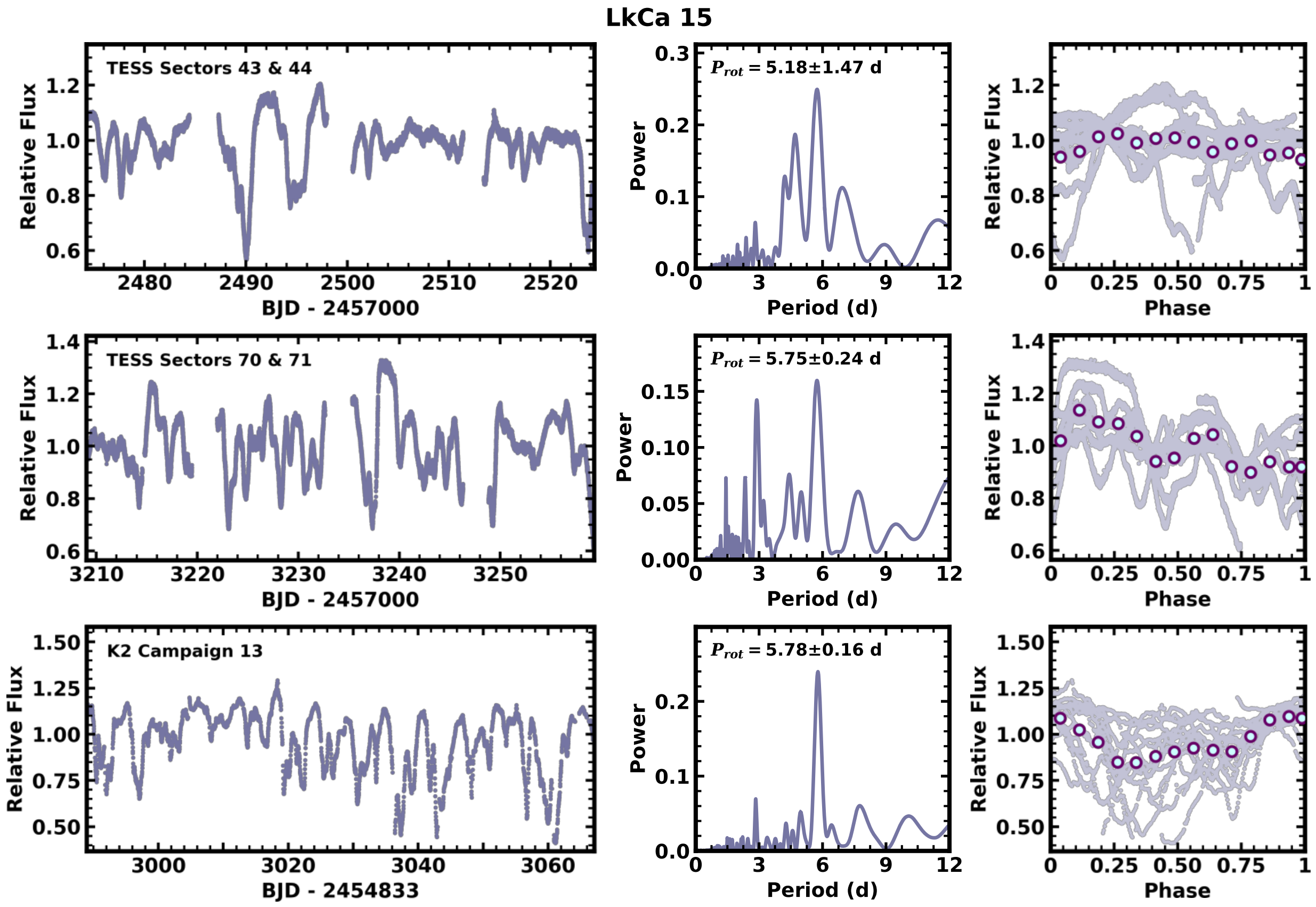

**SD–Figure 1.** *Continued.*

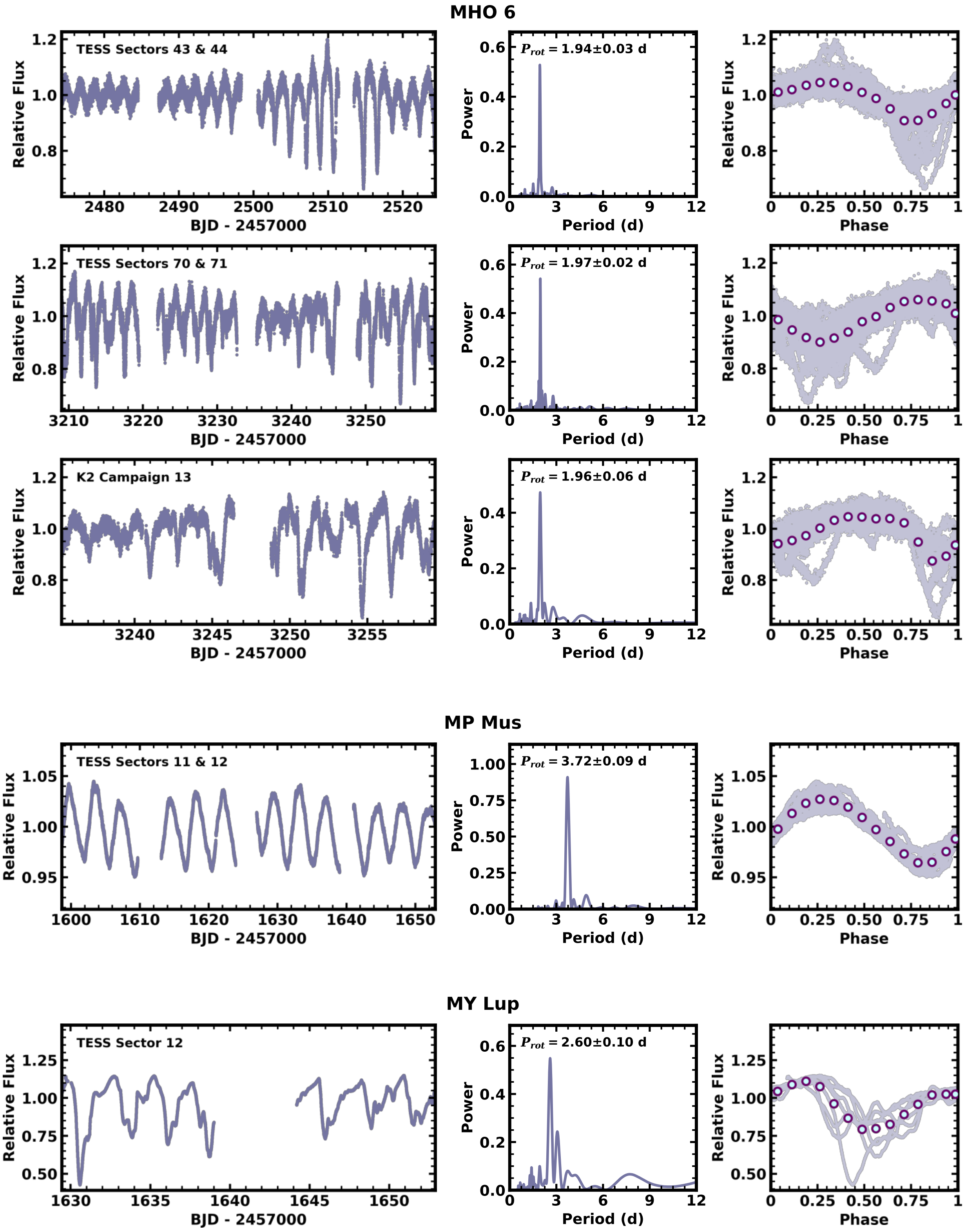

**SD–Figure 1.** *Continued.*

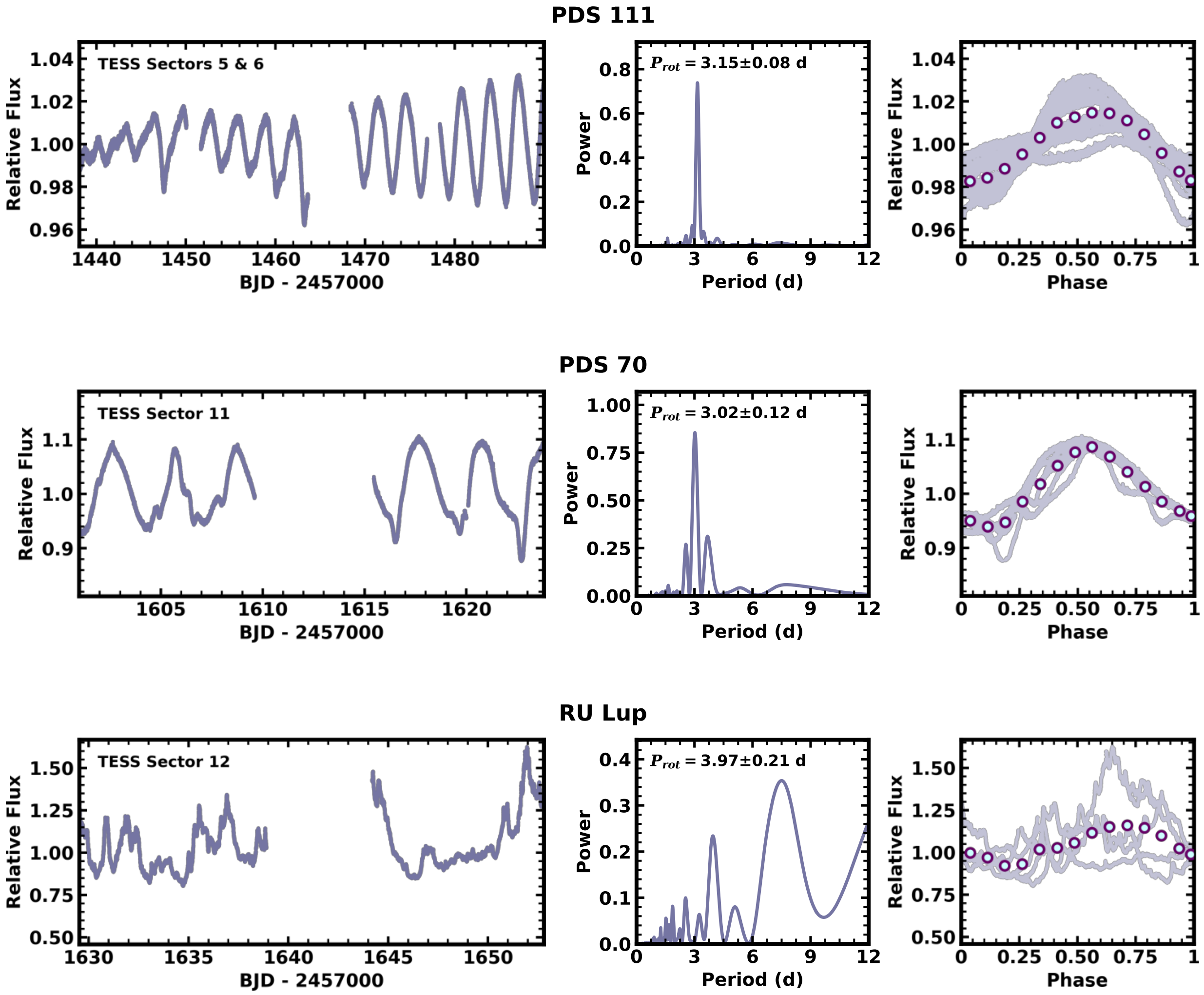

**SD–Figure 1.** *Continued.*

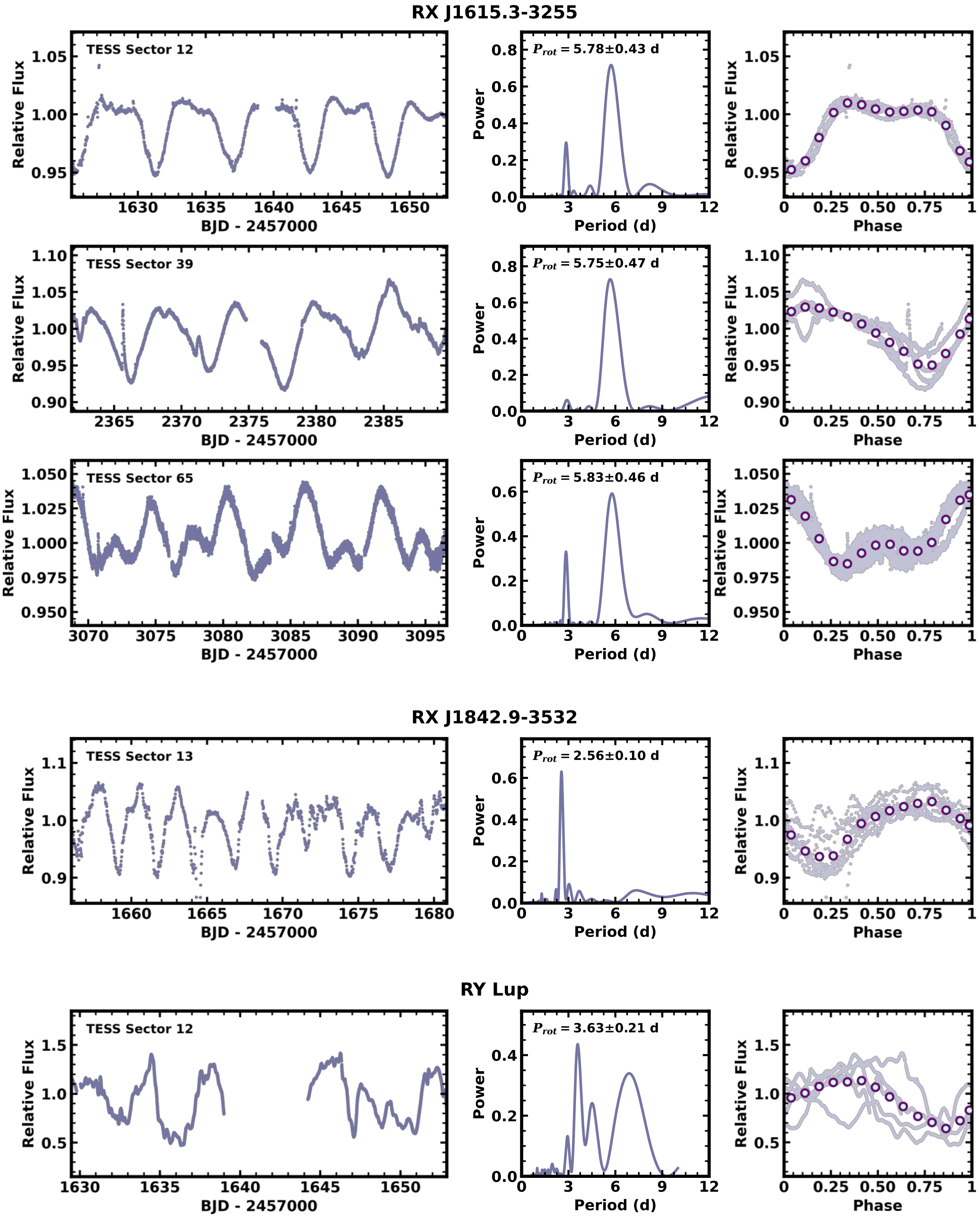

**SD–Figure 1.** *Continued.*

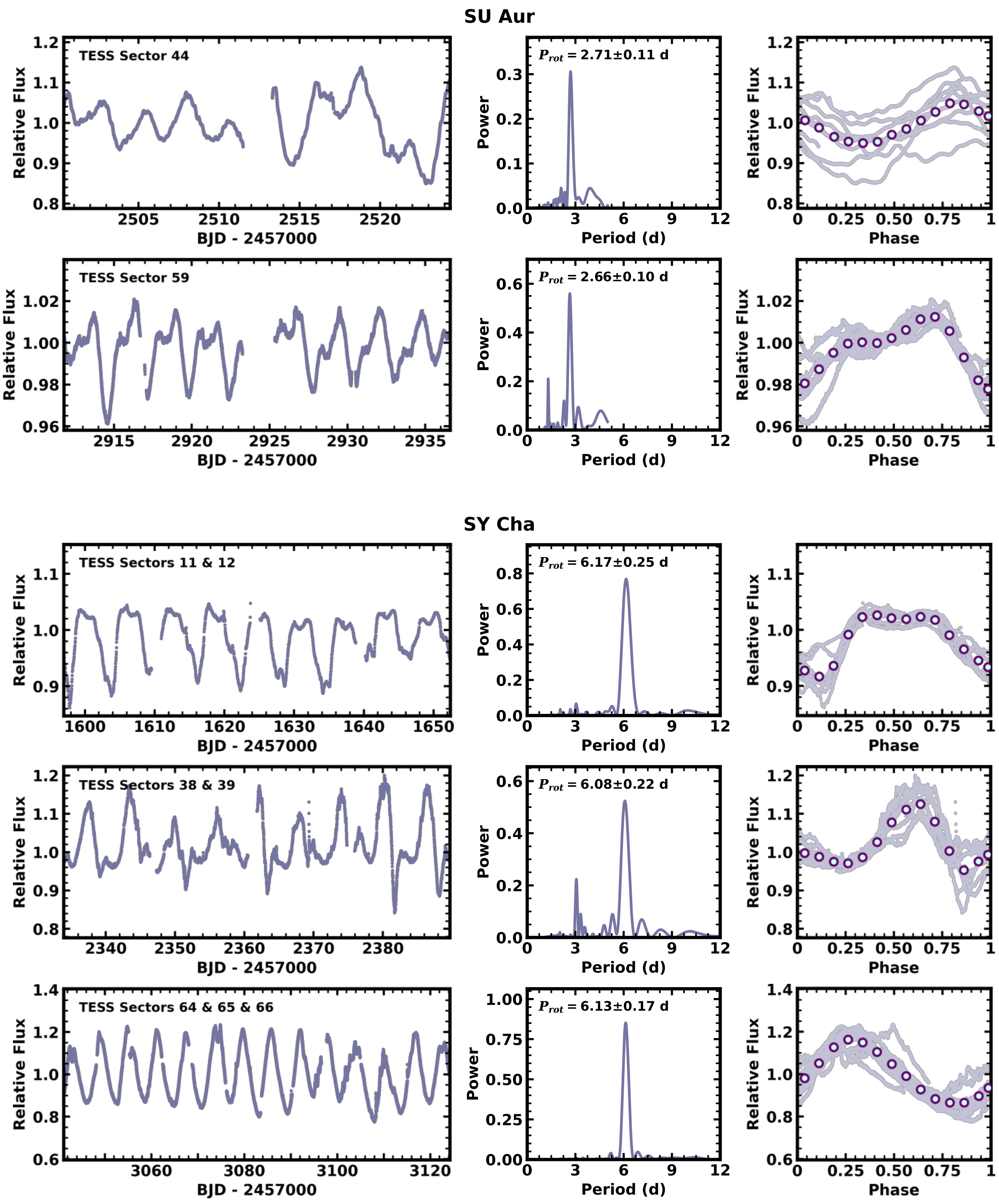

**SD–Figure 1.** *Continued.*

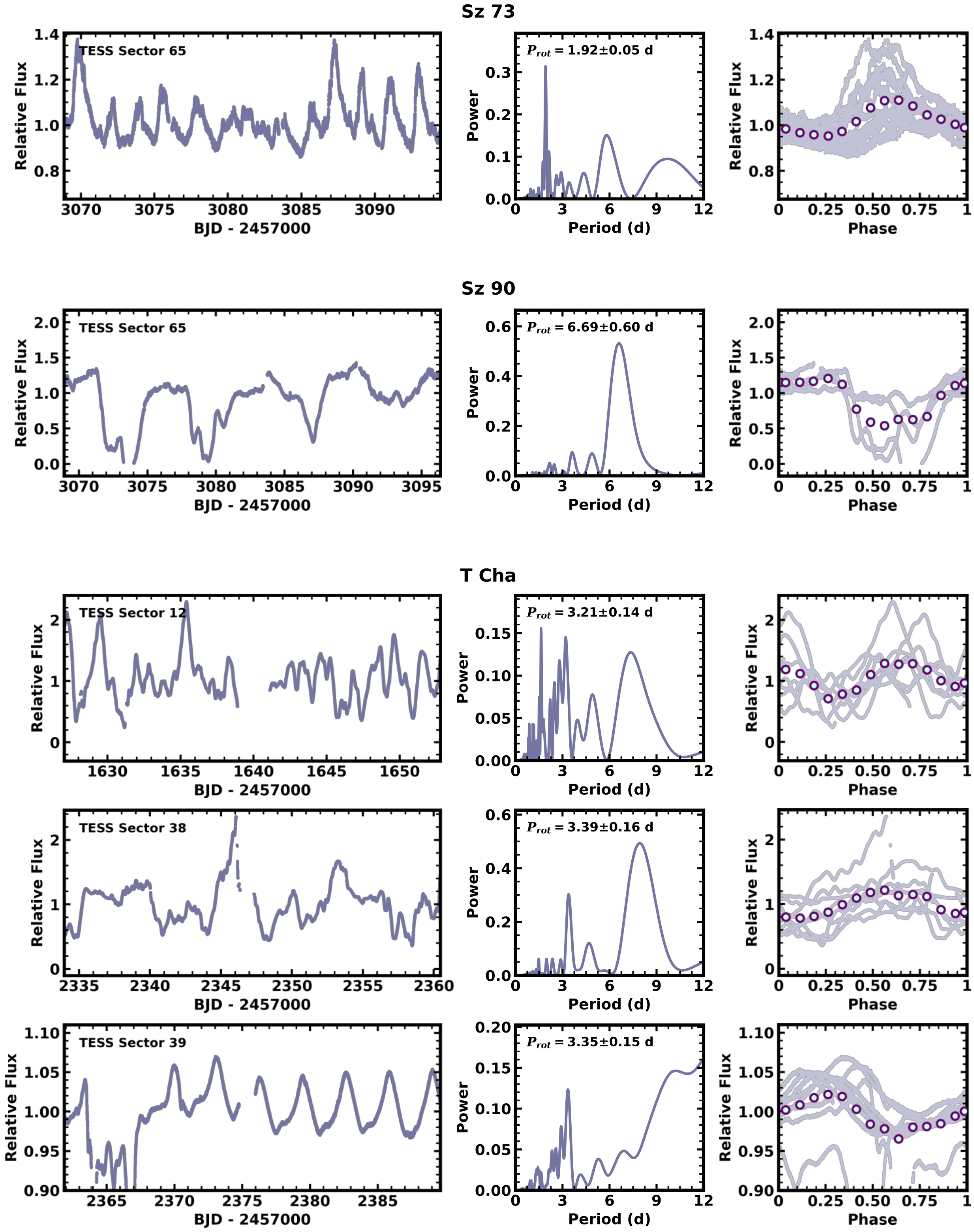

**SD–Figure 1.** *Continued.*

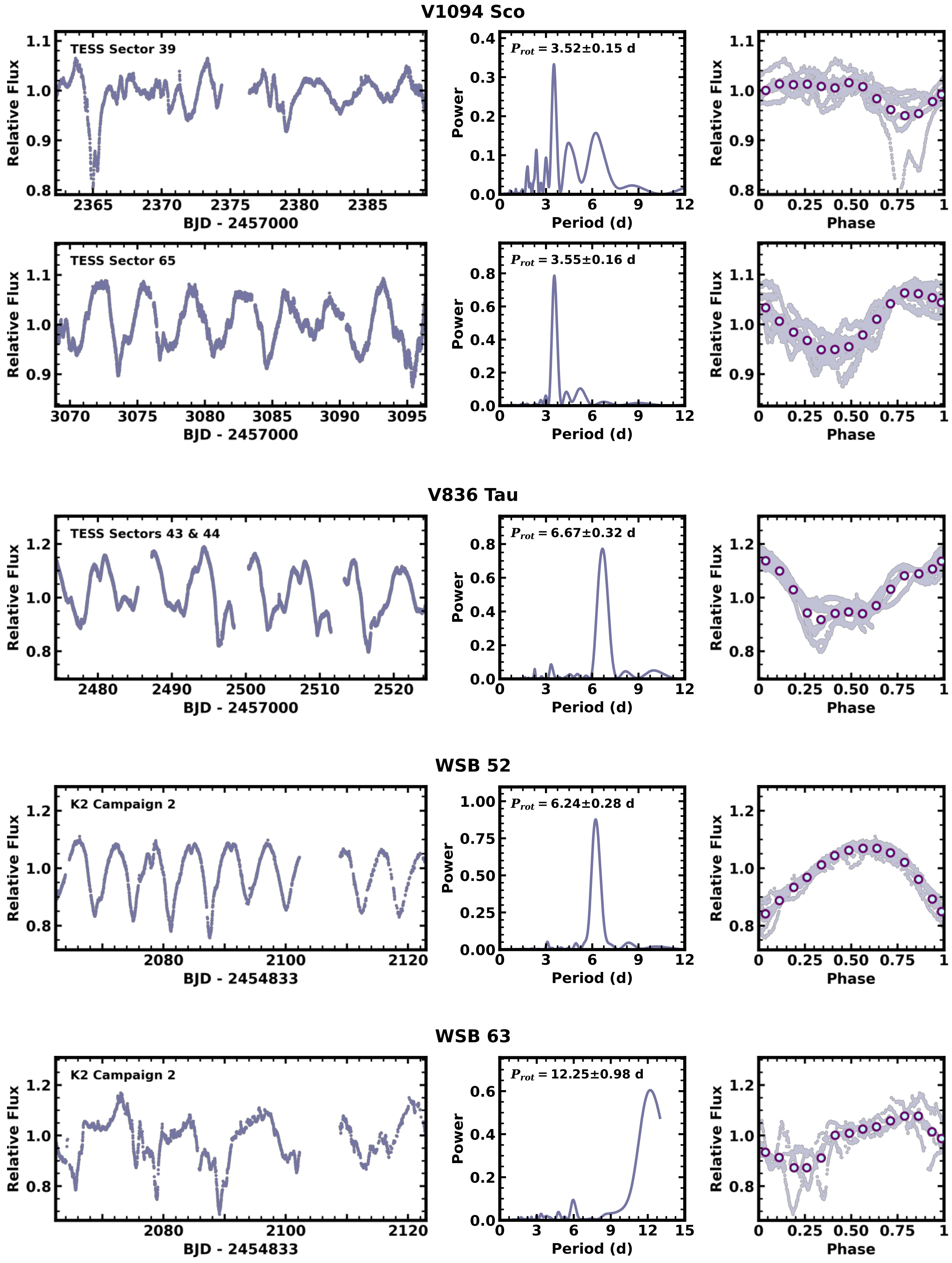

**SD–Figure 1.** *Continued.*

## SD–4: Interpreting the High Incidence of Equator-On Stars

There are a few interpretations for why there is relatively high incidence of equator-on stars. The first is that this could be a result of our adopted Bayesian formalism where very high $v\sin i_*$ relative to $v_{\rm eq}$ tends toward the $\sin i_*$ isotropic prior, $i_* \approx 90°$. To test whether this is a likely cause, we compare the distribution of $\sigma_{v\sin i_*}/v\sin i_*$ for stars with $i_* > 80°$ to stars with $i_* < 80°$. The subsample should show a consistent shift toward broader $v\sin i_*$ if the high occurrence of equator-on stars is caused by poor constraints on $v\sin i_*$. However, we find that the median precision of $v\sin i_*$ for stars with $i_* > 80°$ is 0.05$\pm$0.05 compared to 0.03$\pm$0.04 for stars with $i_* < 80°$. Poorly constrained projected rotational velocities in this case are therefore not likely to be the driver of the over-representation of equator-on stars.

Overestimating $P_{\rm rot}$ as a result of differential rotation could also produce a systematically high incidence of stars that are interpreted as being edge-on. We account for differential rotation effects by applying an additional error term to the uncertainty of the adopted rotation periods, although it is predicted that differential rotation tends to only affect highly convective stars by just a few percent[12,13]. Provided that the stars in our sample have deeper convective zones than the Sun (which is likely the case for young late-type stars), our treatment of differential rotation effects (see Methods in main article) is likely conservative. However, there remains a broad distribution of absolute shear among low mass stars[14–16], and other forms of differential rotation may be possible such as antisolar rotation and cylindrical rotation[17], for example.

An alternative explanation for the high rate of equator-on stars is that there is an observational bias toward this orientation. The methods used to measure $P_{\rm rot}$ in our sample rely on time-dependent brightness modulations from spots coming into and out of view of the observer as the star rotates. Stars with large spot sizes and high inclinations exhibit modulations with greater amplitudes than stars with small spot sizes and near pole-on orientations, producing stronger periodic signals that are more likely to be detected. Our selection criteria requires that the stellar rotation period is known, which may therefore inherently bias the sample toward favoring equator-on stars.

The reliance on $v\sin i_*$ measurements to determine stellar inclination also favors stars with high inclinations. Traditionally, $v\sin i_*$ is measured through the identification of spectral line broadening that results from the simultaneous blue- and red-shifted hemispheres moving toward and away from the observer, respectively, as the star rotates. This effect is diminished for stars with low inclinations, requiring higher resolution to confidently distinguish weak rotational broadening from microturbulence and macroturbulence. As a result, $v\sin i_*$ is more likely to be measured for stars with high inclinations.

## SD–5: Misalignment Rates and Probabilities

We determine whether a system is misaligned based on the measurement precision determined from its individual $\Delta i$ probability distribution. Specifically, the $\Delta i$ MAP value must be greater than 0° by at least two times the lower ("$-$") uncertainty, indicating with >95% confidence that $\Delta i$ is inconsistent with alignment. With this criterion, we identify 16/49 misaligned systems. The corresponding occurrence rate of $33^{+7}_{-6}\%$—based on the *overall* misalignment rate measured in our sample—is what we adopt in this work.

It is also useful to consider how common progressively more extreme misalignments are, and how this misalignment rate declines when considering increasing misalignment thresholds. To analyze this, we determine the number of misaligned systems with $\Delta i$ at or above increasing values of $\Delta i$: $10°$, $20°$, and $40°$. We find that all 16 systems that have been identified as misaligned happen to have $\Delta i$ ≥10°, 11 are misaligned by ≥20°, and 6 are misaligned by ≥40°, corresponding to underlying proportions of $33^{+7}_{-6}\%$, $22^{+6}_{-5}\%$, and $12^{+5}_{-4}\%$, respectively (see SD–Table 1).

**SD–Table 1**. Fractions and probabilities of misaligned systems with $\Delta i$ greater than $10°$, $20°$, and $40°$.

| $\geq \Delta i$ | **Misalignment Rate** | |
|---|---|---|
| $10°$ | 16/49 | $33^{+7}_{-6}\%$ |
| $20°$ | 11/49 | $22^{+6}_{-5}\%$ |
| $40°$ | 6/49 | $12^{+5}_{-4}\%$ |

Among misaligned stars, we see a range in minimum obliquities between ~10° and ~60°. For clarity, we emphasize that the lower end of this range is not a definitive lower limit of $\Delta i$ for misaligned systems, but is instead an effect of the characteristic precision of $\Delta i$ across the sample, which is ~10°. It is therefore unsurprising that no significant ($>2\sigma$) misalignments were identified with $\Delta i \lesssim 10°$. For example, the Sun's obliquity is below this range, yet it is considered misaligned by this measure. It is therefore possible that this type of more modest primordial misalignment (on the scale of a few degrees) could be even more common than the misalignment rate measured here.

There are a few motivations for choosing the $10°$, $20°$, and $40°$ thresholds to characterize misalignment rates. The 10° threshold represents the average precision of $\Delta i$ across the entire sample, and the 20° and 40° thresholds are the 2× and 4× multiples of this precision. The scale of these thresholds is also physically motivated by theoretical studies of chaotic cloud core collapse[18], broken disks[19], and late accretion infall onto Class II stars through accretion streamers[20]. Such events are predicted to strongly affect the orientation of the outer disk, often resulting in large obliquity angles of several tens of degrees[20,21]. Characterizing the misalignment rate with minimum $\Delta i$ thresholds therefore offers useful comparison points for theoretical studies.

## SD–6: Inner-Outer Disk Misalignments in The Sample

The range of observed misalignments between the inner and outer regions of protoplanetary disks spans a few degrees to as much as 70° (e.g., refs 22,23). Although hints of inner-outer disk misalignments have been resolved in scattered light for a handful of disks in our sample (see Supplementary Discussion), their misalignments do not exceed 20° and are less severe than misalignments inferred from shadows[24] and dipper stars[25]. There are also a few objects in our sample that appear to show inner-outer disk misalignments via dips in the light curve (e.g., AA Tau[5,26,27]; DoAr 25[25]; MY Lup, RY Lup, Sz 90,[28]), shadowing effects (e.g., DoAr 44,[24]; TW Hya,[29,30]), and from scattered-light imaging of the inner disk (e.g., CI Tau,[31]; RY Lup,[23]).

Many objects in the sample with evidence of inner-outer disk misalignment show no evidence that the outer disk is misaligned: RY Lup[28], LkCa 15, GM Aur[23], AA Tau[27], SY Cha[32], MY Lup, Sz 90[28], and DoAr 25[25]. Four systems show evidence of inner-outer disk misalignment as well as star-outer disk misalignment: DoAr 44[24], RX J1615.3-3255[33], CI Tau[31], TW Hya[34]. However, whether the inner disks of these systems are aligned or misaligned with the central star is not known.

# Supplementary Material

**SM–Table 1. Adopted Protoplanetary Disk Inclinations.** Adopted protoplanetary disk inclinations and their literature sources.

| Identifier [a] | $i_{disk}$ (°) | Reference |
|---|---|---|
| 2M J0412 | 15.8±0.8 | 35 |
| 2M J0420 | 38.24±0.25 | 36 |
| 2M J0433 | 57.57±0.07 | 36 |
| 2M J0434 | 68.54±0.23 | 37 |
| 2M J0436 | 53.42±1.45 | 37 |
| 2M J1100 | 4.4±3.1 | 38 |
| 2M J1608 | 72.4±0.6 | 38 |
| AA Tau | 59.1±0.3 | 39 |
| AS 209 | 35.0±0.1 | 40 |
| BP Tau | 38.2±0.5 | 41 |
| CI Tau | 50.0±0.1 | 41 |
| CIDA-7 | 31.3±0.3 | 36 |
| CR Cha | 31.0±1.4 | 42 |
| CX Tau | 55.1±1.0 | 43 |
| DE Tau | 34±1 | 44 |
| DN Tau | 35.2±0.6 | 41 |
| DoAr 25 | 67.4±0.2 | 40 |
| DoAr 44 | 21.8±0.9 | 45 |
| Elias 2-24 | 29.0±0.3 | 40 |
| EX Lup | 32.4±0.9 | 46 |
| FP Tau | 66.0±4.0 | 47 |
| FT Tau | 35.5±0.4 | 41 |
| GM Aur | 53.21±0.01 | 48 |
| GW Lup | 38.7±0.3 | 40 |
| IM Lup | 48.4±0.1 | 49 |
| IP Tau | 45.2±0.9 | 41 |
| IQ Tau | 62.1±0.5 | 41 |
| LkCa 15 | 50.16±0.03 | 50 |
| MHO 6 | 64.56±0.03 | 36 |

**SM–1**. *Continued.*

| Identifier [a] | $i_{disk}$ (°) | Reference |
|---|---|---|
| MP Mus | 32±1 | 51 |
| MY Lup | 73.2±0.1 | 40 |
| PDS 111 | 58.2±0.1 | 52 |
| PDS 70 | 51.7±0.1 | 53 |
| RU Lup | 18.8±1.6 | 40 |
| RX J1615 | 46.5±0.5 | 54 |
| RX J1842 | 32±5 [b] | 22 [c] |
| RX J1852 | 30±5 [b] | 22 [c] |
| RY Lup | 67.7±0.6 | 38 |
| SU Aur | 53.0±1.5 | 55 |
| SY Cha | 51.7±1.2 [b] | 56 |
| Sz 114 | 21.3±1.3 | 40 |
| Sz 73 | 49.7±4.0 | 49 |
| Sz 90 | 61.3±5.3 | 49 |
| T Cha | 73±5 [b] | 22 [c] |
| TW Hya | 5.8±2.9 | 57 |
| V1094 Sco | 56.3±0.2 | 58 |
| V836 Tau | 43.1±0.8 | 41 |
| WSB 52 | 54.4±0.3 | 40 |
| WSB 63 | 67.3±0.5 | 59 |

[a] Full names of abbreviated identifiers are listed on Page 1.

[b] No uncertainties were reported in the original publication of the data, so we adopt the value determined by the reference listed here.

[c] Authors adopt a sweeping uncertainty of 5° on $i_{disk}$ to account for varying measurement methods of original authors (see text in referenced work).

**SM–Table 2. Compiled Stellar Properties: Spectral Type, $T_{eff}$, and $M_*$**
Compiled values of stellar spectral type, $T_{eff}$, and $M_*$ for each object and their respective sources. Adopted values are displayed in bold.

| Identifier | Spectral Type | Reference | $T_{eff}$ (K) | Reference | $M_*$ ($M_\odot$) | Reference |
|---|---|---|---|---|---|---|
| 2M J0412 | M4.3 | 37 | 3328±86 | 60 | 0.38 | 61 |
| | ··· | ··· | 3362±123 | 61 | ··· | ··· |
| | ··· | ··· | 3350±5 | 62 | ··· | ··· |
| | ··· | ··· | 3328±4 | 62 | ··· | ··· |
| | ··· | ··· | 3366±14 | 62 | ··· | ··· |
| | **M4.3** | **Adopted** | **3337±3** | **Adopted** | **0.38±0.38** | **Adopted** |
| 2M J0420 | M5.25 | 63 | 2826±157 | 61 | 0.54±0.02 | 61 |
| | ··· | ··· | 2810 | 64 | 0.02 | 65 |
| | ··· | ··· | 3091 | 66 | 0.12±0.02 | 67 |
| | ··· | ··· | 2820 | 68 | 0.5 | 69 |
| | ··· | ··· | 3095±33 | 67 | 0.16 | 70 |
| | ··· | ··· | 3580 | 69 | 0.10±0.04 | 71 |
| | ··· | ··· | 2943 | 71 | 0.15 | 72 |
| | ··· | ··· | 3092 | 72 | ··· | ··· |
| | **M5.3** | **Adopted** | **3083±32** | **Adopted** | **0.25±0.01** | **Adopted** |
| 2M J0433 | M5.2 | 73 | 3161 | 74 | 0.05 | 73 |
| | ··· | ··· | 2950 | 65 | 0.15 | 65 |
| | ··· | ··· | 3143±4 | 62 | 0.14±0.04 | 71 |
| | ··· | ··· | 3020 | 68 | ··· | ··· |
| | ··· | ··· | 2878±164 | 75 | ··· | ··· |
| | ··· | ··· | 3027 | 71 | ··· | ··· |
| | ··· | ··· | 3090 | 76 | ··· | ··· |
| | **M5.2** | **Adopted** | **3142±4** | **Adopted** | **0.13±0.03** | **Adopted** |
| 2M J0434 | M4.3 | 37 | 3295±89 | 60 | 0.3 | 37 |
| | ··· | ··· | 3472±124 | 61 | 0.44 | 61 |
| | ··· | ··· | 3291±6 | 62 | ··· | ··· |
| | ··· | ··· | 3295±5 | 62 | ··· | ··· |
| | **M4.3** | **Adopted** | **3293±3** | **Adopted** | **0.37±0.10** | **Adopted** |
| 2M J0436 | M2.7 | 37 | 3400 | 77 | 0.58 | 37 |
| | ··· | ··· | 3560±143 | 61 | 0.49 | 61 |
| | ··· | ··· | 3610±107 | 60 | ··· | ··· |
| | **M2.7** | **Adopted** | **3592±85** | **Adopted** | **0.54±0.06** | **Adopted** |

**SM–Table 2.** *Continued.*

| Identifier | Spectral Type | Reference | $T_{eff}$ (K) | Reference | $M_*$ ($M_\odot$) | Reference |
|---|---|---|---|---|---|---|
| 2M J1100 | M3.7 | 78 | 3418±122 | 61 | 0.41 | 61 |
| | ⋯ | ⋯ | 3311±76 | 79 | 0.25±0.05 | 80 |
| | ⋯ | ⋯ | 3270 | 81 | 0.46±0.06 | 82 |
| | ⋯ | ⋯ | ⋯ | ⋯ | 0.24±0.07 | 81 |
| | **M3.7** | **Adopted** | **3340±64** | **Adopted** | **0.32±0.03** | **Adopted** |
| 2M J1608 | K2.0 | 83 | 4800 | 84 | 1.4±0.1 | 84 |
| | ⋯ | ⋯ | 4900 | 85 | 1.8 | 85 |
| | ⋯ | ⋯ | 4797±145 | 86 | 1.58±0.18 | 81 |
| | ⋯ | ⋯ | 4898 | 87 | ⋯ | ⋯ |
| | **K2.0** | **Adopted** | **4856±53** | **Adopted** | **1.44±0.09** | **Adopted** |
| AA Tau | K7.0 | 88 | 4350 | 22 | 0.68 | 22 |
| | ⋯ | ⋯ | 3970 | 89 | 0.53 | 27 |
| | ⋯ | ⋯ | 3800 | 90 | 0.69±0.07 | 89 |
| | ⋯ | ⋯ | 3783±219 | 61 | 0.61 | 73 |
| | ⋯ | ⋯ | ⋯ | ⋯ | 0.55 | 73 |
| | ⋯ | ⋯ | ⋯ | ⋯ | 0.57 | 73 |
| | ⋯ | ⋯ | ⋯ | ⋯ | 0.57 | 61 |
| | **K7.0** | **Adopted** | **3879±172** | **Adopted** | **0.62±0.04** | **Adopted** |
| AS 209 | K4.0 | 91 | 4266 | 92 | 1.2 | 92 |
| | ⋯ | ⋯ | 4300±300 | 87 | 1.09 | 93 |
| | ⋯ | ⋯ | 4395 | 94 | 1.4 | 94 |
| | ⋯ | ⋯ | 3887±122 | 61 | 1.0 | 95 |
| | ⋯ | ⋯ | 4350 | 95 | 0.8 | 96 |
| | ⋯ | ⋯ | 4265 | 87 | 1.18 | 97 |
| | ⋯ | ⋯ | 4600 | 98 | 1.1±0.1 | 98 |
| | **K4.0** | **Adopted** | **4118±87** | **Adopted** | **1.1±0.09** | **Adopted** |

**SM–Table 2.** *Continued.*

| Identifier | Spectral Type | Reference | $T_{eff}$ (K) | Reference | $M_*$ ($M_\odot$) | Reference |
|---|---|---|---|---|---|---|
| BP Tau | K7.0 | 99 | 3717±17 | 100 | 0.48 | 100 |
| | ··· | ··· | 4138±27 | 101 | 0.49 | 27 |
| | ··· | ··· | 4055±112 | 102 | 0.8 | 103 |
| | ··· | ··· | 4142±141 | 61 | 0.75 | 101 |
| | ··· | ··· | 3970 | 89 | 0.54 | 104 |
| | ··· | ··· | ··· | ··· | 0.7 | 105 |
| | ··· | ··· | ··· | ··· | 0.67±0.06 | 89 |
| | ··· | ··· | ··· | ··· | 0.45 | 106 |
| | **K7.0** | **Adopted** | **3843±14** | **Adopted** | **0.66±0.06** | **Adopted** |
| CI Tau | K7.0 | 99 | 4200 | 90 | 1.29 | 100 |
| | ··· | ··· | 4277 | 41 | 0.68±0.07 | 89 |
| | ··· | ··· | 4100±190 | 87 | 0.92±0.02 | 47 |
| | ··· | ··· | 3807±131 | 61 | 0.89±0.19 | 41 |
| | ··· | ··· | ··· | ··· | 0.9 | 107 |
| | **K7.0** | **Adopted** | **4049±74** | **Adopted** | **0.9±0.02** | **Adopted** |
| CIDA-7 | M4.75 | 108 | 2950 | 64 | 0.06 | 73 |
| | ··· | ··· | 3161 | 109 | 0.11 | 65 |
| | ··· | ··· | 2957±157 | 61 | 0.67±0.02 | 61 |
| | ··· | ··· | 2960 | 68 | 0.14±0.14 | 71 |
| | ··· | ··· | 3027 | 71 | ··· | ··· |
| | ··· | ··· | 2955 | 68 | ··· | ··· |
| | **M4.7** | **Adopted** | **2997±77** | **Adopted** | **0.51±0.02** | **Adopted** |
| CR Cha | K0.0 | 110 | 4900±100 | 102 | 1.9±0.2 | 111 |
| | ··· | ··· | 4800±230 | 112 | 1.2 | 113 |
| | ··· | ··· | 5164±84 | 61 | 1.66±0.38 | 112 |
| | ··· | ··· | ··· | ··· | 1.62±0.34 | 114 |
| | **K0.0** | **Adopted** | **5036±61** | **Adopted** | **1.8±0.16** | **Adopted** |

**SM–Table 2.** *Continued.*

| Identifier | Spectral Type | Reference | $T_{eff}$ (K) | Reference | $M_*$ ($M_\odot$) | Reference |
|---|---|---|---|---|---|---|
| CX Tau | M2.5 | 63 | 3520±180 | 115 | 0.37±0.02 | 47 |
| | ··· | ··· | 3685±62 | 116 | 0.35 | 104 |
| | ··· | ··· | 3696±62 | 116 | 0.38 | 73 |
| | ··· | ··· | 4325 | 117 | 0.39 | 73 |
| | ··· | ··· | 3489 | 71 | 0.39 | 73 |
| | ··· | ··· | 3580 | 118 | 0.39 | 73 |
| | ··· | ··· | 3490 | 67 | 0.33±0.08 | 117 |
| | ··· | ··· | ··· | ··· | 0.33±0.04 | 71 |
| | **M2.5** | **Adopted** | **3681±42** | **Adopted** | **0.37±0.01** | **Adopted** |
| DE Tau | M2.3 | 73 | 3499±57 | 100 | 0.42±0.11 | 89 |
| | ··· | ··· | 3463±158 | 115 | 0.26 | 27 |
| | ··· | ··· | 3620 | 119 | 0.37 | 73 |
| | ··· | ··· | 3688±122 | 120 | 0.38 | 73 |
| | ··· | ··· | 3690±7 | 62 | 0.37 | 73 |
| | ··· | ··· | 3412 | 106 | 0.38 | 73 |
| | ··· | ··· | 3580 | 118 | 0.45±0.45 | 71 |
| | ··· | ··· | 3722 | 71 | ··· | ··· |
| | **M2.3** | **Adopted** | **3685±7** | **Adopted** | **0.37±0.05** | **Adopted** |
| DN Tau | M0.3 | 115 | 3823±150 | 100 | 0.55 | 100 |
| | ··· | ··· | 3648±144 | 115 | 0.38 | 27 |
| | ··· | ··· | 3770 | 89 | 0.50±0.11 | 89 |
| | ··· | ··· | 3806 | 41 | 0.52±0.12 | 41 |
| | ··· | ··· | 3900 | 90 | 0.56 | 121 |
| | ··· | ··· | 3821±108 | 60 | 0.64 | 122 |
| | ··· | ··· | 3955±138 | 101 | 0.65±0.05 | 123 |
| | ··· | ··· | 3898 | 71 | ··· | ··· |
| | ··· | ··· | 3890 | 124 | ··· | ··· |
| | ··· | ··· | 3850 | 118 | ··· | ··· |
| | ··· | ··· | 3950±50 | 123 | ··· | ··· |
| | **M0.3** | **Adopted** | **3883±32** | **Adopted** | **0.6±0.04** | **Adopted** |

**SM–Table 2.** *Continued.*

| Identifier | Spectral Type | Reference | $T_{eff}$ (K) | Reference | $M_*$ ($M_\odot$) | Reference |
|---|---|---|---|---|---|---|
| DoAr 25 | K4.0 | 125 | 4300±300 | 87 | 1.09 | 93 |
| | ··· | ··· | 4600 | 125 | 0.95±0.12 | 87 |
| | ··· | ··· | 4350 | 126 | 0.65 | 127 |
| | ··· | ··· | 4450 | 119 | 1.11 | 126 |
| | ··· | ··· | 4202 | 72 | 0.8 | 72 |
| | ··· | ··· | 4200 | 45 | 0.8±0.2 | 128 |
| | ··· | ··· | 4205 | 129 | 1.12 | 97 |
| | **K4.0** | **Adopted** | **4326±144** | **Adopted** | **0.92±0.09** | **Adopted** |
| DoAr 44 | K3.0 | 130 | 4780 | 22 | 1.58 | 93 |
| | ··· | ··· | 4730 | 131 | 1.4 | 22 |
| | ··· | ··· | 5800±220 | 87 | 1.25 | 132 |
| | ··· | ··· | 4900 | 133 | 0.97±0.19 | 133 |
| | ··· | ··· | 4600±120 | 134 | 1.2±0.2 | 134 |
| | ··· | ··· | 3904±122 | 61 | 1.3 | 131 |
| | ··· | ··· | 3941±190 | 135 | 1.29 | 97 |
| | ··· | ··· | 4760 | 45 | 1.5±0.1 | 98 |
| | ··· | ··· | 4800 | 98 | ··· | ··· |
| | **K3.0** | **Adopted** | **4614±48** | **Adopted** | **1.34±0.07** | **Adopted** |
| Elias 2-24 | K5.5 | 127 | 3890±173 | 136 | 1.09 | 93 |
| | ··· | ··· | 4057 | 137 | 0.78±0.12 | 87 |
| | ··· | ··· | 4483 | 138 | 1.15 | 139 |
| | ··· | ··· | 4571 | 140 | 1.66 | 140 |
| | ··· | ··· | 4276 | 72 | 1.1 | 72 |
| | **K5.5** | **Adopted** | **4055±138** | **Adopted** | **0.86±0.11** | **Adopted** |
| EX Lup | M0.5 | 141 | 3801 | 142 | 0.56±0.08 | 143 |
| | ··· | ··· | 3970 | 144 | 0.5 | 86 |
| | ··· | ··· | 3970 | 145 | 0.83 | 146 |
| | ··· | ··· | 3859±62 | 86 | 0.55 | 146 |
| | ··· | ··· | 3941±199 | 147 | ··· | ··· |
| | ··· | ··· | 4015±125 | 61 | ··· | ··· |
| | **M0.5** | **Adopted** | **3898±46** | **Adopted** | **0.57±0.07** | **Adopted** |

**SM–Table 2.** *Continued.*

| **Identifier** | **Spectral Type** | **Reference** | $T_{\mathrm{eff}}$ **(K)** | **Reference** | $M_*$ **($M_\odot$)** | **Reference** |
|---|---|---|---|---|---|---|
| FP Tau | M2.5 | 73 | 2530±257 | 115 | 0.37±0.02 | 47 |
| | ⋯ | ⋯ | 3270 | 109 | 0.40±0.01 | 148 |
| | ⋯ | ⋯ | 3640±102 | 60 | 0.35 | 104 |
| | ⋯ | ⋯ | 3670±122 | 61 | 0.39 | 73 |
| | ⋯ | ⋯ | 3597 | 62 | 0.39 | 73 |
| | ⋯ | ⋯ | 3640 | 62 | 0.38 | 73 |
| | ⋯ | ⋯ | 3311±152 | 79 | 0.39 | 73 |
| | ⋯ | ⋯ | ⋯ | ⋯ | 0.39 | 73 |
| | **M2.5** | **Adopted** | **3508±63** | **Adopted** | **0.39±0.01** | **Adopted** |
| FT Tau | M2.8 | 73 | 3407±73 | 100 | 0.25 | 100 |
| | ⋯ | ⋯ | 3444 | 41 | 0.34±0.11 | 41 |
| | ⋯ | ⋯ | 3451±98 | 60 | 0.45±0.09 | 61 |
| | ⋯ | ⋯ | 3490±115 | 61 | 0.4±0.01 | 124 |
| | ⋯ | ⋯ | 3451 | 62 | ⋯ | ⋯ |
| | ⋯ | ⋯ | 3467 | 124 | ⋯ | ⋯ |
| | **M2.8** | **Adopted** | **3453±11** | **Adopted** | **0.40±0.01** | **Adopted** |
| GM Aur | K6.0 | 115 | 4350 | 22 | 1.19 | 100 |
| | ⋯ | ⋯ | 4564±76 | 100 | 1.01 | 22 |
| | ⋯ | ⋯ | 4350 | 92 | 1.14±0.02 | 148 |
| | ⋯ | ⋯ | 4350 | 133 | 1.36±0.36 | 133 |
| | ⋯ | ⋯ | 4287±35 | 149 | 0.95±0.13 | 149 |
| | ⋯ | ⋯ | 4800±400 | 87 | 1.1±0.1 | 71 |
| | ⋯ | ⋯ | 4730 | 118 | 1.0 | 150 |
| | ⋯ | ⋯ | 4566 | 71 | 1.3 | 69 |
| | ⋯ | ⋯ | 4202 | 72 | 0.82 | 72 |
| | ⋯ | ⋯ | 4189±127 | 61 | ⋯ | ⋯ |
| | **K6.0** | **Adopted** | **4332±30** | **Adopted** | **1.13±0.02** | **Adopted** |
| GW Lup | M2.3 | 73 | 3632±167 | 151 | 0.42±0.11 | 151 |
| | ⋯ | ⋯ | 3600±170 | 87 | 0.37 | 104 |
| | ⋯ | ⋯ | 3751±122 | 61 | 0.6±0.1 | 151 |
| | ⋯ | ⋯ | ⋯ | ⋯ | 0.41 | 73 |
| | ⋯ | ⋯ | ⋯ | ⋯ | 0.43±0.12 | 81 |
| | ⋯ | ⋯ | ⋯ | ⋯ | 0.46±0.14 | 87 |
| | **M2.3** | **Adopted** | **3682±85** | **Adopted** | **0.41±0.03** | **Adopted** |

**SM–Table 2.** *Continued.*

| Identifier | Spectral Type | Reference | $T_{\mathrm{eff}}$ (K) | Reference | $M_*$ ($M_\odot$) | Reference |
|---|---|---|---|---|---|---|
| IM Lup | K5.0 | 83 | 4266 | 92 | 1.1 | 92 |
| | ··· | ··· | 4146±95 | 86 | 0.89±0.22 | 148 |
| | ··· | ··· | 4400±200 | 87 | 0.67 | 104 |
| | ··· | ··· | 4365 | 87 | 1.0±0.2 | 152 |
| | ··· | ··· | 4144 | 147 | 0.78 | 73 |
| | ··· | ··· | 4009±200 | 135 | 1.2±0.4 | 153 |
| | ··· | ··· | 4350 | 83 | 0.7 | 86 |
| | ··· | ··· | 3767 | 154 | 0.51±0.07 | 83 |
| | ··· | ··· | 3850 | 155 | 0.95 | 72 |
| | ··· | ··· | 4000 | 98 | 0.7±0.1 | 98 |
| | **K5.0** | **Adopted** | **4158±74** | **Adopted** | **0.65±0.05** | **Adopted** |
| IP Tau | M0.5 | 73 | 3850 | 22 | 0.54 | 22 |
| | ··· | ··· | 3686±183 | 115 | 0.51 | 100 |
| | ··· | ··· | 3763 | 41 | 0.52±0.14 | 41 |
| | ··· | ··· | 3900±90 | 87 | 0.59 | 73 |
| | ··· | ··· | 3788±112 | 60 | 0.6±0.1 | 61 |
| | ··· | ··· | 3870±136 | 61 | 0.56±0.06 | 156 |
| | ··· | ··· | 3792±36 | 157 | 0.5 | 150 |
| | ··· | ··· | 3898 | 156 | ··· | ··· |
| | ··· | ··· | 3850 | 118 | ··· | ··· |
| | **M0.5** | **Adopted** | **3813±26** | **Adopted** | **0.55±0.03** | **Adopted** |
| IQ Tau | M1.0 | 73 | 3811±65 | 100 | 0.55 | 100 |
| | ··· | ··· | 3690 | 41 | 0.46±0.1 | 89 |
| | ··· | ··· | 3802 | 87 | 0.5±0.14 | 41 |
| | ··· | ··· | 3956±221 | 61 | 0.51 | 73 |
| | ··· | ··· | 3612 | 115 | 0.62±0.10 | 61 |
| | ··· | ··· | 3704±32 | 157 | 0.67±0.09 | 157 |
| | ··· | ··· | 3788 | 62 | 0.74±0.02 | 157 |
| | ··· | ··· | 3764 | 62 | 0.6 | 122 |
| | ··· | ··· | 3780 | 118 | ··· | ··· |
| | **M1.0** | **Adopted** | **3730±26** | **Adopted** | **0.69±0.02** | **Adopted** |

**SM–Table 2.** *Continued.*

| Identifier | Spectral Type | Reference | $T_{eff}$ (K) | Reference | $M_*$ ($M_\odot$) | Reference |
|---|---|---|---|---|---|---|
| LkCa 15 | K5.5 | 73 | 4730 | 22 | 1.4±0.11 | 23 |
| | ⋯ | ⋯ | 4588±73 | 100 | 1.32 | 22 |
| | ⋯ | ⋯ | 4156±123 | 115 | 1.71 | 100 |
| | ⋯ | ⋯ | 4400±300 | 87 | 1.09 | 100 |
| | ⋯ | ⋯ | 4900 | 133 | 1.24±0.33 | 133 |
| | ⋯ | ⋯ | 4365 | 87 | 1.14±0.03 | 148 |
| | ⋯ | ⋯ | 4194±130 | 61 | 1.1 | 103 |
| | ⋯ | ⋯ | 4201 | 158 | 0.82 | 73 |
| | ⋯ | ⋯ | 4350 | 118 | 0.89 | 73 |
| | ⋯ | ⋯ | 4730 | 69 | 0.87 | 73 |
| | ⋯ | ⋯ | 4492±50 | 159 | 0.87 | 73 |
| | ⋯ | ⋯ | ⋯ | ⋯ | 1.2 | 69 |
| | **K5.5** | **Adopted** | **4461±36** | **Adopted** | **1.16±0.03** | **Adopted** |
| MHO 6 | M5.0 | 73 | 3162±145 | 79 | 0.19±0.01 | 82 |
| | ⋯ | ⋯ | 2980±157 | 61 | 0.09 | 73 |
| | ⋯ | ⋯ | 2950 | 64 | 0.14±0.04 | 71 |
| | ⋯ | ⋯ | 3159±87 | 60 | 0.61±0.02 | 61 |
| | ⋯ | ⋯ | 3161 | 160 | ⋯ | ⋯ |
| | ⋯ | ⋯ | 3027 | 71 | ⋯ | ⋯ |
| | **M5.0** | **Adopted** | **3103±57** | **Adopted** | **0.21±0.0** | **Adopted** |
| MP Mus | K1.0 | 91 | 5000±100 | 51 | 1.3±0.08 | 51 |
| | ⋯ | ⋯ | 4593±124 | 61 | 1.52 | 161 |
| | ⋯ | ⋯ | 4585±8 | 101 | 1.2±0.2 | 162 |
| | ⋯ | ⋯ | 5000 | 98 | 1.15 | 101 |
| | ⋯ | ⋯ | 5228 | 163 | 1.1 | 164 |
| | ⋯ | ⋯ | ⋯ | ⋯ | 1.4±0.1 | 98 |
| | **K1.0** | **Adopted** | **4588±7** | **Adopted** | **1.32±0.06** | **Adopted** |

**SM–Table 2.** *Continued.*

| Identifier | Spectral Type | Reference | $T_{\rm eff}$ (K) | Reference | $M_*$ ($M_\odot$) | Reference |
|---|---|---|---|---|---|---|
| MY Lup | K0.0 | 83 | 5100 | 83 | 1.0 | 85 |
| | ··· | ··· | 4968±200 | 86 | 1.23±0.2 | 148 |
| | ··· | ··· | 5129 | 87 | 1.15±0.06 | 83 |
| | ··· | ··· | 4620±139 | 61 | 1.1 | 86 |
| | ··· | ··· | 5250 | 69 | 1.3±0.1 | 98 |
| | ··· | ··· | 4300±50 | 165 | ··· | ··· |
| | ··· | ··· | 5200 | 98 | ··· | ··· |
| | **K0.0** | **Adopted** | **4619±37** | **Adopted** | **1.14±0.04** | **Adopted** |
| PDS 111 | G2.0 | 52 | 5432 | 166 | 0.95±0.12 | 61 |
| | ··· | ··· | 5433±135 | 61 | 1.19±0.10 | 135 |
| | ··· | ··· | 5340±224 | 135 | 1.2±0.1 | 52 |
| | ··· | ··· | 5900±125 | 52 | ··· | ··· |
| | **G2.0** | **Adopted** | **5634±84** | **Adopted** | **1.13±0.06** | **Adopted** |
| PDS 70 | K7.0 | 154 | 4060 | 22 | 0.8±0.1 | 162 |
| | ··· | ··· | 3800 | 167 | 0.64±0.08 | 61 |
| | ··· | ··· | 3927 | 168 | 0.7 | 154 |
| | ··· | ··· | 4000 | 162 | ··· | ··· |
| | ··· | ··· | 4117±122 | 61 | ··· | ··· |
| | ··· | ··· | 4030±98 | 60 | ··· | ··· |
| | ··· | ··· | 3890 | 154 | ··· | ··· |
| | **K7.0** | **Adopted** | **4016±60** | **Adopted** | **0.7±0.1** | **Adopted** |
| RU Lup | K7.0 | 151 | 4037±96 | 86 | 0.63±0.17 | 169 |
| | ··· | ··· | 4060±200 | 83 | 0.8±0.2 | 49 |
| | ··· | ··· | 4100±190 | 87 | 0.55 | 104 |
| | ··· | ··· | 3590 | 170 | 1.25 | 101 |
| | ··· | ··· | 4857±227 | 61 | 1.15±0.15 | 151 |
| | ··· | ··· | 4488±333 | 101 | 0.7 | 86 |
| | ··· | ··· | 4000 | 98 | 0.6 | 150 |
| | ··· | ··· | 3950 | 170 | 0.7±0.2 | 98 |
| | **K7.0** | **Adopted** | **4125±69** | **Adopted** | **0.85±0.08** | **Adopted** |

**SM–Table 2.** *Continued.*

| Identifier | Spectral Type | Reference | $T_{\rm eff}$ (K) | Reference | $M_*$ ($M_\odot$) | Reference |
|---|---|---|---|---|---|---|
| RX J1615 | K5.0 | 171 | 4060 | 133 | 1.16±0.16 | 133 |
| | ··· | ··· | 4000 | 162 | 0.6±0.1 | 162 |
| | ··· | ··· | 4590 | 172 | 1.28 | 172 |
| | ··· | ··· | 4400 | 98 | 0.8±0.1 | 98 |
| | ··· | ··· | 4343±124 | 61 | ··· | ··· |
| | **K5.0** | **Adopted** | **4329±113** | **Adopted** | **0.78±0.06** | **Adopted** |
| RX J1842 | K3.0 | 73 | 4780 | 22 | 1.14 | 22 |
| | ··· | ··· | 4900 | 133 | 0.93±0.16 | 133 |
| | ··· | ··· | 4800±480 | 22 | 0.67±0.082 | 61 |
| | ··· | ··· | 4289±127 | 61 | 0.8 | 96 |
| | ··· | ··· | 4995 | 163 | 1.99 | 73 |
| | ··· | ··· | 4759 | 163 | 1.2 | 69 |
| | **K3.0** | **Adopted** | **4619±82** | **Adopted** | **0.74±0.07** | **Adopted** |
| RX J1852 | K4.0 | 73 | 4780 | 22 | 1.05 | 22 |
| | ··· | ··· | 4850 | 84 | 1.0±0.1 | 84 |
| | ··· | ··· | 4900 | 133 | 1.04±0.19 | 133 |
| | ··· | ··· | 4800±480 | 22 | 0.68±0.08 | 61 |
| | ··· | ··· | 4365±128 | 61 | 0.93 | 73 |
| | ··· | ··· | 4759 | 163 | ··· | ··· |
| | **K4.0** | **Adopted** | **4729±57** | **Adopted** | **0.88±0.05** | **Adopted** |
| RY Lup | K4.0 | 141 | 4780 | 22 | 1.4±0.2 | 22 |
| | ··· | ··· | 4900±230 | 87 | 1.5 | 85 |
| | ··· | ··· | 4898 | 87 | 1.27 | 104 |
| | ··· | ··· | 4651±218 | 61 | 1.4 | 86 |
| | ··· | ··· | 5082±118 | 86 | 1.1±0.1 | 173 |
| | **K4.0** | **Adopted** | **4896±62** | **Adopted** | **1.24±0.07** | **Adopted** |
| SU Aur | G4 | 73 | 5414±171 | 100 | 2.22 | 100 |
| | ··· | ··· | 4457±211 | 61 | 2.0 | 173 |
| | ··· | ··· | 5860 | 118 | 1.7 | 174 |
| | ··· | ··· | 5945 | 174 | 2.07 | 73 |
| | ··· | ··· | 5860 | 175 | 3.09 | 176 |
| | ··· | ··· | 5800 | 177 | 2.0 | 175 |
| | **G4.0** | **Adopted** | **5726±54** | **Adopted** | **2.18±0.48** | **Adopted** |

**SM–Table 2.** *Continued.*

| Identifier | Spectral Type | Reference | $T_{\mathrm{eff}}$ (K) | Reference | $M_*$ ($M_\odot$) | Reference |
|---|---|---|---|---|---|---|
| SY Cha | K2.0 | 141 | 3792 | 32 | 0.78 | 178 |
| | ··· | ··· | 4287±108 | 179 | 1.3 | 179 |
| | ··· | ··· | 3917 | 180 | 0.67 | 181 |
| | ··· | ··· | 4224±129 | 61 | 0.8 | 180 |
| | ··· | ··· | 4060 | 81 | 0.85 | 81 |
| | ··· | ··· | 3988±66 | 137 | ··· | ··· |
| | ··· | ··· | 3700±62 | 182 | ··· | ··· |
| | **K2.0** | **Adopted** | **3931±38** | **Adopted** | **0.88±0.24** | **Adopted** |
| Sz 114 | M5.0 | 83 | 3175±73 | 83 | 1.8±0.4 | 49 |
| | ··· | ··· | 3162 | 87 | 0.17±0.04 | 148 |
| | ··· | ··· | 3469±123 | 61 | 0.19±0.07 | 81 |
| | ··· | ··· | 3134±35 | 86 | 0.3±0.08 | 151 |
| | ··· | ··· | 2900 | 183 | 0.2 | 86 |
| | ··· | ··· | 3370 | 143 | ··· | ··· |
| | ··· | ··· | 3600±50 | 165 | ··· | ··· |
| | **M5.0** | **Adopted** | **3279±25** | **Adopted** | **0.20±0.03** | **Adopted** |
| Sz 73 | K8.5 | 73 | 4060±187 | 83 | 0.8±0.2 | 49 |
| | ··· | ··· | 4074 | 87 | 0.76 | 73 |
| | ··· | ··· | 3538±123 | 61 | 0.48 | 61 |
| | ··· | ··· | 3980±33 | 86 | 0.85 | 151 |
| | ··· | ··· | ··· | ··· | 0.7 | 86 |
| | **K8.5** | **Adopted** | **3953±31** | **Adopted** | **0.74±0.12** | **Adopted** |
| Sz 90 | K7.0 | 184 | 3900±187 | 83 | 0.8±0.2 | 49 |
| | ··· | ··· | 4022±52 | 185 | 0.56±0.09 | 61 |
| | ··· | ··· | 3723±131 | 61 | 0.51±0.34 | 186 |
| | ··· | ··· | 3507±82 | 186 | 0.7 | 86 |
| | **K7.0** | **Adopted** | **3860±40** | **Adopted** | **0.6±0.1** | **Adopted** |

**SM–Table 2.** *Continued.*

| Identifier | Spectral Type | Reference | $T_{eff}$ (K) | Reference | $M_*$ ($M_\odot$) | Reference |
|---|---|---|---|---|---|---|
| T Cha | G8.0 | 22 | 5570 | 22 | 1.12 | 22 |
| | ··· | ··· | 5600±560 | 22 | 1.3 | 187 |
| | ··· | ··· | 5520 | 187 | 1.5 | 188 |
| | ··· | ··· | 5400 | 188 | 1.76 | 189 |
| | ··· | ··· | 5309 | 189 | 1.4 | 190 |
| | ··· | ··· | 5600 | 190 | 0.75 | 161 |
| | ··· | ··· | ··· | ··· | 0.9 | 164 |
| | **G8.0** | **Adopted** | **5485±119** | **Adopted** | **1.25±0.35** | **Adopted** |
| TW Hya | M0.5 | 73 | 4205 | 22 | 0.7 | 27 |
| | ··· | ··· | 4000 | 162 | 0.61 | 104 |
| | ··· | ··· | 4060 | 133 | 0.79±0.17 | 133 |
| | ··· | ··· | 4100±280 | 87 | 0.8±0.1 | 162 |
| | ··· | ··· | 4073 | 87 | 0.6±0.1 | 191 |
| | ··· | ··· | 3800±100 | 191 | 0.64±0.082 | 61 |
| | ··· | ··· | 4097±134 | 61 | 0.69 | 73 |
| | ··· | ··· | 3800±100 | 191 | 0.69 | 73 |
| | ··· | ··· | 4110 | 192 | 0.69 | 73 |
| | ··· | ··· | 3783±107 | 136 | 0.68 | 73 |
| | ··· | ··· | ··· | ··· | 0.68 | 73 |
| | ··· | ··· | ··· | ··· | 0.68 | 73 |
| | ··· | ··· | ··· | ··· | 0.69 | 73 |
| | ··· | ··· | ··· | ··· | 0.75 | 101 |
| | ··· | ··· | ··· | ··· | 0.77 | 193 |
| | ··· | ··· | ··· | ··· | 0.80±0.05 | 194 |
| | **M0.5** | **Adopted** | **3986±30** | **Adopted** | **0.72±0.03** | **Adopted** |
| V1094 Sco | K6.0 | 83 | 4100±410 | 195 | 0.86±0.18 | 148 |
| | ··· | ··· | 4205±193 | 83 | 0.83 | 196 |
| | ··· | ··· | 3967±133 | 61 | 0.7±0.2 | 197 |
| | ··· | ··· | 4255 | 198 | 0.6±0.1 | 61 |
| | ··· | ··· | 4429±83 | 86 | 0.78 | 198 |
| | ··· | ··· | 4018 | 154 | 1.1 | 86 |
| | **K6.0** | **Adopted** | **4264±60** | **Adopted** | **0.69±0.07** | **Adopted** |

**SM–Table 2.** *Continued.*

| Identifier | Spectral Type | Reference | $T_{\mathrm{eff}}$ (K) | Reference | $M_*$ ($M_\odot$) | Reference |
|---|---|---|---|---|---|---|
| V836 Tau | M1.0 | 73 | 3333±150 | 100 | 0.58 | 61 |
| | ··· | ··· | 3313±150 | 100 | 0.8 | 103 |
| | ··· | ··· | 3760±159 | 115 | 0.48±0.13 | 41 |
| | ··· | ··· | 3734 | 41 | 0.75±0.1 | 67 |
| | ··· | ··· | 4060 | 199 | 0.54 | 101 |
| | ··· | ··· | 4073 | 87 | 0.7 | 150 |
| | ··· | ··· | 3787±124 | 61 | ··· | ··· |
| | ··· | ··· | 3740±47 | 101 | ··· | ··· |
| | **M1.0** | **Adopted** | **3699±38** | **Adopted** | **0.65±0.07** | **Adopted** |
| WSB 52 | K5.0 | 73 | 3833±197 | 136 | 0.45 | 93 |
| | ··· | ··· | 3220±20 | 200 | 0.44 | 61 |
| | ··· | ··· | 3469±124 | 61 | 0.5 | 201 |
| | ··· | ··· | 3514 | 202 | ··· | ··· |
| | ··· | ··· | 3750 | 201 | ··· | ··· |
| | ··· | ··· | 3715 | 87 | ··· | ··· |
| | **K5.0** | **Adopted** | **3242±19** | **Adopted** | **0.46±0.03** | **Adopted** |
| WSB 63 | M1.5 | 127 | 3583±134 | 136 | 0.39 | 72 |
| | ··· | ··· | 3632 | 72 | 0.58 | 95 |
| | ··· | ··· | 3690±103 | 60 | 0.66 | 139 |
| | ··· | ··· | 3560 | 95 | 0.6 | 203 |
| | ··· | ··· | 3580 | 203 | 0.35±0.10 | 204 |
| | ··· | ··· | 3550±150 | 205 | ··· | ··· |
| | **M1.5** | **Adopted** | **3598±32** | **Adopted** | **0.43±0.07** | **Adopted** |

**SM–Table 3. Compiled Stellar Properties: $v\sin i_*$, $P_{\mathrm{rot}}$, and $R_*$**

Compiled values of stellar $v\sin i_*$, $P_{\mathrm{rot}}$, and $R_*$ for each object and their respective sources. Adopted values and, if applicable, inflated uncertainties, are highlighted in bold. Rotation period sources with the ' ‡' symbol were reduced MIT Quick-Look Pipeline lightcurve. Values shown with the ' ** ' symbol were the only entry in the compilation with no quoted uncertainty and are therefore not incorporated in the calculation of the weighted mean, but included in this table for completeness and transparency.

| Identifier | $v\sin i_*$ (km $s^{-1}$) | Reference | $P_{\mathrm{rot}}$ (d) | Reference | $R_*$ ($R_\odot$) | Reference |
|---|---|---|---|---|---|---|
| 2M J0412 | 6.63 | 62 | 5.86±0.2 | K2 (This Work) | 0.99 | 61 |
| | 5.71 | 62 | 5.63±0.16 | TESS (This Work) | ··· | ··· |
| | 6.98 | 62 | ··· | ··· | ··· | ··· |
| | **6.4±0.7** | **Adopted** | **5.72±0.21** | **Adopted** | **0.99±0.17** | **Adopted** |
| 2M J0420 | 14.2±0.8 | 68 | 2.24±0.08 | TESS (This Work) | 0.87 | 65 |
| | ··· | ··· | ··· | ··· | 0.49 | 109 |
| | ··· | ··· | ··· | ··· | 1.22 | 66 |
| | **14.2±0.8** | **Adopted** | **2.24±0.08** | **Adopted** | **0.86±0.37** | **Adopted** |
| 2M J0433 | 14.99** | 62 | 5.81±0.17 | K2 (This Work) | 1.08 | 73 |
| | 15.97±0.50 | 68 | 5.96±0.19 | TESS† (This Work) | 1.96 | 65 |
| | ··· | ··· | ··· | ··· | 1.38 | 74 |
| | **15.9±0.8** | **Adopted** | **5.88±0.22** | **Adopted** | **1.48±0.45** | **Adopted** |
| 2M J0434 | 9.10 | 62 | 5.98±0.21 | TESS† (This Work) | 0.9 | 61 |
| | 8.31 | 62 | 6.06±0.23 | K2§ (This Work) | ··· | ··· |
| | **8.7±0.6** | **Adopted** | **6.02±0.24** | **Adopted** | **0.90±0.16** | **Adopted** |
| 2M J0436 | 16.4 | 62 | 2.71±0.05 | TESS (This Work) | 0.98±0.06 | 60 |
| | **16.4±0.8** | **Adopted** | **2.71±0.14** | **Adopted** | **0.98±0.08** | **Adopted** |
| 2M J1100 | 5.4±0.8 | 78 | 2.79±0.06 | TESS (This Work) | 0.91 | 61 |
| | **5.4±0.8** | **Adopted** | **2.79±0.07** | **Adopted** | **0.91±0.16** | **Adopted** |
| 2M J1608 | < 8.0 | 206 | 6.21±0.52 | TESS (This Work) | 1.66 | 61 |
| | ··· | 116 | 6.24±0.13 | 207 | ··· | ··· |
| | ··· | ··· | 6.33** | 208 | ··· | ··· |
| | **< 8.0** | **Adopted** | **6.24±0.24** | **Adopted** | **1.66±0.29** | **Adopted** |

**SM–Table 3.** *Continued.*

| **Identifier** | $\boldsymbol{v\sin i_*}$ **(km s$^{-1}$)** | **Reference** | $\boldsymbol{P_{\rm rot}}$ **(d)** | **Reference** | $\boldsymbol{R_*}$ **($\boldsymbol{R_\odot}$)** | **Reference** |
|---|---|---|---|---|---|---|
| AA Tau | 12.8±1.1 | 78 | 8.06±0.43 | K2[‡] (This Work) | 2.4±1.64 | 60 |
| | 11.5±0.5 | 209 | 8.22±0.3 | TESS (This Work) | ⋯ | ⋯ |
| | 12.3±0.9 | 90 | 8.22 | 102 | ⋯ | ⋯ |
| | 12.5±2.3 | 115 | 8.2 | 210 | ⋯ | ⋯ |
| | 12.7±2.0 | 211 | 8.2±0.3 | 27 | ⋯ | ⋯ |
| | 13.1±1.8 | 90 | 8.18 | 209 | ⋯ | ⋯ |
| | 11.0 | 212 | ⋯ | ⋯ | ⋯ | ⋯ |
| | 11.4 | 213 | ⋯ | ⋯ | ⋯ | ⋯ |
| | 11.3±0.7 | 26 | ⋯ | ⋯ | ⋯ | ⋯ |
| | **11.6±0.3** | **Adopted** | **8.20±0.34** | **Adopted** | **2.40±1.64** | **Adopted** |
| AS 209 | 10±3 | 214 | 8.6 | 121 | 2.01 | 61 |
| | 11.5±0.6 | 91 | 8.43 | 215 | ⋯ | ⋯ |
| | 9.7±0.1 | 216 | ⋯ | ⋯ | ⋯ | ⋯ |
| | **10.4±0.37** | **Adopted** | **8.52±0.39** | **Adopted** | **2.01±0.35** | **Adopted** |
| BP Tau | 13.1±1.6 | 78 | 8.7±0.51 | TESS (This Work) | 1.41±0.14 | 61 |
| | 9.0±0.5 | 217 | 7.6±0.1 | 218 | ⋯ | ⋯ |
| | 7.8±2.0 | 211 | 8.31 | 219 | ⋯ | ⋯ |
| | 9.1±2.0 | 90 | 7.7±1.0 | 220 | ⋯ | ⋯ |
| | 8.5±0.7 | 90 | 7.6 | 210 | ⋯ | ⋯ |
| | 9.9±2.3 | 115 | 7.65 | 217 | ⋯ | ⋯ |
| | 10[**] | 150 | 7.0±0.1 | 90 | ⋯ | ⋯ |
| | **9.1±0.4** | **Adopted** | **7.34±0.29** | **Adopted** | **1.41±0.17** | **Adopted** |
| CI Tau | 10.1±0.7 | 90 | 9.02±0.46 | K2 (This Work) | 1.68 | 61 |
| | 13±2 | 78 | 10.53±0.89 | TESS (This Work) | ⋯ | ⋯ |
| | 12.0±1.7 | 78 | 9.0±0.05 | 221 | ⋯ | ⋯ |
| | 12.5±1.9 | 115 | ⋯ | ⋯ | ⋯ | ⋯ |
| | 9.5±0.5 | 221 | ⋯ | ⋯ | ⋯ | ⋯ |
| | 11.0 | 212 | ⋯ | ⋯ | ⋯ | ⋯ |
| | 10.4 | 213 | ⋯ | ⋯ | ⋯ | ⋯ |
| | 10.08±0.14 | 222 | ⋯ | ⋯ | ⋯ | ⋯ |
| | **10.2±0.3** | **Adopted** | **9.0±0.41** | **Adopted** | **1.68±0.29** | **Adopted** |

**SM–Table 3.** *Continued.*

| Identifier | $v\sin i_*$ (km s$^{-1}$) | Reference | $P_{rot}$ (d) | Reference | $R_*$ ($R_\odot$) | Reference |
|---|---|---|---|---|---|---|
| CIDA-7 | 19.1±0.4 | 68 | 1.9±0.02 | TESS (This Work) | 0.71±0.02 | 61 |
| | ⋯ | ⋯ | 1.89±0.02 | K2 (This Work) | ⋯ | ⋯ |
| | **19.1±1.0** | **Adopted** | **1.90±0.02** | **Adopted** | **0.71±0.05** | **Adopted** |
| CR Cha | 35±2 | 78 | 2.3±0.2 | 210 | 2.25 | 61 |
| | 38.1±0.4 | 223 | ⋯ | ⋯ | ⋯ | ⋯ |
| | 38.0±1.5 | 211 | ⋯ | ⋯ | ⋯ | ⋯ |
| | 36.1±1.0 | 224 | ⋯ | ⋯ | ⋯ | ⋯ |
| | 35±5 | 210 | ⋯ | ⋯ | ⋯ | ⋯ |
| | 34.5** | 112 | ⋯ | ⋯ | ⋯ | ⋯ |
| | 37.0±2.2 | 225 | ⋯ | ⋯ | ⋯ | ⋯ |
| | **36.5±0.9** | **Adopted** | **2.3±0.2** | **Adopted** | **2.25±0.39** | **Adopted** |
| CX Tau | 19.8±0.6 | 78 | 3.31±0.08 | TESS (This Work) | 1.39 | 61 |
| | 18.2±0.8 | 213 | 3.3** | 219 | ⋯ | ⋯ |
| | 21.41±0.33 | 157 | ⋯ | ⋯ | ⋯ | ⋯ |
| | 20 | 212 | ⋯ | ⋯ | ⋯ | ⋯ |
| | 20.5±2.5 | 115 | ⋯ | ⋯ | ⋯ | ⋯ |
| | 19.1 | 211 | ⋯ | ⋯ | ⋯ | ⋯ |
| | 21.45±0.261 | 68 | ⋯ | ⋯ | ⋯ | ⋯ |
| | **20.1±0.4** | **Adopted** | **3.3±0.1** | **Adopted** | **1.39±0.24** | **Adopted** |
| DE Tau | 9.7±0.3 | 78 | 5.97±0.2 | TESS (This Work) | 1.62 | 61 |
| | 8.38±0.54 | 226 | 7.6±4.0 | 227 | ⋯ | ⋯ |
| | 10.6±1.8 | 90 | ⋯ | ⋯ | ⋯ | ⋯ |
| | 9.4±2.2 | 115 | ⋯ | ⋯ | ⋯ | ⋯ |
| | 10.33±0.36 | 157 | ⋯ | ⋯ | ⋯ | ⋯ |
| | 8.89 | 62 | ⋯ | ⋯ | ⋯ | ⋯ |
| | 10.0 | 213 | ⋯ | ⋯ | ⋯ | ⋯ |
| | 7.6 | 227 | ⋯ | ⋯ | ⋯ | ⋯ |
| | **9.5±0.3** | **Adopted** | **5.97±0.27** | **Adopted** | **1.62±0.28** | **Adopted** |

**SM–Table 3.** *Continued.*

| **Identifier** | **$v\sin i_*$ (km s$^{-1}$)** | **Reference** | **$P_{\rm rot}$ (d)** | **Reference** | **$R_*$ ($R_\odot$)** | **Reference** |
|---|---|---|---|---|---|---|
| DN Tau | 12.3±0.6 | 78 | 6.38±0.29 | TESS (This Work) | 3.55 | 61 |
| | 10.9±2.1 | 115 | 6.0 | 228 | ··· | ··· |
| | 10.8±1.8 | 90 | 6.6±0.1 | 218 | ··· | ··· |
| | 9.8±0.7 | 90 | 6.2 | 215 | ··· | ··· |
| | 11.01±0.21 | 157 | 6.32 | 121 | ··· | ··· |
| | 9.8±0.9 | 229 | 6.33±0.05 | 123 | ··· | ··· |
| | 8.1±2.0 | 228 | 6.25±0.1 | 123 | ··· | ··· |
| | 9.0±1.0 | 123 | 6.22±0.1 | 123 | ··· | ··· |
| | ··· | ··· | 6.17±0.1 | 123 | ··· | ··· |
| | **10.7±0.3** | **Adopted** | **6.31±0.21** | **Adopted** | **3.55±0.62** | **Adopted** |
| DoAr 25 | 15.9±0.5 | 157 | 9.02±0.53 | K2 (This Work) | 2.51±0.16 | 60 |
| | **15.9±0.8** | **Adopted** | **9.02±0.67** | **Adopted** | **2.51±0.20** | **Adopted** |
| DoAr 44 | 15.7±0.4 | 223 | 2.96±0.02 | 134 | 1.58 | 61 |
| | 17.0±1.1 | 134 | 3.4** | 230 | ··· | ··· |
| | 16.7±1.1 | 157 | ··· | ··· | ··· | ··· |
| | 17.5±1.0 | 132 | ··· | ··· | ··· | ··· |
| | **16.3±0.5** | **Adopted** | **2.96±0.05** | **Adopted** | **1.58±0.28** | **Adopted** |
| Elias 2-24 | 13.6±2.9 | 136 | 6.58±0.25 | K2§ (This Work) | 3.98 | 93 |
| | 15.78 | 116 | ··· | ··· | 4.2 | 201 |
| | 17.2 | 231 | ··· | ··· | ··· | ··· |
| | **16.2±0.9** | **Adopted** | **6.58±0.4** | **Adopted** | **4.09±0.20** | **Adopted** |
| EX Lup | 7.5±1.9 | 86 | 7.51±0.78 | TESS (This Work) | 1.42 | 61 |
| | 4.4±2.0 | 232 | 7.4 | 233 | ··· | ··· |
| | ··· | ··· | 7.0 | 234 | ··· | ··· |
| | ··· | ··· | 7.2 | 234 | ··· | ··· |
| | ··· | ··· | 7.417±0.001 | 235 | ··· | ··· |
| | ··· | ··· | 7.42±0.02 | 146 | ··· | ··· |
| | **6.0±1.4** | **Adopted** | **7.42±0.28** | **Adopted** | **1.42±0.25** | **Adopted** |

**SM–Table 3.** *Continued.*

| Identifier | $v\sin i_*$ (km s$^{-1}$) | Reference | $P_{\rm rot}$ (d) | Reference | $R_*$ ($R_\odot$) | Reference |
|---|---|---|---|---|---|---|
| FP Tau | 32±2 | 78 | 2.2±0.04 | TESS (This Work) | 1.38 | 61 |
| | 31±4.2 | 115 | 2.19 | 219 | ··· | ··· |
| | 33.47±0.64 | 157 | 2.19 | 236 | ··· | ··· |
| | 27.9±4.2 | 211 | ··· | ··· | ··· | ··· |
| | 32.28 | 62 | ··· | ··· | ··· | ··· |
| | 33.14 | 62 | ··· | ··· | ··· | ··· |
| | 26.6±1.6 | 213 | ··· | ··· | ··· | ··· |
| | **31.0±0.8** | **Adopted** | **2.19±0.03** | **Adopted** | **1.38±0.20** | **Adopted** |
| FT Tau | 10.2±0.56 | 68 | 6.03±0.21 | TESS (This Work) | 1.5±0.09 | 60 |
| | 11.47** | 62 | 5.24** | 219 | ··· | ··· |
| | **10.2±0.6** | **Adopted** | **6.03±0.28** | **Adopted** | **1.5±0.11** | **Adopted** |
| GM Aur | 14.8±0.9 | 78 | 6.0±0.21 | TESS (This Work) | 1.49 | 61 |
| | 14.4±2.0 | 115 | 5.3±0.2 | 220 | ··· | ··· |
| | 12.4±1.8 | 213 | 6.04±0.15 | 149 | ··· | ··· |
| | 12.59±1.02 | 226 | 6.02 | 121 | ··· | ··· |
| | 14.9±0.3 | 149 | 6.1 | 215 | ··· | ··· |
| | 13.7±1.7 | 90 | 5.94±0.11 | 149 | ··· | ··· |
| | 13.5±0.8 | 90 | 6.03±0.09 | 149 | ··· | ··· |
| | 14.06** | 157 | ··· | ··· | ··· | ··· |
| | 12.8±2.3 | 212 | ··· | ··· | ··· | ··· |
| | **14.0±0.4** | **Adopted** | **6.0±0.19** | **Adopted** | **1.49±0.26** | **Adopted** |
| GW Lup | < 8.0 | 86 | 4.12±0.3 | TESS (This Work) | 1.35 | 61 |
| | **< 8.0** | **Adopted** | **4.12±0.31** | **Adopted** | **1.35±0.24** | **Adopted** |
| IM Lup | 13.4±0.4 | 223 | 7.45±0.77 | TESS (This Work) | 2.24 | 61 |
| | 17.1±1.4 | 86 | 7.31±0.18 | 207 | ··· | ··· |
| | 15±2 | 237 | 7.42** | 238 | ··· | ··· |
| | 15.98±0.40 | 239 | 7.25±0.03 | 239 | ··· | ··· |
| | ··· | ··· | 7.38±0.4 | 239 | ··· | ··· |
| | **14.8±0.5** | **Adopted** | **7.25±0.27** | **Adopted** | **2.24±0.39** | **Adopted** |

**SM–Table 3.** *Continued.*

| Identifier | $v\sin i_*$ (km s$^{-1}$) | Reference | $P_{\rm rot}$ (d) | Reference | $R_*$ ($R_\odot$) | Reference |
|---|---|---|---|---|---|---|
| IP Tau | 12.3±0.8 | 78 | 5.67±0.15 | TESS (This Work) | 1.12±0.14 | 61 |
| | 11.5±2.5 | 115 | 5.59 | 219 | ··· | ··· |
| | 9.73±0.86 | 226 | 5.7 | 240 | ··· | ··· |
| | 10.3±1.8 | 90 | ··· | ··· | ··· | ··· |
| | 10.8±1.1 | 90 | ··· | ··· | ··· | ··· |
| | 10.6±0.4 | 90 | ··· | ··· | ··· | ··· |
| | 12.47** | 62 | ··· | ··· | ··· | ··· |
| | 12.67±0.81 | 157 | ··· | ··· | ··· | ··· |
| | **11.1±0.3** | **Adopted** | **5.65±0.18** | **Adopted** | **1.12±0.16** | **Adopted** |
| IQ Tau | 14.4±0.3 | 78 | 6.65±0.21 | K2 (This Work) | 2.25±0.14 | 60 |
| | 11.5±1.0 | 213 | 6.25±0.05 | 241 | ··· | ··· |
| | 13.5±1.8 | 211 | 6.9 | 219 | ··· | ··· |
| | 14.3±2.5 | 115 | 6.44 | 242 | ··· | ··· |
| | 12.5±2.0 | 220 | 5.89 | 240 | ··· | ··· |
| | 12.4** | 212 | ··· | ··· | ··· | ··· |
| | 15.59±0.36 | 157 | ··· | ··· | ··· | ··· |
| | **14.1±0.4** | **Adopted** | **6.29±0.21** | **Adopted** | **2.25±0.18** | **Adopted** |
| LkCa 15 | 13.9±1.2 | 78 | 5.78±0.16 | K2 (This Work) | 1.61 | 61 |
| | 13.6±0.8 | 90 | 5.73±0.23 | TESS (This Work) | ··· | ··· |
| | 12.5±1.3 | 213 | 5.85 | 243 | ··· | ··· |
| | 14.6±1.8 | 90 | 5.76 | 244 | ··· | ··· |
| | 15.4±2.3 | 115 | 5.86 | 227 | ··· | ··· |
| | 16.6±0.2 | 245 | 5.7±0.05 | 159 | ··· | ··· |
| | 16.82±0.15 | 246 | 5.55±0.71 | 159 | ··· | ··· |
| | 13.82±0.5 | 159 | 5.76±0.74 | 159 | ··· | ··· |
| | ··· | ··· | 5.63±0.66 | 159 | ··· | ··· |
| | ··· | ··· | 5.41±0.64 | 159 | ··· | ··· |
| | ··· | ··· | 5.56±0.65 | 159 | ··· | ··· |
| | **14.8±0.4** | **Adopted** | **5.76±0.18** | **Adopted** | **1.61±0.28** | **Adopted** |
| MHO 6 | 20.801 | 62 | 1.92±0.02 | K2 (This Work) | 0.99±0.06 | 60 |
| | 21.062 | 62 | 1.96±0.02 | TESS (This Work) | ··· | ··· |
| | ··· | ··· | 1.94** | 242 | ··· | ··· |
| | **20.9±1.1** | **Adopted** | **1.94±0.02** | **Adopted** | **0.99±0.08** | **Adopted** |

**SM–Table 3.** *Continued.*

| Identifier | $v\sin i_*$ (km s$^{-1}$) | Reference | $P_{\rm rot}$ (d) | Reference | $R_*$ ($R_\odot$) | Reference |
|---|---|---|---|---|---|---|
| MP Mus | 14.25±0.74 | 161 | 3.72±0.09 | TESS (This Work) | 1.53 | 61 |
| | 13.0±0.4 | 223 | 3.75** | 164 | ··· | ··· |
| | 14.2±1.42 | 164 | ··· | ··· | ··· | ··· |
| | 13.96±0.69 | 161 | ··· | ··· | ··· | ··· |
| | 14.2±0.8 | 91 | ··· | ··· | ··· | ··· |
| | **13.8±0.3** | **Adopted** | **3.72±0.12** | **Adopted** | **1.53±0.27** | **Adopted** |
| MY Lup | 29.1±2.0 | 86 | 2.6±0.1 | TESS (This Work) | 1.66 | 61 |
| | ··· | ··· | 2.5** | 238 | ··· | ··· |
| | **29.1±2.0** | **Adopted** | **2.6±0.11** | **Adopted** | **1.66±0.29** | **Adopted** |
| PDS 111 | 33** | 166 | 3.15±0.08 | TESS (This Work) | 1.88±0.11 | 61 |
| | 21±0.3 | 52 | ··· | ··· | ··· | ··· |
| | **21.0±1.1** | **Adopted** | **3.15±0.17** | **Adopted** | **1.88±0.17** | **Adopted** |
| PDS 70 | 17.27±0.1 | 246 | 3.02±0.12 | TESS (This Work) | 1.19±0.11 | 61 |
| | 16.0±0.5 | 247 | 3.007±0.001 | 248 | ··· | ··· |
| | 17.16±0.16 | 245 | ··· | ··· | ··· | ··· |
| | **16.8±0.5** | **Adopted** | **3.01±0.05** | **Adopted** | **1.19±0.14** | **Adopted** |
| RU Lup | 8.5±4.8 | 86 | 3.97±0.21 | TESS (This Work) | 1.8 | 61 |
| | 9.0±0.9 | 170 | 3.7±0.2 | 249 | ··· | ··· |
| | 8.63 | 239 | 3.65±0.05 | 250 | ··· | ··· |
| | 7.33 | 239 | 3.71±0.1 | 251 | ··· | ··· |
| | **8.5±0.6** | **Adopted** | **3.68±0.08** | **Adopted** | **1.80±0.31** | **Adopted** |
| RX J1615 | 13±2 | 252 | 5.79±0.26 | TESS † (This Work) | 1.52 | 61 |
| | 17±2 | 252 | 5.73±0.14 | 207 | ··· | ··· |
| | **15.0±1.4** | **Adopted** | **5.74±0.21** | **Adopted** | **1.52±0.27** | **Adopted** |
| RX J1842 | 26.7±2.9 | 253 | 2.56±0.1 | TESS † (This Work) | 1.32±0.12 | 61 |
| | **26.7±2.9** | **Adopted** | **2.56±0.13** | **Adopted** | **1.32±0.15** | **Adopted** |
| RX J1852 | 19.5±0.5 | 254 | 2.82 | 208 | 1.2±0.1 | 61 |
| | 23.05±3.59 | 253 | ··· | ··· | ··· | ··· |
| | **19.7±0.9** | **Adopted** | **2.82±0.57** | **Adopted** | **1.20±0.13** | **Adopted** |

**SM–Table 3.** *Continued.*

| Identifier | $v\sin i_*$ (km s$^{-1}$) | Reference | $P_{\rm rot}$ (d) | Reference | $R_*$ ($R_\odot$) | Reference |
|---|---|---|---|---|---|---|
| RY Lup | 16.3±5.3 | 86 | 3.63±0.21 | TESS (This Work) | 1.63 | 61 |
| | 25±4.6 | 255 | 3.8 | 255 | ⋯ | ⋯ |
| | 22.4±3.0 | 211 | 3.75 | 215 | ⋯ | ⋯ |
| | ⋯ | ⋯ | 3.62 | 255 | ⋯ | ⋯ |
| | ⋯ | ⋯ | 3.74 | 256 | ⋯ | ⋯ |
| | **21.9±2.3** | **Adopted** | **3.72±0.1** | **Adopted** | **1.63±0.28** | **Adopted** |
| SU Aur | 59±2 | 78 | 2.68±0.08 | TESS (This Work) | 3.45 | 61 |
| | 59±1 | 257 | 2.7 | 258 | ⋯ | ⋯ |
| | 66.2±4.6 | 212 | 2.98±0.4 | 177 | ⋯ | ⋯ |
| | 66.1±15.8 | 255 | 2.78 | 227 | ⋯ | ⋯ |
| | 69.0 | 259 | ⋯ | ⋯ | ⋯ | ⋯ |
| | 65.0 | 260 | ⋯ | ⋯ | ⋯ | ⋯ |
| | 61±2 | 261 | ⋯ | ⋯ | ⋯ | ⋯ |
| | 59.5±2.0 | 261 | ⋯ | ⋯ | ⋯ | ⋯ |
| | **61.3±1.6** | **Adopted** | **2.72±0.06** | **Adopted** | **3.45±0.60** | **Adopted** |
| SY Cha | 12.7±2.9 | 179 | 6.13±0.12 | TESS † (This Work) | 1.63 | 61 |
| | 12.4±0.4 | 78 | 6.13 | 208 | ⋯ | ⋯ |
| | 16.9±2.0 | 211 | 6.1 | 255 | ⋯ | ⋯ |
| | 15.4±0.4 | 231 | 6.13 | 262 | ⋯ | ⋯ |
| | 13.5±0.9 | 137 | 6.07 | 238 | ⋯ | ⋯ |
| | 15.2±3.0 | 211 | 5.97 | 238 | ⋯ | ⋯ |
| | ⋯ | ⋯ | 6.2 | 263 | ⋯ | ⋯ |
| | **13.6±0.4** | **Adopted** | **6.11±0.21** | **Adopted** | **1.63±0.28** | **Adopted** |
| Sz 114 | < 8.0 | 86 | 2.0±0.02 | 207 | 1.48 | 61 |
| | **< 8.0** | **Adopted** | **2.00±0.03** | **Adopted** | **1.48±0.26** | **Adopted** |
| Sz 73 | 31±3 | 86 | 1.92±0.05 | TESS (This Work) | 1.4±0.3 | 83 |
| | **31.0±3.0** | **Adopted** | **1.92±0.05** | **Adopted** | **1.4±0.3** | **Adopted** |
| Sz 90 | <8.0 | 86 | 6.69±0.6 | TESS† (This Work) | 1.06±0.14 | 61 |
| | **8.0±0.0** | **Adopted** | **6.69±0.64** | **Adopted** | **1.06±0.16** | **Adopted** |

**SM–Table 3.** *Continued.*

| Identifier | $v\sin i_*$ (km s$^{-1}$) | Reference | $P_{rot}$ (d) | Reference | $R_*$ ($R_\odot$) | Reference |
|---|---|---|---|---|---|---|
| T Cha | 39±3 | 164 | 3.31±0.1 | TESS (This Work) | 1.8 | 187 |
| | 40±2.7 | 223 | 3.2±0.2 | 264 | 1.3 | 188 |
| | 33.99±0.61 | 161 | 3.2 | 263 | 2.18 | 189 |
| | 37±2 | 187 | 3.24 | 265 | 2.56±0.13 | 266 |
| | 48±10 | 267 | ⋯ | ⋯ | 1.4 | 268 |
| | 45±2.88 | 269 | ⋯ | ⋯ | ⋯ | ⋯ |
| | **37.7±1.0** | **Adopted** | **3.23±0.06** | **Adopted** | **2.48±0.31** | **Adopted** |
| TW Hya | 4±1 | 270 | 2.2 | 271 | 1.05±0.11 | 61 |
| | 5.0 | 272 | 3.56 | 273 | ⋯ | ⋯ |
| | 6.2±1.3 | 223 | 2.8±0.04 | 274 | ⋯ | ⋯ |
| | 6±1.2 | 91 | 3.7±0.1 | 275 | ⋯ | ⋯ |
| | 6.2 | 192 | 3.61±0.01 | 194 | ⋯ | ⋯ |
| | 5±2 | 276 | ⋯ | ⋯ | ⋯ | ⋯ |
| | 8.4±2 | 136 | ⋯ | ⋯ | ⋯ | ⋯ |
| | 4.6±0.7 | 277 | ⋯ | ⋯ | ⋯ | ⋯ |
| | 7±3 | 278 | ⋯ | ⋯ | ⋯ | ⋯ |
| | 3±1 | 194 | ⋯ | ⋯ | ⋯ | ⋯ |
| | **4.9±0.4** | **Adopted** | **3.56±0.07** | **Adopted** | **1.05±0.13** | **Adopted** |
| V1094 Sco | 21.8 | 252 | 3.53±0.11 | TESS† (This Work) | 1.72 | 279 |
| | 20.0 | 252 | 3.55** | 280 | 1.38 | 198 |
| | 22.4±1.9 | 280 | ⋯ | ⋯ | 1.9 | 195 |
| | ⋯ | ⋯ | ⋯ | ⋯ | 2.63±0.63 | 83 |
| | ⋯ | ⋯ | ⋯ | ⋯ | ⋯ | ⋯ |
| | **21.4±1.1** | **Adopted** | **3.53±0.13** | **Adopted** | **1.81±0.61** | **Adopted** |
| V836 Tau | 13.4±1.1 | 281 | 6.67±0.32 | TESS (This Work) | 1.42 | 61 |
| | 10.51±0.77 | 226 | 7.0 | 243 | ⋯ | ⋯ |
| | 11.8±2.4 | 115 | 6.77 | 242 | ⋯ | ⋯ |
| | 12.1±1.8 | 90 | 6.76±0.04 | 282 | ⋯ | ⋯ |
| | 12.5±0.9 | 90 | 6.8±0.1 | 90 | ⋯ | ⋯ |
| | 12.1±1.0 | 229 | 7.0 | 210 | ⋯ | ⋯ |
| | **11.9±0.4** | **Adopted** | **6.77±0.24** | **Adopted** | **1.42±0.25** | **Adopted** |

**SM–Table 3.** *Continued.*

| Identifier | $v\sin i_*$ (km s$^{-1}$) | Reference | $P_{rot}$ (d) | Reference | $R_*$ ($R_\odot$) | Reference |
|---|---|---|---|---|---|---|
| WSB 52 | 18.2±2.9 | 136 | 6.24±0.28 | K2 (This Work) | 1.86 | 93 |
| | 16.9±0.5 | 200 | ··· | ··· | 1.9 | 201 |
| | 18.21±0.42 | 157 | ··· | ··· | 1.71 | 97 |
| | 18.17** | 62 | ··· | ··· | ··· | ··· |
| | 22±1.9 | 283 | ··· | ··· | ··· | ··· |
| | **17.9±0.6** | **Adopted** | **6.24±0.35** | **Adopted** | **1.82±0.10** | **Adopted** |
| WSB 63 | 9.65** | 116 | 12.31±1.04 | K2 (This Work) | 1.88±0.11 | 60 |
| | 10.4±2.1 | 136 | ··· | ··· | ··· | ··· |
| | **10.4±2.1** | **Adopted** | **12.31±1.28** | **Adopted** | **1.88±0.14** | **Adopted** |

† Reduced with the MIT Quick-Look Pipeline.

‡ Reduced with the EPIC Variability Extraction and Removal for Exoplanet Science Targets pipeline.

§ Reduced with the K2 Systematics Correction pipeline.

## SM–Figure 1: Probability Distributions of $i_{\rm disk}$, $i_*$, and $\Delta i$

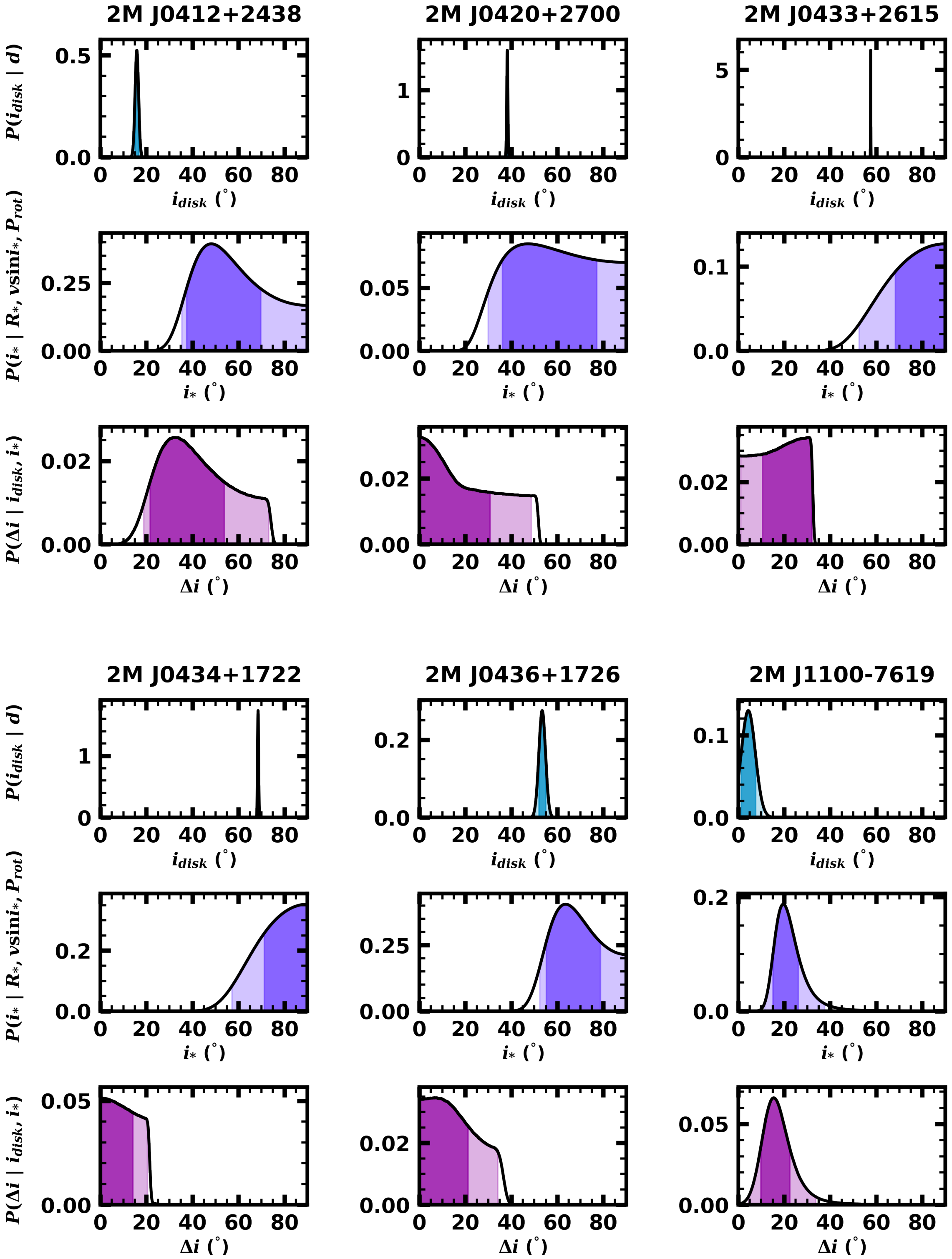

**SM–Figure 1.** A gallery of the posterior distributions of $i_{\rm disk}$, $i_*$, and Δi for each object in the sample. Distributions that correspond to a single object are grouped into a 3-panel column below the respective object's identifier (caption continued on next page).

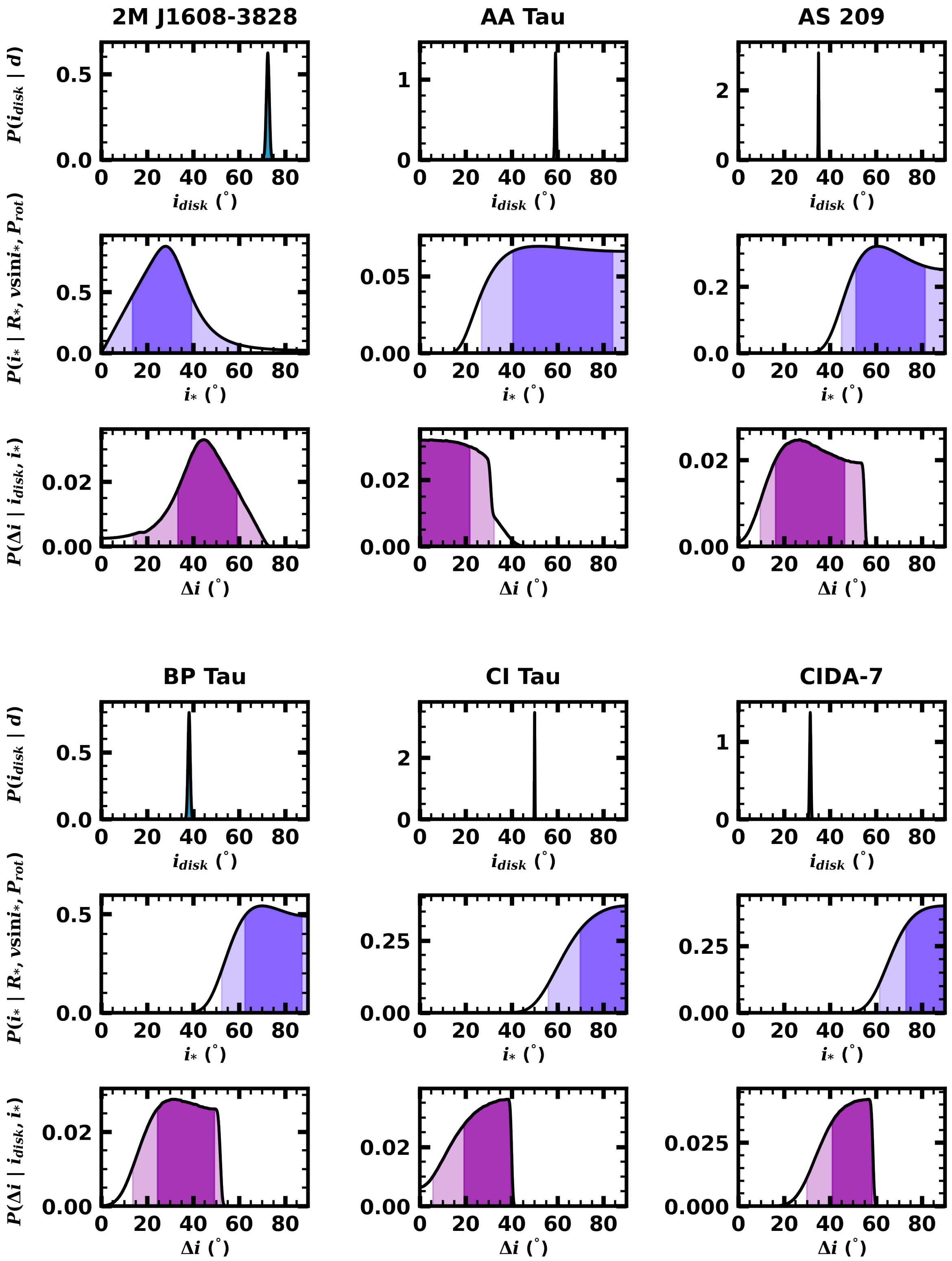

**SM–Figure 1.** *Continued*. For each object, from top to bottom, we display the probability distribution of $i_{disk}$ (blue), the posterior distribution of $i_*$ (purple), and the resulting distribution of Δi (magenta). The shaded darker and lighter regions under the solid curves indicate the 1σ and 2σ highest density intervals, respectively.

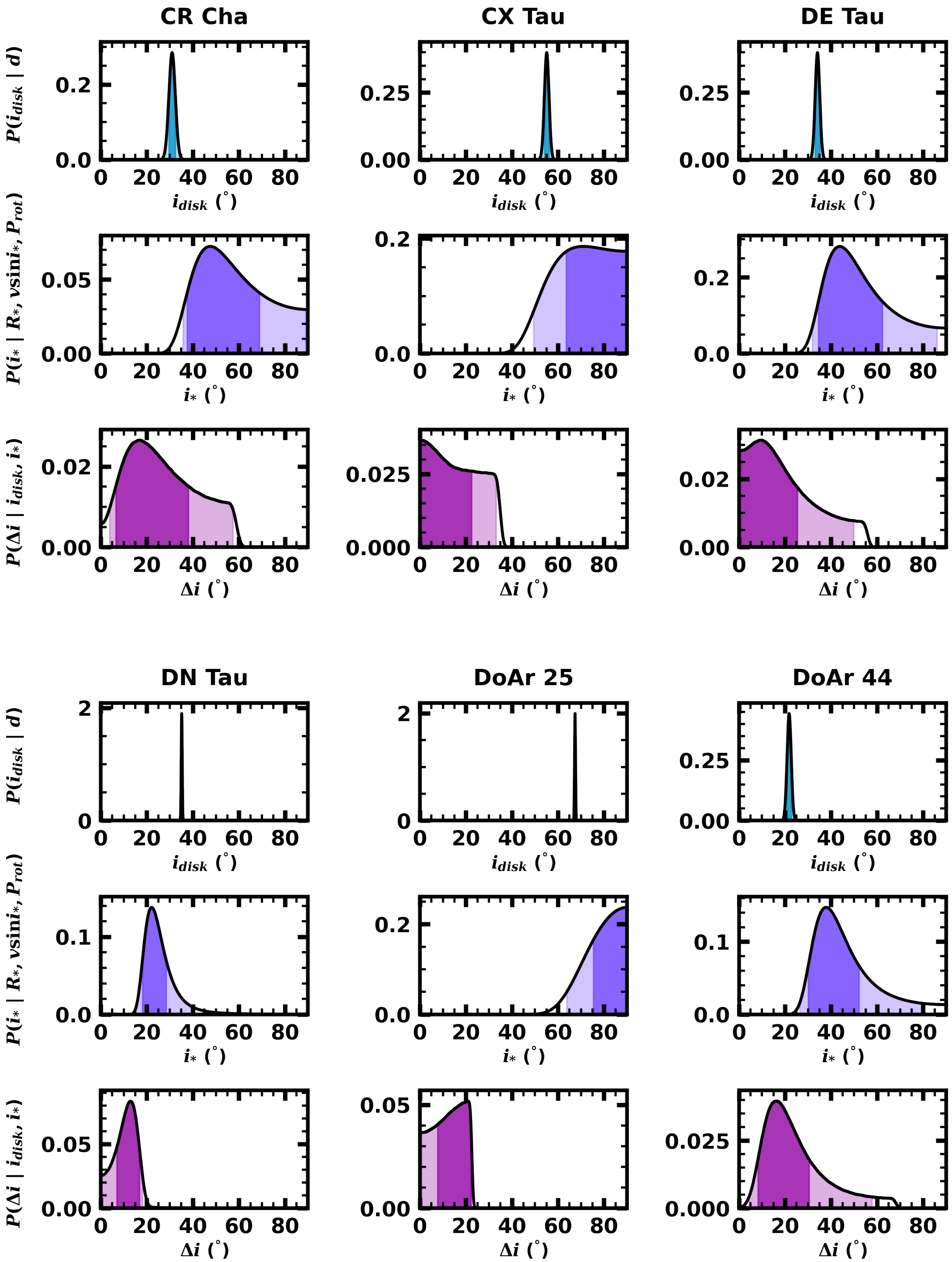

**SM–Figure 1.** *Continued.*

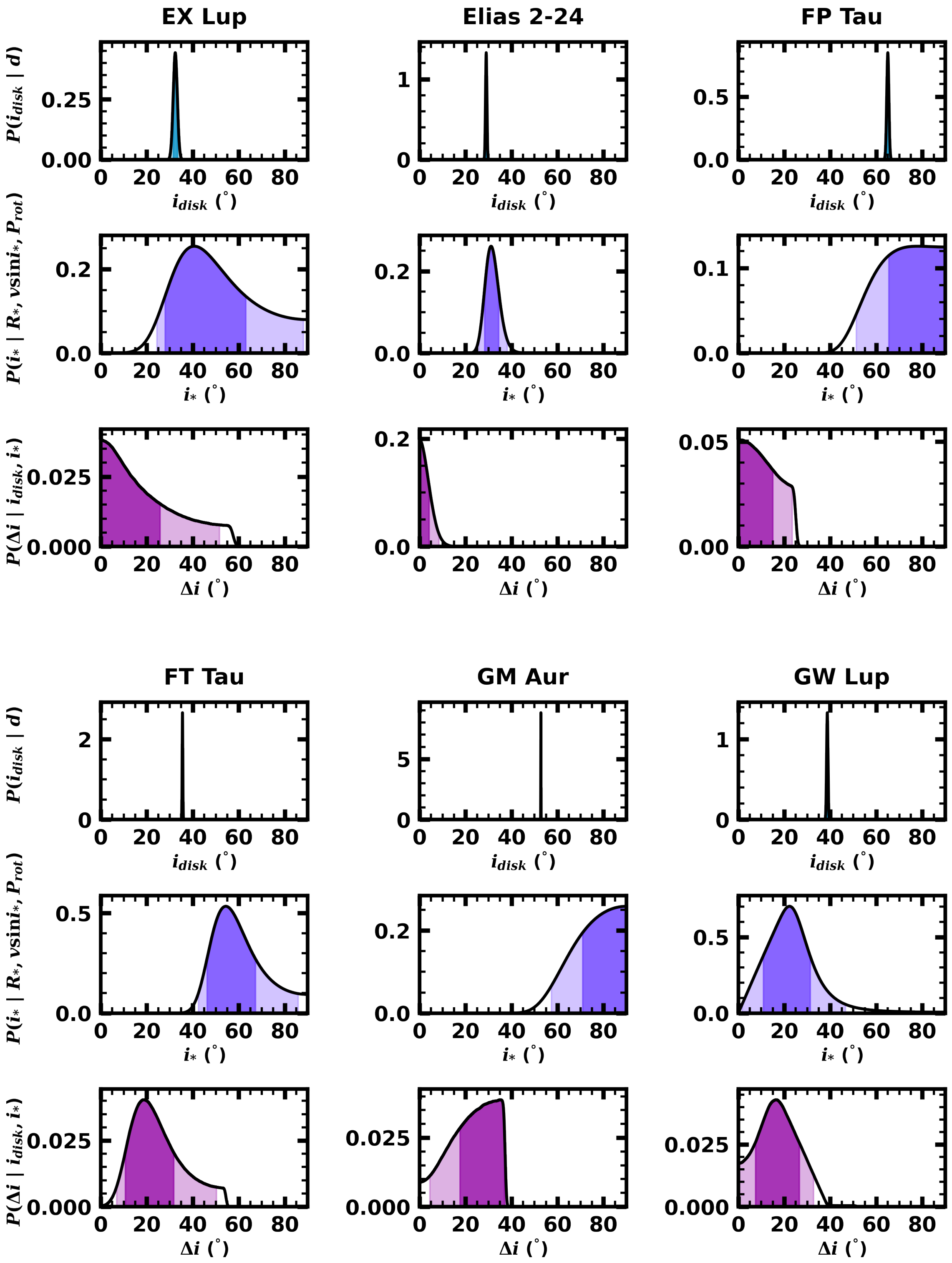

**SM–Figure 1.** *Continued.*

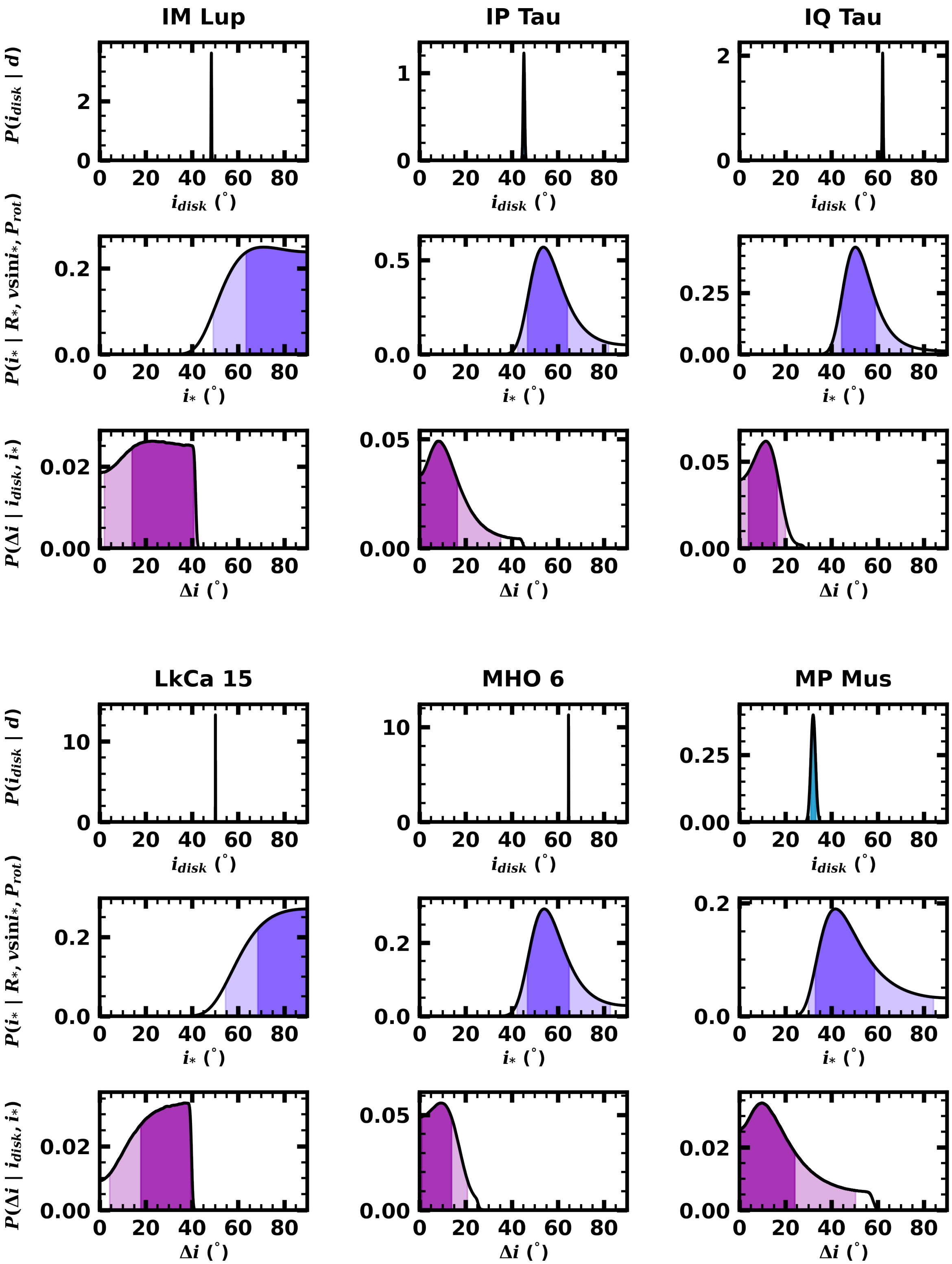

**SM–Figure 1.** *Continued.*

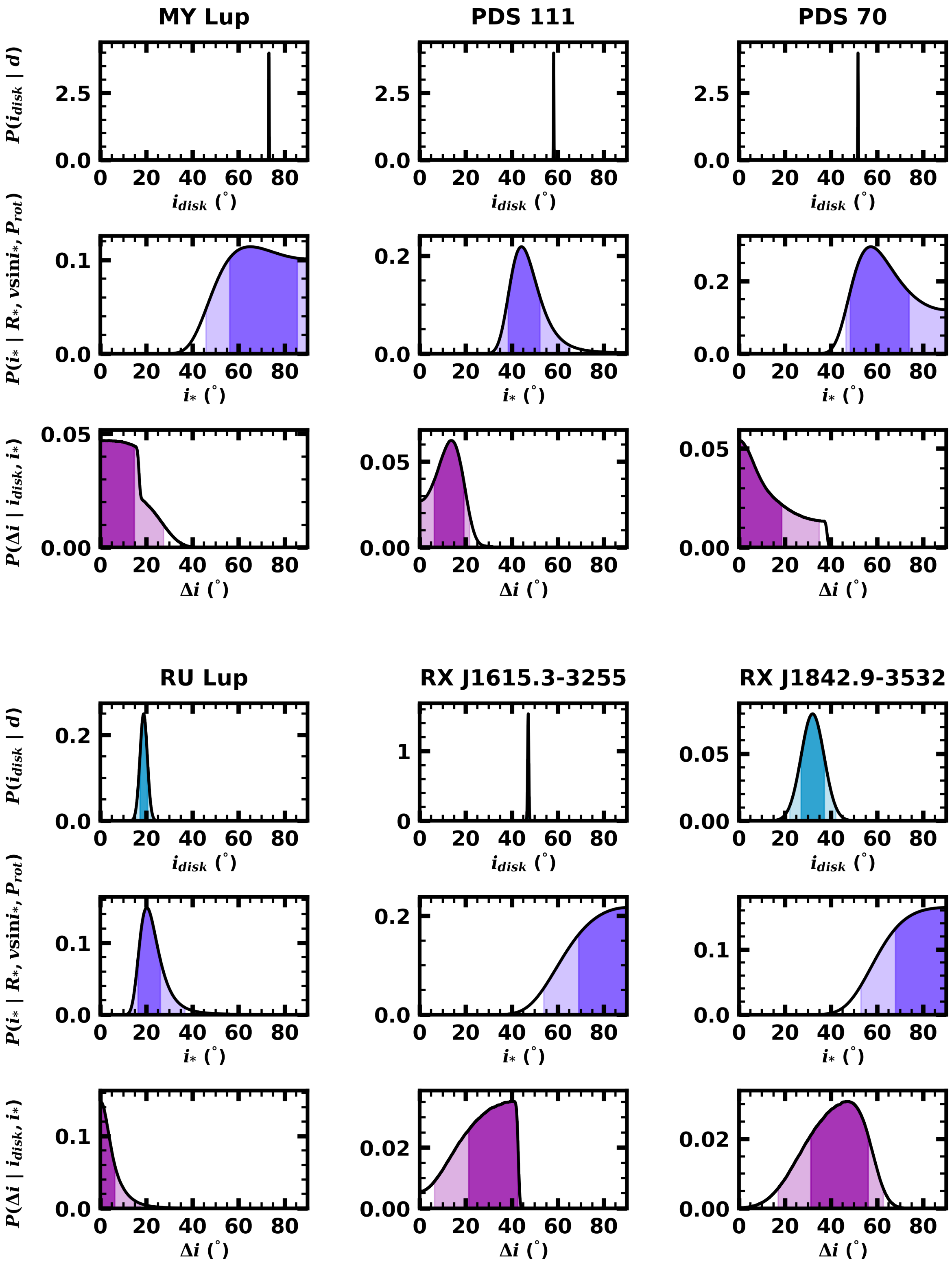

**SM–Figure 1.** *Continued.*

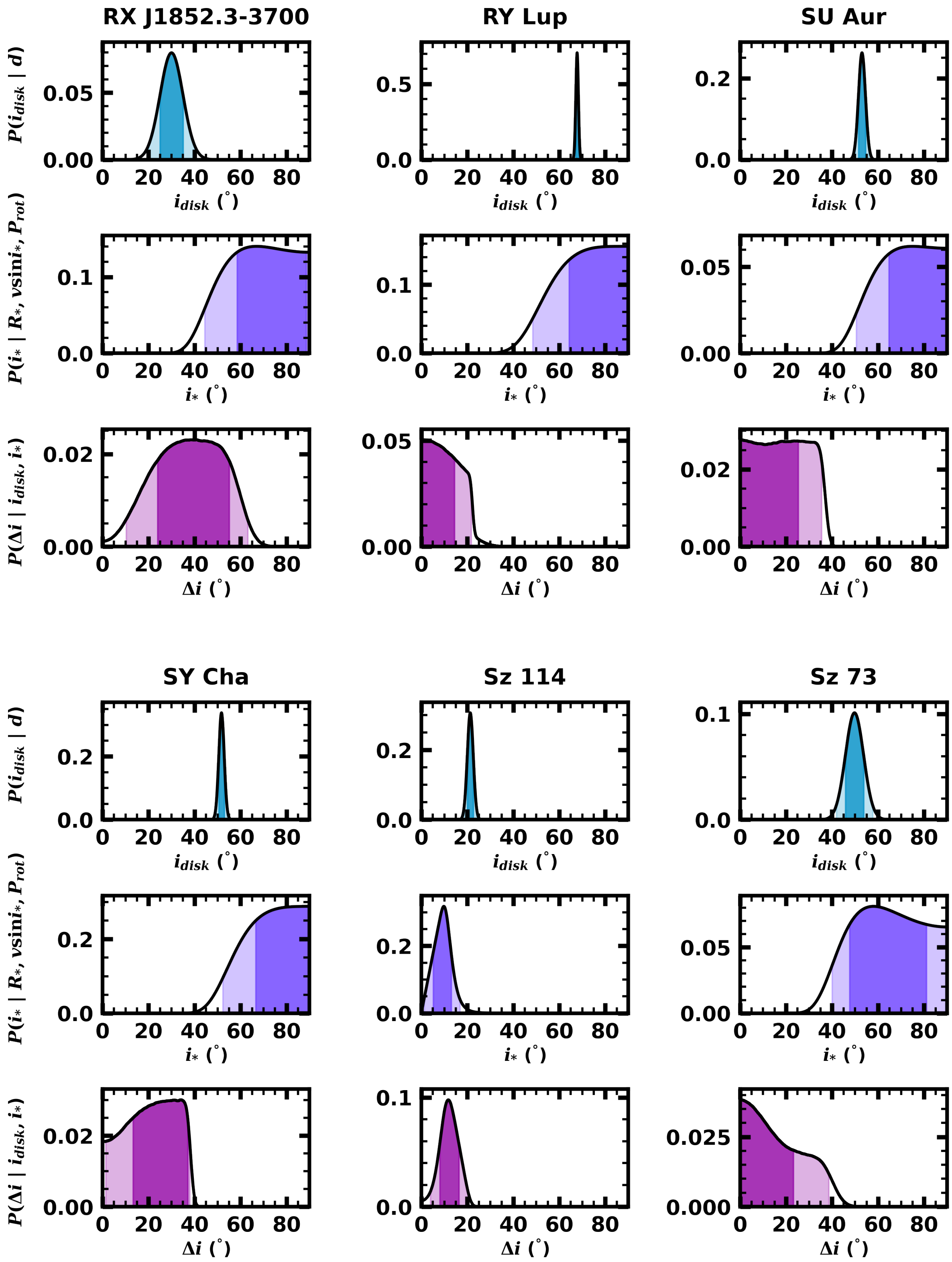

**SM–Figure 1.** *Continued.*

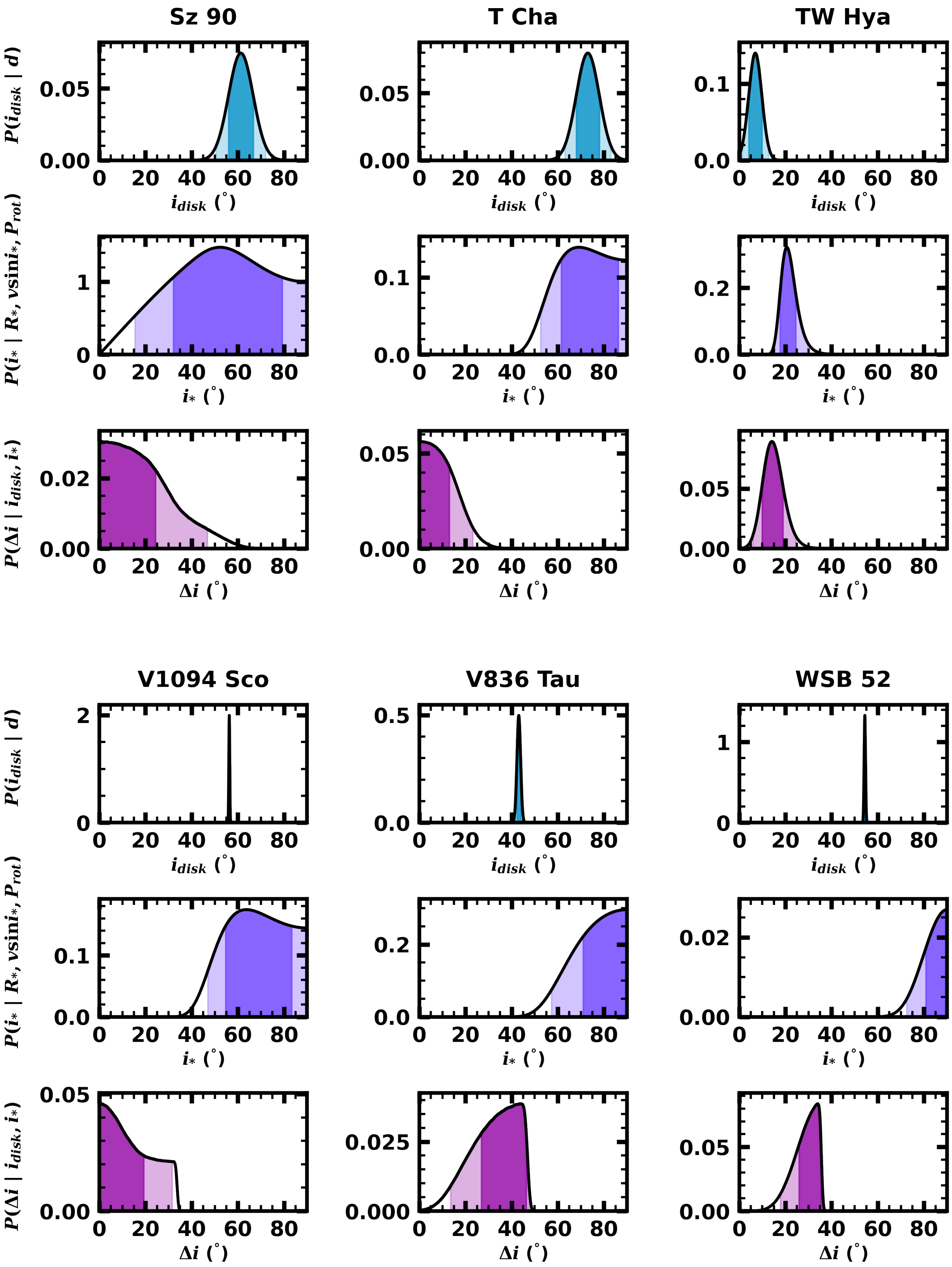

**SM–Figure 1.** *Continued.*

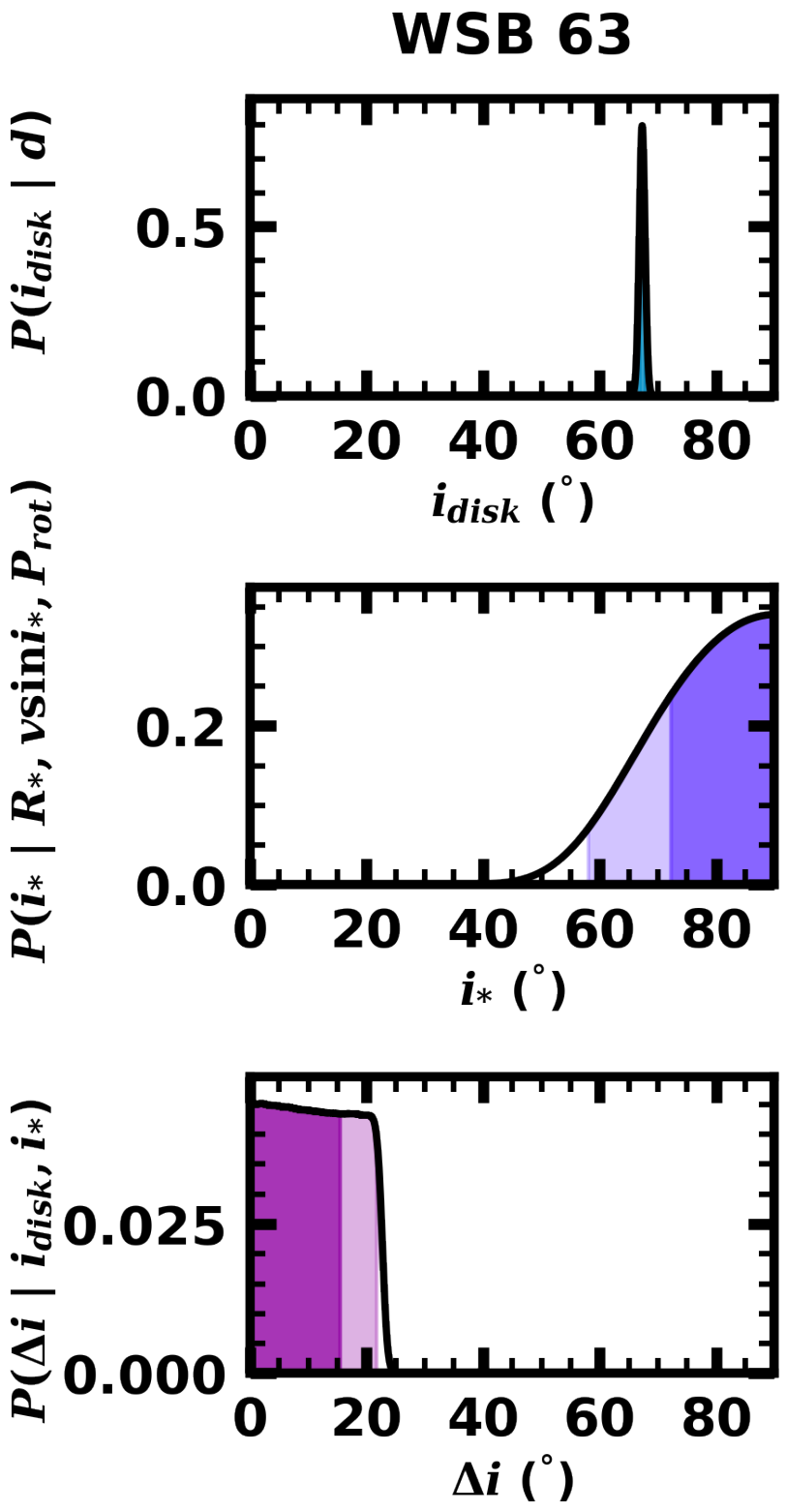

**SM–Figure 1.** *Continued.*

# References for Supplementary Information